\documentclass[ twocolumn,
aps,prd,   
               preprintnumbers,numbers,sort&compress,nofootinbib,
                            showpacs,
               colorlinks,
               linkcolor=blue,   
               citecolor=blue]{revtex4-1}
    \newcommand{\exclude}[1]{}

\usepackage{graphicx,amsmath,amssymb,bm}
\usepackage{psfrag}
\usepackage{feynmp}
\usepackage{hyperref}
\usepackage{enumitem}

\def\<{\langle}
\def\>{\rangle}
\def\d{\partial}
\def\+{\dagger}

\def\U1A{U(1)$_{\rm A}$}

 \def\<{\langle}
\def\>{\rangle}
\def\d{\partial}
\def\+{\dagger}

\def\ra{\rangle}
\def\la{\langle}
\def\U1A{U(1)$_{\rm A}$}

\newcommand{\be}{\begin{eqnarray}}
\newcommand{\ee}{\end{eqnarray}}
\newcommand{\beq}{\begin{equation}}
\newcommand{\eeq}{\end{equation}}

\begin{document}
 
\title{ Axion field and the 
   quark nugget's formation    at the QCD phase transition
}

\author{ Xunyu Liang}
\author{Ariel Zhitnitsky}
\affiliation{Department of Physics and Astronomy, University of
  British Columbia, Vancouver,  Canada}


\begin{abstract}
We study a   testable 
dark matter (DM) model outside of the standard WIMP paradigm in which the 
observed ratio $\Omega_{\rm dark} \simeq   \Omega_{\rm visible}$ for 
visible and dark matter 
densities  finds its natural explanation  as a result
of their common QCD origin when both types of matter (DM and visible) are formed at the QCD phase transition and both are proportional to $\Lambda_{\rm QCD}$. 
Instead of conventional ``baryogenesis" mechanism we advocate a paradigm when the  
``baryogenesis'' is actually a charge separation process which always occur in the presence of the $\cal{CP}$ odd axion field $a(x)$.
In this scenario  the global baryon number of the Universe remains 
zero, while  the unobserved anti-baryon charge is hidden in form of heavy nuggets, similar to Witten's strangelets and compromise   
the DM of the Universe.  

In the present work we study  in great details a possible  formation mechanism of  such macroscopically large heavy objects. 
We argue that the nuggets will be inevitably  produced   during the QCD phase transition as a result of Kibble-Zurek  mechanism 
  on formation of the topological defects during a phase transition. Relevant topological defects in our scenario are   the closed bubbles made of the $N_{\rm DW}=1$ axion domain walls. These bubbles,    in general,   accrete the baryon (or anti baryon) charge,  which eventually result in  formation of the nuggets and anti-nuggets carrying a huge baryon (anti-baryon) charge. A typical  size and the baryon charge  of these macroscopically large objects is mainly  determined  by the axion mass $m_a$. 
 However,    the main consequence of the model,  $\Omega_{\rm dark} \approx   \Omega_{\rm visible}$ is insensitive to   the axion mass which may  assume any value   within the observationally allowed window $10^{-6} {\rm eV} \lesssim  m_a \lesssim  10^{-3}{\rm eV}$.  We also estimate the baryon to entropy ratio $\eta\equiv   {n_B}/{n_{\gamma}}\sim 10^{-10}$ within this scenario. Finally, we  comment on implications of these results to  the  axion search experiments, including microwave cavity  and the Orpheus experiments.

\end{abstract}
\vspace{0.1in}


\maketitle
\section{Introduction}\label{introduction}
The origin of the observed asymmetry between matter and 
antimatter is one of the largest open questions 
in cosmology. The nature of the  dark matter (DM) is another open question in cosmology.
In this paper  we advocate an idea  that these two,   apparently unrelated, problems are in fact two sides of the same coin.
Furthermore, both mysterious effects  are originated at one and the same cosmological epoch  from one and the same QCD physics. Normally, it is assumed that the majority of dark matter is represented by  
a new fundamental field coupled only weakly to the 
standard model particles, these models may then be tuned 
to match the observed dark matter properties. We take a different 
perspective and consider the possibility that the dark matter 
is in fact composed of well known quarks and antiquarks but in 
a new high density phase, similar to the Witten's strangelets, see original work \cite{Witten:1984rs} and some related studies \cite{old}.

There are few  new crucial elements in  proposal  \cite{Zhitnitsky:2002qa,Oaknin:2003uv}, 
 in comparison with previous studies   \cite{Witten:1984rs, old}. First of all,    the nuggets could be made of matter as well as antimatter in our framework
 as a result of separation of charges, see few comments below. Secondly,  
the stability of the DM nuggets is provided by the axion domain walls with extra pressure, in contrast with original studies when stability is assumed to be achieved even  in vacuum, at zero external pressure.
Finally, an overall coherent baryon asymmetry in the entire Universe is  a result of the strong CP violation due to the fundamental $\theta$ parameter in QCD which is assumed to be nonzero at the beginning of the QCD phase transition.  This source of strong CP violation is no longer 
available at the present epoch as a result of the axion dynamics, see original papers \cite{axion,KSVZ,DFSZ} and 
 recent reviews \cite{vanBibber:2006rb, Sikivie:2008, Asztalos:2006kz,Raffelt:2006cw,Sikivie:2009fv,Rybka:2014cya,Rosenberg:2015kxa,Graham:2015ouw} on the subject. We highlight the basics ideas  of this framework in the present Introduction, while  we elaborate on these   new crucial elements  in details in section
  \ref{formation}.

 It is generally assumed that the universe 
began in a symmetric state with zero global baryonic charge 
and later (through some baryon number violating process) 
evolved into a state with a net positive baryon number. As an 
alternative to this scenario we advocate a model in which 
``baryogenesis'' is actually a charge separation process 
in which the global baryon number of the universe remains 
zero. In this model the unobserved antibaryons come to comprise 
the dark matter.  A connection between dark matter and 
baryogenesis is made particularly compelling by the 
similar energy densities of the visible and dark matter 
with $\Omega_{\rm dark} \simeq 5\cdot \Omega_{\rm visible}$. If these processes 
are not fundamentally related the two components could 
exist at vastly different scales. 

In the model  \cite{Zhitnitsky:2002qa,Oaknin:2003uv} baryogenesis occurs at 
the QCD phase transition. Both quarks and antiquarks are 
thermally abundant in the primordial plasma but, in 
addition to forming conventional baryons, some fraction 
of them are bound into heavy nuggets of quark matter in a 
colour superconducting phase. Nuggets of both matter and 
antimatter are formed as a result of the dynamics of the axion domain walls as originally proposed   in refs.\cite{Zhitnitsky:2002qa,Oaknin:2003uv}. 
A number of very hard dynamical questions in strongly coupled QCD which are related  to the nuggets's formation have  not been studied in any details in the original papers. The main goal of the present work   is to make the first step in the direction to address these hard questions. 
   
If fundamental $\theta$ parameter were identically zero    at the QCD phase transition in early universe, an equal number of  nuggets made of matter and antimatter would be formed.  It  would result in vanishing of the visible baryon density at the  present epoch. However, the  fundamental CP violating processes associated 
with the $\theta$ term in QCD (which  is assumed to be  small, but still non-zero at the very beginning of the 
QCD phase transition) result in the preferential formation of 
anti-nuggets over the nuggets. This preference  is essentially determined by the dynamics of  coherent axion field   ${\theta}(x)$ at the initial stage of the nugget's formation.  The resulting asymmetry is not sensitive to a small magnitude of the axion field ${\theta}(x)$ at the QCD phase transition as long as it remains  coherent on the scale of the Universe, see section \ref{asymmetry} for the details.  

The remaining antibaryons in the plasma then 
annihilate away leaving only the baryons whose antimatter 
counterparts are bound in the excess of anti-nuggets and thus 
unavailable to annihilate. All asymmetry effects   are order of one, irrespectively to the magnitude of $\theta(x)$ at the moment of formation. This  is precisely  the main reason of why the  visible and dark matter 
densities must be the same order of magnitude 
\be
\label{Omega}
 \Omega_{\rm dark} \approx \Omega_{\rm visible}
\ee
   as they both  proportional to the same fundamental $\Lambda_{\rm QCD} $ scale,  and they both are originated at the same  QCD epoch.  In particular, if one assumes that the nuggets and anti-nuggets saturate the dark matter density than 
  the observed 
matter to dark matter ratio $\Omega_{\rm dark} \simeq 5\cdot \Omega_{\rm visible}$ corresponds to a specific proportion  when  number  of anti-nuggets is larger than number of nuggets 
  by a factor of $\sim$ 3/2 at the end of nugget's formation. This would 
result in a matter content with baryons, quark nuggets 
and antiquark nuggets in an approximate  ratio 
\be
\label{ratio1}
|B_{\rm visible}|: |B_{\rm nuggets}|: |B_{\rm antinuggets}|\simeq 1:2:3, 
\ee
 with  no net baryonic charge. 
 If these processes 
are not fundamentally related the two components $\Omega_{\rm dark}$ and $\Omega_{\rm visible}$  could easily 
exist at vastly different scales.

Though the QCD phase diagram at $\theta\neq 0$ as a function of $T$ and $\mu$  is basically unknown, it is well understood that $\theta$ is in fact the angular variable, and therefore supports various types of the domain walls, including the so-called $ N_{DW}=1$ domain walls when $\theta$ interpolates between one and the same physical vacuum state $\theta\rightarrow\theta+2\pi$. Furthermore, it is expected that the closed bubbles made of these $ N_{DW}=1$ axion domain walls are also produced during the QCD phase transition
with a typical wall tension $\sigma_a\sim m_{a}^{-1}$ where $m_a $ is the axion mass.  Precisely this scale determines the size and the baryon charge of the nuggets, see (\ref{sigma}), (\ref{Baryon}) below. 

The collapse of these close bubbles is halted due to the Fermi pressure acting  inside of the bubbles.  The crucial  element which stops the collapse of the bubbles from complete annihilation is the presence of the QCD substructure inside the axion domain wall.  This substructure forms  immediately after the QCD phase transition  as discussed in \cite{Zhitnitsky:2002qa}.  
The equilibrium of the obtained system has been analyzed in \cite{Zhitnitsky:2002qa} for a specific axion domain wall  tension 
within the observationally allowed window $10^{-6} {\rm eV}\leq m_a\leq  10^{-3}{\rm eV}$ consistent with the recent constraints  \cite{vanBibber:2006rb, Sikivie:2008, Asztalos:2006kz,Raffelt:2006cw,Sikivie:2009fv,Rybka:2014cya,Rosenberg:2015kxa,Graham:2015ouw}.
It has been also argued in \cite{Zhitnitsky:2002qa}  that the equilibrium is typically  achieved when the Fermi pressure inside the nuggets falls into the region when the colour superconductivity (CS) indeed  sets in\footnote{\label{N=1}There is no  requirement on the first order phase transition (in contrast with  original proposal \cite{Witten:1984rs}) for the bubble formation in this framework  because the $ N_{DW}=1$ axion  domain walls are formed  irrespectively to the order of the phase transition.  Needless to say that the  phase diagram in general and the order of the phase transition in particular at $\theta\neq 0$ are  still unknown because of the  longstanding   ``sign problem" in the QCD lattice simulations at $\theta\neq 0$, see few comments and related references in Conclusion.}.

The  size and the baryon charge of the nuggets  scale  with the axion mass as follows
\be
\label{sigma}
\sigma_a\sim m_a^{-1}, ~~~R\sim \sigma_a, ~~~~B\sim \sigma_a^3.
\ee
  Therefore,  when   the axion mass 
$m_a$ varies within the observationally allowed window $10^{-6} {\rm eV} \lesssim  m_a \lesssim  10^{-3}{\rm eV}$ the   nuggets parameters also vary as follows
\be
\label{Baryon}
 10^{-6} {\rm cm}\lesssim R \lesssim 10^{-3}{\rm cm}, ~~~ 10^{23}\lesssim B\lesssim 10^{32},
\ee
where the lowest axion mass $m_a\simeq 10^{-6} {\rm eV}$ approximately\footnote{There is no one-to-one correspondence between 
the axion mass $m_a$ and the baryon charge of the nuggets $ B$ because for each given $m_a$ there is an extended  window of 
stable solutions describing different nugget's sizes  \cite{Zhitnitsky:2002qa}.}  corresponds to the largest possible nuggets with
$\la B\ra \simeq  10^{32}$. Variation of the axion mass by three orders of magnitude results in variation of the nugget's baryon charge by nine orders of magnitude according to relation (\ref{sigma}). 
The corresponding allowed  region  is essentially  unconstrained by present  experiments, see details in section \ref{nuggets} below.

The fact that the CS may be realized in nature in the cores of neutron stars has been known for sometime \cite{Alford:2007xm,Rajagopal:2000wf}. A new element which  was  advocated in proposal \cite{Zhitnitsky:2002qa}    is that a similar dense environment can be realized in nature due to  the axion domain wall pressure    playing  a role of a ``squeezer", similar to the gravity pressure   in the neutron star physics.

Another fundamental ratio (along with $\Omega_{\rm dark} \approx  \Omega_{\rm visible}$   discussed above)
is the baryon to entropy ratio at present time
\be
\label{eta}
\eta\equiv\frac{n_B-n_{\bar{B}}}{n_{\gamma}}\simeq \frac{n_B}{n_{\gamma}}\sim 10^{-10}.
\ee
If the nuggets were not present after the phase transition the conventional baryons 
and anti-baryons would continue to annihilate each other until the temperature 
reaches $T\simeq 22$ MeV when density would be 9 orders of magnitude smaller 
than observed (\ref{eta}). This annihilation catastrophe, normally thought   to be  resolved   as a result of  ``baryogenesis" as formulated by Sakharov\cite{Sakharov}, see also review \cite{Dolgov:1997qr}. In this framework  the ratio (\ref{eta}) is highly sensitive to many  specific details of the models such as  the  spectrum of the system in general and the coupling constants and the  strength of CP violation in particular, see e.g. review\cite{Dolgov:1997qr}.

\begin{figure}
\centering
\includegraphics[width=0.8\linewidth]{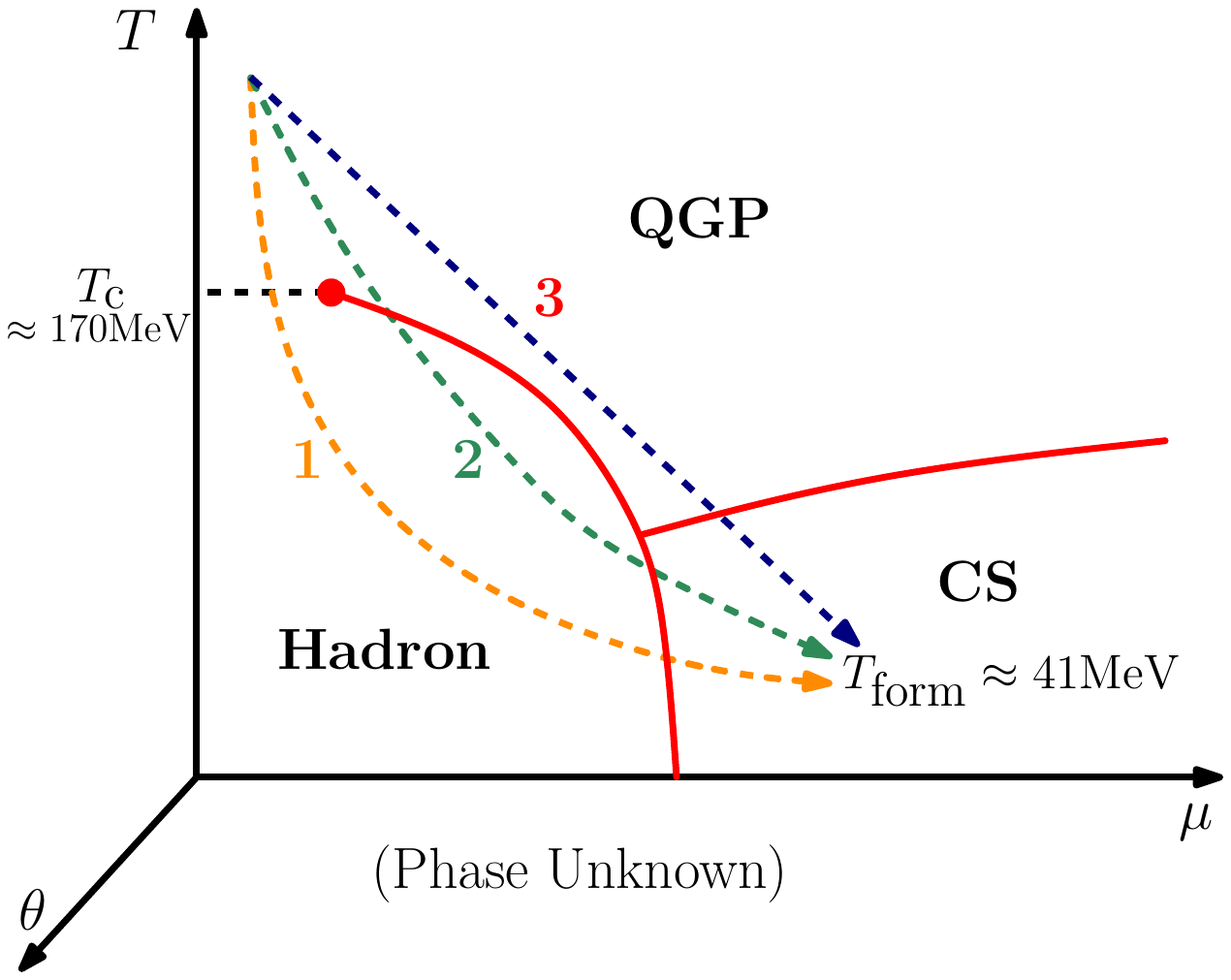}
\caption{ The conjectured phase diagram. Possible cooling paths are denoted as path 1, 2 or  3. The phase diagram is in fact much more complicated as the dependence on the third essential parameter, the $\theta$ is not shown as it is largely unknown. Therefore, the paths should be thought as lines  in three dimensional parametrical space, not as lines on two-dimensional $(\mu , T)$  slice  at $\theta=0$ as shown on the present plot.  It is assumed that the final destination after the nuggets are formed is the region with $T_{\rm form}\approx 41$ MeV, $\mu >\mu_c$ and $\theta\approx 0$, corresponding to the presently observed ratio (\ref{eta}), see text for the details.  }
\label{fig:phase_diagram}
\end{figure}

In our proposal (in contrast with conventional frameworks on baryogenesis) this ratio is determined by a single parameter with a typical  QCD scale, the formation temperature $T_{\rm form}$. This temperature is defined by a moment in  evolution of the Universe  when the nuggets and anti-nuggets basically have completed  their formation and not much annihilation would occur at lower  temperatures $T \leq T_{\rm form}$.     
The exact magnitude of  temperature $T_{\rm form}\sim \Lambda_{\rm QCD}$ in our proposal is determined by many factors: transmission/reflection coefficients, evolution of the nuggets, expansion of the universe, cooling rates, evaporation rates, viscosity of the environment, the  dynamics of the axion domain wall network, etc. All these effects are, in general, equally contribute  to $T_{\rm form}$ at the QCD scale. Technically, the corresponding effects are hard    to compute from the first principles as even basic properties of the  QCD phase diagram at nonzero $\theta\neq 0$ are still unknown\footnote{\label{theta}The basic consequence (\ref{Omega}) as well as (\ref{eta}) of this proposal  are largely insensitive to the   absolute value
of   initial magnitude of the $\theta$ parameter. In other words,     a fine tuning of the initial  $\theta$ parameter  is not required in this mechanism.  The same comment (on ``insensitivity" of the initial conditions) also applies to efficiency of the  nugget's formation. This is because the baryon density  at the present time is 10 orders of magnitude lower than the  particle density at the QCD phase transition epoch according to the observations (\ref{eta}). Therefore, even a sufficiently low efficiency of the nugget's formation (still larger than $10^{-7}$, see estimates in section \ref{asymmetry2})  cannot drastically  modify the generic  relations  (\ref{Omega}), (\ref{eta})  due to  a long evolution  which  eventually washes out any sensitivity to the initial conditions. The only  crucial parameter  which  determines the final outcome (\ref{Omega}),  (\ref{eta})   is the formation temperature $T_{\rm form}$ as  estimated below.}. We plot three different conjectured cooling paths on Fig. \ref{fig:phase_diagram}.

However, the estimate of $T_{\rm form}$ up to factor 2 is quite a simple exercise as  $T_{\rm form}$  must be  proportional to the gap $\Delta\sim 100$ MeV when colour superconducting  (CS) phase   sets in  inside the nuggets.  The observed ratio (\ref{eta}) corresponds to $T_{\rm form}\simeq 40$ MeV, see \cite{Oaknin:2003uv} for the  details.
This temperature   indeed represents a  typical QCD scale,  slightly below the critical temperature $T_{CS}\simeq 0.6 \Delta\simeq 60$ MeV,   according to  standard  estimates  on colour superconductivity, see  reviews  \cite{Alford:2007xm,Rajagopal:2000wf}.

 Unlike conventional dark matter candidates, such as WIMPs (Weakly interacting Massive Particles) the dark-matter/antimatter
nuggets are strongly interacting but macroscopically large.  
They do not contradict any of the many known observational
constraints on dark matter or
antimatter  for three main reasons~\cite{Zhitnitsky:2006vt}:
\begin{itemize} 
\item They carry a huge (anti)baryon charge 
$|B|  \gtrsim 10^{25}$, and so have an extremely tiny number
density; 
\item The nuggets have nuclear densities, so their effective interaction
is small $\sigma/M \sim 10^{-10}$ ~cm$^2$/g,  well below the typical astrophysical
and cosmological limits which are on the order of 
$\sigma/M<1$~cm$^2$/g;
\item They have a large binding energy $\sim \Delta$, 
such that baryon charge  in the
nuggets is not available to participate in big bang nucleosynthesis
(\textsc{bbn}) at $T \approx 1$~MeV. 
\end{itemize} 
To reiterate: the
  weakness of the visible-dark matter 
interaction is achieved 
in this model due to the small geometrical parameter $\sigma/M \sim B^{-1/3}$ 
rather than due to a weak coupling 
of a new fundamental field with standard model particles. In other words, this small effective interaction $\sim \sigma/M \sim B^{-1/3}$ replaces a conventional requirement
of sufficiently weak interactions of the visible matter with  WIMPs. 

 As we already mentioned, this  model when  DM is represented by quark and antiquark nuggets is consistent with fundamental  astrophysical  constraints as highlighted above.   Furthermore, there is    a number of frequency bands where some excess  of emission was  observed, but  not  explained by conventional astrophysical sources. Our comment here is that this  model  may explain some portion, or even entire excess of the observed  radiation in these frequency bands. This phenomenological part of the proposal  is the key ingredient in our advocacy of the model, and may play 
very important role for interpretation of  the present and future observations.  Therefore, we devote  next section \ref{nuggets}  to review   the original results 
\cite{Oaknin:2004mn, Zhitnitsky:2006tu,Forbes:2006ba, Lawson:2007kp,Forbes:2008uf,Forbes:2009wg,Lawson:2010uz,Lawson:2012vk,Lawson:2012zu,Lawson:2013bya}
 where   predictions of the model have been confronted with  the observations in specific frequency bands covering more than eleven orders of magnitude, from radio frequency with $\omega\sim 10^{-4} $ eV to $\gamma$ rays with $\omega\sim 10$ MeV.    
 We also mention   in  section  \ref{nuggets}  some interesting  results   \cite{Abers:2007ji, Gorham:2012hy,Gorham:2015rfa,Lawson:2015xsq,Lawson:2015cla}  which  presently perfectly consistent with the model.  
 However, in future,    the same       studies   with modest improvements 
 will provide a powerful test of the viability of the quark nugget dark matter model. 
 
One should emphasize here  that the corresponding analysis \cite{Oaknin:2004mn, Zhitnitsky:2006tu,Forbes:2006ba, Lawson:2007kp,Forbes:2008uf,Forbes:2009wg,Lawson:2010uz,Lawson:2012vk,Lawson:2012zu,Lawson:2013bya} is determined by conventional physics, and as such all effects are calculable from the first principles. In other words, the model contains no tuneable fundamental parameters, except for a single mean baryon number of a nugget $\la B\ra\sim 10^{25}$ which enters all the computations \cite{Oaknin:2004mn, Zhitnitsky:2006tu,Forbes:2006ba, Lawson:2007kp,Forbes:2008uf,Forbes:2009wg,Lawson:2010uz,Lawson:2012vk,Lawson:2012zu,Lawson:2013bya} as a single normalization factor.  At the same time, the crucial assumptions  of the model,  such as  specific mechanisms on the baryon charge separation and  dynamics of the nugget formation, etc, have never been  explored  in our previous studies. 

We believe that the  phenomenological success \cite{Oaknin:2004mn, Zhitnitsky:2006tu,Forbes:2006ba, Lawson:2007kp,Forbes:2008uf,Forbes:2009wg,Lawson:2010uz,Lawson:2012vk,Lawson:2012zu,Lawson:2013bya}  of the model warrants further theoretical studies  of this framework, in spite of its naively counter-intuitive nature.
 Therefore, the present work should be considered as the first step in this direction where we attempt to develop     the theoretical framework to address (and hopefully answer)   some  of the   hardest questions  about a possible mechanism for the nugget's formation during the QCD phase transition in  strongly  coupled  regime when even the phase diagram at $\theta\neq 0$ as a function of the chemical potential $\mu$ and temperature $T$ is still unknown, see footnote \ref{N=1}. 
 
 The structure of this work is as follows. In section \ref{nuggets} we briefly review the observational constraints on the model.  In   section  \ref{formation} we highlight the basic assumptions and ingredients of this framework, while in  sections  \ref{baryons} and  \ref{time-evolution} we present some  analytical estimates  which strongly  substantiate   the idea that such heavy objects  indeed    can be   formed  and   survive until the present epoch during the QCD phase transition in  early Universe. Section  \ref{analysis} as well as Appendices \ref{appendix:flux} and  \ref{numerics}  are  devoted to a number of technical details which  support our basic claim. 
 
 In section \ref{asymmetry}  we argue  that there will be  
 the preferential formation of  one species of nuggets over another. This preference  is    determined by the 
 dynamics of the  axion field ${\theta}(x)$ which itself is correlated on the   scales of the Universe at the beginning of the nugget's formation.    Finally, in section  \ref{axion-search}  we   comment on implications of our studies  to direct  axion search experiments.
 
To conclude this long introduction: the nuggets  in our framework play the {\it dual} role: they  serve as the DM candidates and they  also explain the observed asymmetry between matter and antimatter. These two crucial elements of the proposal  lead to very generic consequence of the entire  framework expressed by eq. (\ref{Omega}).   This basic generic result is not very sensitive to any specific details of the model, but rather, entirely determined by two fundamental ingredients of the framework: \\$\bullet$  the contribution to $\Omega$ for both types of matter (visible and dark) are proportional to one and the same fundamental scale $\sim \Lambda_{\rm QCD}$;  \\$\bullet$ the preferential formation of  one species of nuggets over another is correlated on huge cosmological scales where $\cal{CP}$ violating axion phase $\theta(x)$  remains   coherent just a moment before  the QCD phase transition. 
 
 The readers interested in the cosmological consequences, rather than in technical computational details may directly jump to section \ref{formation}   where we formulate the basics ingredients of the proposal,  to section \ref{generic} where we explain  the main model-independent consequence (\ref{Omega}) of  this framework, and to section \ref{axion-search} where we make few comments on implications to other axion search experiments, including   microwave cavity \cite{vanBibber:2006rb, Sikivie:2008, Asztalos:2006kz,Rosenberg:2015kxa} and the Orpheus experiments \cite{Rybka:2014cya}.

\section{ Quark (anti) nugget DM    confronting the observations}\label{nuggets}

 While the observable consequences of this model are on average strongly suppressed  
by the low number density  of the quark nuggets $\sim B^{-1/3}$ as explained above, the interaction of these objects 
with the visible matter of the galaxy will necessarily produce observable 
effects. Any such consequences will be largest where the densities 
of both visible and dark matter are largest such as in the 
core of the galaxy or the early universe. In other words, the nuggets behave as  a conventional cold DM in the environment where density of the visible matter is small, while they become interacting and emitting radiation  objects (i.e. effectively become visible matter)  when they are placed in the environment with sufficiently large density.

 The relevant phenomenological features of the resulting nuggets 
 are determined by properties of the so-called electro-sphere  as discussed in original refs. \cite{Oaknin:2004mn, Zhitnitsky:2006tu,Forbes:2006ba, Lawson:2007kp,Forbes:2008uf,Forbes:2009wg,Lawson:2010uz,Lawson:2012vk,Lawson:2012zu,Lawson:2013bya}. 
 These properties are in principle, calculable from first principles using only 
the well established and known properties of QCD and QED. As such 
the model contains no tunable fundamental parameters, except for a single mean baryon number  $\la B\ra$
which itself is determined by the axion mass $m_a$  as we already mentioned.

 A comparison between   emissions with drastically different frequencies 
from the centre  of galaxy is possible because the rate of annihilation events (between visible matter and antimatter DM nuggets) is proportional to 
 the product of the local visible and DM distributions at the annihilation site. 
The observed fluxes for different emissions thus depend through one and the same line-of-sight integral 
\be
\label{flux1}
\Phi \sim R^2\int d\Omega dl [n_{\rm visible}(l)\cdot n_{DM}(l)],
\ee
where $R\sim B^{1/3}$ is a typical size of the nugget which determines the effective cross section of interaction between DM and visible matter. As $n_{DM}\sim B^{-1}$ the effective interaction is strongly suppressed $\sim B^{-1/3}$ as we already mentioned  in the Introduction. The parameter $\la B\ra\sim 10^{25}$  was fixed in this  proposal by assuming that this mechanism  saturates the observed  511 keV line   \cite{Oaknin:2004mn, Zhitnitsky:2006tu}, which resulted from annihilation of the electrons from visible matter and positrons from anti-nuggets. It has been also assumed that the observed dark matter density is saturated by the nuggets and anti-nuggets.  It corresponds to an average baryon charge  $\la B\ra \sim 10^{25}$ for typical  density distributions  $n_{\rm visible}(r),  n_{DM}(r)$ entering (\ref{flux1}). Other emissions from different bands  are expressed in terms of the same integral (\ref{flux1}), and therefore, the  relative\textsc{} intensities  are completely determined by internal structure of the nuggets which is described by conventional nuclear physics and basic QED. We present a short overview of these results below.  

Some  galactic electrons are able to penetrate to a sufficiently 
large depth of the anti-nuggets. These events
 no longer produce  the characteristic positronium decay 
spectrum (511 keV line  with a typical width of order $\sim {\rm few~ keV}$ 
accompanied by the conventional continuum due to $3\gamma$ decay)   but a direct non-resonance $e^-e^+ \rightarrow 2\gamma$ emission spectrum. 
 The transition between the resonance positronium decays and non-resonance 
  regime is determined by conventional physics and allows us to  compute   
the strength and spectrum of the MeV scale emissions relative to 
that of the 511~keV line \cite{Lawson:2007kp,Forbes:2009wg}.  Observations 
by the \textsc{Comptel} satellite indeed show some  excess above the galactic 
background  consistent with our estimates. 

Galactic protons incident on the anti-nugget 
will penetrate some distance into the quark matter before 
annihilating into hadronic jets. This process results in the emission of Bremsstrahlung 
photons at x-ray energies \cite{Forbes:2006ba}. Observations by the 
\textsc{Chandra} observatory apparently  indicate an excess in x-ray emissions from 
the galactic centre.  

Hadronic jets produced 
deeper in the nugget or emitted in the downward direction 
will be completely absorbed. They  eventually emit 
thermal photons with radio frequencies \cite{Forbes:2008uf,Lawson:2012zu}. Again the relative scales of these 
 emissions may be estimated and is found to be in 
agreement with observations.  

 These apparent excess emission sources have been cited 
as possible support for a number of dark matter models 
as well as other exotic astrophysical phenomenon. At present 
however they remain open matters for investigation and, given 
the uncertainties in the galactic spectrum and the wide 
variety of proposed explanations are unlikely to provide 
clear evidence in the near future. Therefore, it would be highly desirable if some   direct detection   of such objects
is found, similar to direct searches of the weakly interacting massive particles (WIMPs). 

While direct searches for WIMPs 
require large sensitivity, a search for very massive  dark 
matter nuggets requires large area detectors. If the dark matter 
consists of quark nuggets at the $B\sim 10^{25}$ scale they 
will have a flux of
\begin{equation}
\label{eq:flux}
\frac{dN}{dA ~ dt} = nv \approx \left( \frac{10^{25}}{B} \right) {\rm km}^{-2} {\rm yr}^{-1}. 
\end{equation}
Though
this flux is far below the sensitivity of conventional dark 
matter searches it is similar to the flux of cosmic rays 
near the GZK limit. As such present and future 
experiments investigating ultrahigh energy cosmic rays 
may also serve as search platforms for dark matter of this type.

It has been suggested  that large scale 
cosmic ray detectors 
may be capable of observing quark (anti-) nuggets passing through the earth's
atmosphere either through the extensive air shower such an event 
would trigger \cite{Lawson:2010uz} or through the geosynchrotron 
emission generated by the large number of secondary particles
\cite{Lawson:2012vk}, see also \cite{Lawson:2013bya} for review. 

It has also been estimated in \cite{Abers:2007ji} that, based on Apollo data, 
nuggets of mass from $\sim$ 10 kg to 1 ton (corresponding to 
$B \sim 10^{28\text{-}30}$) must account for less than an order of
magnitude of the local dark matter.  While our preferred range of
$B\sim 10^{25}$  is somewhat smaller and is not excluded 
by \cite{Abers:2007ji}, we still believe that $B\geq 10^{28}$ is not
completely excluded by the Apollo data, as the corresponding
constraints are based on specific model dependent assumptions about
the nugget mass-distribution. 

\begin{figure}
\centering
\includegraphics[width=1.0\linewidth]{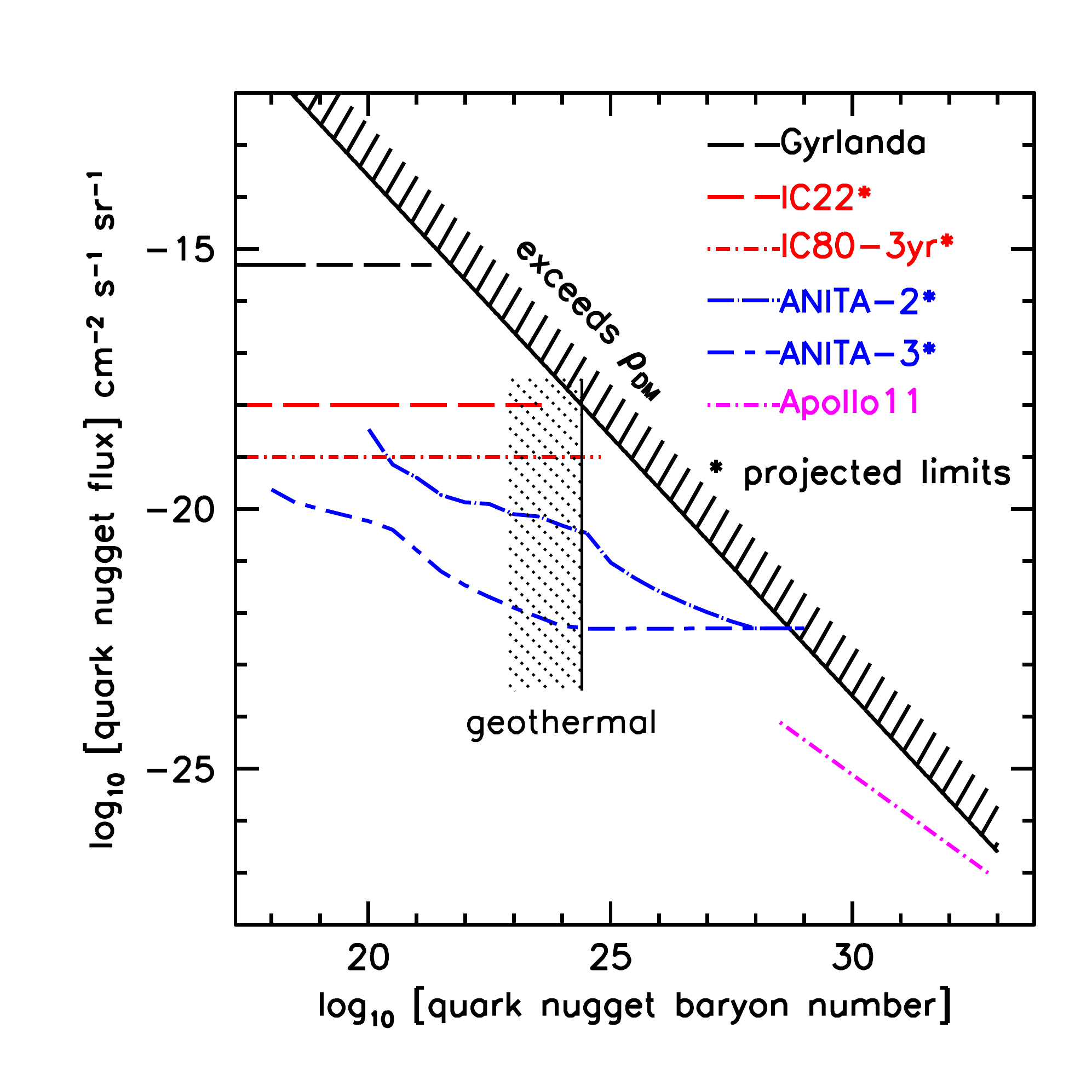}
\caption{Limits on quark nugget mass and fluxes based on current constraints, taken from   \cite{Gorham:2012hy}. 
Our preferable  value $\la B\ra \sim 10^{25}$ is translated  to the axion mass $m_a\simeq 10^{-4}$ eV according to the scaling relation (\ref{sigma}). The corresponding constraints expressed in terms of $m_a$ have important  implication for the direct axion search experiments as  discussed in section \ref{axion-search}. Orpheus experiment ``B" is designed to be    sensitive exactly  to this value of the  axion mass $m_a\simeq 10^{-4}$ eV,   see Fig.\ref{rybka}. }
\label{gorham}
\end{figure}

It has also been suggested that the \textsc{anita} 
experiment may be sensitive to the radio band 
thermal emission generated by these objects as they pass through the 
antarctic ice \cite{Gorham:2012hy}. These experiments may thus be 
capable of adding direct detection capability to the indirect evidence 
discussed above, see Fig.\ref{gorham} taken from  \cite{Gorham:2012hy} which reviews these constarints.

It has been also suggested recently \cite{Gorham:2015rfa} that  the interactions of these (anti-) nuggets  with normal matter in the Earth and Sun will lead to annihilation and an associated neutrino flux. 
Furthermore, it has been claimed  \cite{Gorham:2015rfa} that the antiquark nuggets cannot account for more than 20$\%$ of the dark matter flux based on constraints for the neutrino flux in 20-50 MeV range where sensitivity of the underground neutrino detectors such as SuperK have their highest signal-to-noise ratio. 

However, the claim  \cite{Gorham:2015rfa} was based on assumption that 
the annihilation of visible baryons with antiquark nuggets  generate  the neutrino spectrum    similar to conventional   baryon- antibaryon  annihilation  spectrum when the  large number of produced  pions   eventually decay to  muons 
and consequently to highly energetic neutrinos  in the 20-50 MeV energy range. Precisely these highly energetic neutrinos play the crucial role in analysis \cite{Gorham:2015rfa}. However, in   most CS phases  the lightest pseudo Goldstone mesons (the pions and Kaons) have masses in the 5-20 MeV range \cite{Alford:2007xm,Rajagopal:2000wf} in huge contrast with hadronic confined phase where $m_{\pi}\sim 140$ MeV.  
 Therefore, such light pseudo Goldstone mesons in CS phase  cannot produce  highly energetic neutrinos in the 20-50 MeV energy range and  thus  are not subject to the SuperK constraints \cite{Lawson:2015cla}.

$\bullet$ We conclude this brief overview on  observational constraints of the model with the following remark.  This model which has  a single  fundamental parameter  (the  mean baryon number of a nugget $\la B\ra \sim 10^{25}$, corresponding to the axion mass $m_a\simeq 10^{-4}$ eV), and which      enters  all the computations is   consistent with all known astrophysical, cosmological, satellite  and ground based constraints as highlighted above. Furthermore, in a number of cases the predictions of the model are very close to the presently available  limits, and very modest improving of those constraints may lead to a discovery of the nuggets.  Even more than that: there is a number of frequency bands where some excess of emission was observed, and this model may explain some portion, or even entire excess of the observed radiation in these frequency bands.
 
 In the light of  this (quite optimistic) 
 assessment  of the  observational constraints   of this model  it is  quite obvious    that   further and deeper studies  of this model    are worthwhile to pursue. The relevant developments may   include, but not limited, to such hard problems   as formation mechanisms during the QCD phase transition in early Universe,  even though many  key elements for proper addressing those questions at $\theta\neq 0, \mu\neq 0, T\neq 0$ are still  largely unknown in strongly coupled QCD as shown on Fig.\ref{fig:phase_diagram}. This work is the first step in the direction to explore a possible mechanism of formation of the nuggets.  
       
\section{Formation of the nuggets: the   crucial  ingredients of the proposal.}\label{formation}

 {\bf 1.} First important element of this proposal  is the presence of the topological objects, the axion domain walls \cite{Sikivie}.  As we already mentioned the  $\theta$ parameter is   the angular variable, and therefore supports various types of the domain walls, including the so-called $ N_{DW}=1$ domain walls when $\theta$ interpolates between one and the same physical vacuum state with the same energy $\theta\rightarrow\theta+2\pi n $.    The axion domain walls may form at the same moment when the axion potential get tilted, i.e. at  the moment $T_a$ when the axion field starts to roll due to the misalignment mechanism. The tilt becomes much more pronounced  at the phase transition when the chiral condensate forms at $T_c$.
In general one should expect that the $ N_{DW}=1$ domain walls form once the axion potential is sufficiently tilted, i.e. anywhere between $T_a$ and $T_c$.

One should comment here that it is normally assumed that for the topological defects to be formed the Peccei-Quinn (PQ)  phase transition must occur after inflation. This argument is absolutely correct for a generic type of  domain walls with  $ N_{DW}\neq 1$.  The conventional argument is    based on  the fact that  few physically  {\it different vacua} with the same energy must be present inside of the same horizon for the domain walls to be formed. The $ N_{DW}=1$ domain walls
are unique and very special in the  sense that $\theta$ interpolates between  {\it one and the same} physical vacuum state.   Such $ N_{DW}=1$ domain walls can be formed even if the PQ phase transition occurred before inflation and a unique physical vacuum occupies entire Universe, see some elaboration of this point  at the end of this section.

 It has been  realized many years after    \cite{Sikivie}       that  the  walls, in general, demonstrate a  sandwich-like  substructure
on  the QCD scale $\Lambda_{QCD}^{-1}\simeq $ fm.
The arguments supporting the 
QCD scale substructure inside the axion domain walls are based on analysis \cite{FZ}
of QCD in the large $N$ limit with inclusion of the  $\eta'$ field\footnote{The  $\eta'$ field 
has special  property that it enters the effective Lagrangian  in unique combination $[\theta- \eta'(x)]$  where  the $\theta$ parameter in the present context  plays the role of   the axion dynamical field $\theta(x)$.  
A similar structure is known to occur in CS phase as well. The corresponding domain walls in CS phase    have been also  constructed \cite{Son:2000fh}.} and  independent
analysis \cite{SG} of supersymmetric models
where a similar $\theta$ vacuum structure occurs. 

One should remark here that the described structure is classically stable configuration. In particular, the  $\eta'$ field   cannot decay to $2\gamma$ simply due to the kinematical reasons when the $\eta'$ field   is off-shell, and  cannot be expressed as a superposition of on-shell free particles. It can only decay through the tunnelling, and therefore,  such $ N_{DW}=1$ domain walls are  formally metastable rather than absolutely stable configurations.

 {\bf 2.} Second important    element  is that  in addition to this known QCD  substructures \cite{FZ, SG,Son:2000fh} of the axion domain walls expressed in terms of the  $\eta'$ and gluon    fields, there is  another substructure with a similar QCD scale which  carries  the baryon charge.   Precisely this novel feature of the  domain walls which was not explored previously  in the literature will play a key role in our proposal  because exactly  this new effect will be eventually responsible for the accretion  of the baryon charge by the nuggets. Both,     the quarks and anti-quarks   can  accrete on a given closed domain wall making eventually the quark nuggets or anti-nuggets, depending on the sign of the baryon charge.   The   sign is chosen  randomly such that equal number of quark and antiquark nuggets are formed if the external environment is CP even, which is the case when   fundamental  $\theta=0$. One can interpret this phenomenon as a {\it local spontaneous symmetry breaking effect}, when on the scales of order  the  correlation length $\xi $ the nuggets may acquire the positive or negative  baryon charge with equal probability, as  discussed in great details   in next section \ref{baryons}. 
  
 {\bf 3.} Next important ingredient of the proposal is the Kibble-Zurek mechanism which gives a generic picture of formation of the topological defects
 during a phase transition, see original papers \cite{KZ},  review \cite{KZ-review} and the textbook \cite{Vilenkin}.  In our context the Kibble-Zurek mechanism suggests that once the axion potential is sufficiently tilted  the  $ N_{DW}=1$ domain walls form. The potential becomes much more pronounced   when the chiral condensate forms at $T_c$. After some time after $T_a$ 
  the system is dominated by a single, percolated,  highly folded and crumpled  domain  wall of very complicated topology. In addition, there will be a  finite portion of the closed walls (bubbles) with typical size of order correlation length $\xi(T)$, which is defined as an average  distance between folded domain walls at temperature $T$. It is known that the probability of finding closed walls with very large size $R\gg \xi$ is exponentially small. Furthermore, numerical simulations suggest  \cite{Vilenkin} that approximately $ 87\%$ of the total wall area belong to the percolated large cluster, while the rest is represented by relatively small    closed bubbles with sizes $R\sim \xi$. 
 
 The key point for our proposal is there existence of these finite closed bubbles made of the axion domain walls\footnote{The presence of such closed bubbles in numerical simulations in context of the axion domain wall has been mentioned in \cite{Sikivie:2008}, where it was argued that these bubbles would oscillate and emit the gravitational waves. However, we could not find any further details  on the fate of these closed bubbles in the literature.}.
  One should remark here that  these closed bubbles  had been formed sometime after $T_a$ when original $\theta$ parameter has not settled yet to its minim  value.  It implies that the domain wall evolution starts at the time when    $\theta$ parameter is not yet zero\footnote{\label{T_a}This $\theta$ parameter in our work is defined as the value of $\theta$ at the moment when the domain walls form.  
It  is  not exactly    the same value as the misalignment angle which normally enters all the computations due to the conventional misalignment mechanism \cite{Wantz:2009it,misalignment}. This is because the temperature   when the domain walls form   and the temperature $T_a$ when the axion field starts to roll do not exactly coincide though  both effects are due to the same axion tilted potential. The crucial point is that the $\theta$ parameter, as defined above, could be numerically small, nevertheless, it preserves its coherence over entire Universe,  see item {\bf 5} below and section \ref{asymmetry} for the details.}.   Normally it is assumed that these closed bubbles collapse as a result of the domain wall pressure, and do not play any significant  role in dynamics of the system. However, as  we already mentioned in Introduction 
 the collapse of these closed bubbles is halted due to the Fermi pressure acting inside of the bubbles.  Therefore, they may survive and serve as the dark matter candidates.
 
 The  percolated network  of the domain walls will decay to the axion in conventional way as discussed   in \cite{Chang:1998tb,Wantz:2009it,Kawasaki:2014sqa}.  
 Those axions (along with the axions produced by  the conventional misalignment mechanism \cite{Wantz:2009it,misalignment})  
 will contribute to the dark matter density today.  The corresponding contribution to dark matter density is highly sensitive to the axion mass as $\Omega_{\rm dark}\sim m_a^{-1}$. It may saturate the observed dark matter density if $m_a\simeq 10^{-6}$ eV \cite{vanBibber:2006rb, Sikivie:2008, Asztalos:2006kz,Raffelt:2006cw,Sikivie:2009fv,Rybka:2014cya,Rosenberg:2015kxa,Graham:2015ouw}, while it may contribute very little to  $\Omega_{\rm dark}$ if the  axion mass is slightly heavier than $m_a\simeq 10^{-6}$ eV. In contrast, in our framework an approximate relation $\Omega_{\rm dark}\approx\Omega_{\rm visible}$ holds irrespectively to the axion mass $m_a$.
   
   We shall not elaborate  on the production and spectral properties of these  axions in the present work.  
 Instead, the  focus of the present paper is  the dynamics of the 
  closed bubbles, which is normally ignored in computations of the axion production. Precisely these closed bubbles, according to this proposal,  will eventually become the stable nuggets and may serve as the dark matter candidates.  
  
  As we already mentioned     the nugget's contribution to  $\Omega_{\rm dark}$ is not very sensitive to the axion mass, but rather,  is determined by the formation temperature $T_{\rm form}$ as explained in Introduction, see also footnote \ref{theta} with few important comments on this.  
    The time  evolution  of these nuggets after their formation is the subject of section \ref{time-evolution}.

{\bf 4.} There existence 
of CS phase in QCD represents the  next   crucial
element of our scenario. The   CS has been an active area of research for quite sometime,  
 see   review papers \cite{Alford:2007xm,Rajagopal:2000wf} on the subject.
 The  CS phase is realized when quarks are squeezed to the density which 
is  few times
nuclear density. It has been known that this regime may be realized in nature in neutron stars
interiors and in the violent events associated with collapse of massive stars or collisions 
of neutron stars, so it is important for astrophysics.  

 The force which squeezes quarks in neutron stars is gravity; the force which does
an analogous  job in early universe  during the QCD phase transition is 
a violent collapse  of a bubble of size $R\sim \xi (T)$  formed from  the axion domain wall as described  in item {\bf 3} above.
 If  number density of quarks trapped inside of the bubble (in the bulk)
 is sufficiently large, the collapse 
 stops due to the internal Fermi pressure.  In this case the system
in the bulk  may reach the equilibrium with   the ground state  being in a CS phase. As we advocate  in section \ref{time-evolution} 
this is very plausible fate of a relatively large size bubbles of size $R\sim \xi (T)$ made of the axion domain walls which were produced after the QCD phase transition. 
  
 {\bf 5.} If $\theta$ vanishes, then equal number of nuggets and anti-nuggets would form. 
 However,   the  $\cal{CP}$ violating $\theta$ parameter (the axion field), which is defined as value of $\theta$ at the moment of domain wall formation   generically is not zero, though it might be numerically quite small. Precisely the dynamics of the coherent axion field $\theta(x)$  leads to  preferences in formation of one   species   of nuggets, as discussed in section \ref{asymmetry}. This sign-preference   is correlated  on the scales   where the axion field $\theta(x)$ is   coherent, i.e.   on the scale of the entire Universe at the moment of the domain wall formation. 
 In other words, we assume that the PQ phase transition happened before inflation. 
 One should emphasize that this assumption on coherence of the axion field on very large scales is consistent with formation of $N_{DW}=1$ domain walls, see item {\bf 1} above. This coherence   obviously cannot be satisfied for a generic type of the domains walls with $N_{DW}\neq 1$ when  $N_{DW}$ {\it  physically distinct} vacuum states with the same energy must be present in the system. 
 
 There are few arguments supporting this claim. First of all, 
one should remind that  the axion domain wall with $N_{DW}=1$ corresponds to the configuration when  $\theta$ field interpolates between $\theta=0$ and $\theta=   2\pi$. It implies that the axion field, describing the domain wall, interpolates between topologically distinct but physically identical and unique  vacuum state.  We present few strong arguments below suggesting that the topological sectors  must be always present in the system everywhere in space, and inflation  does not remove different topological sectors from the system. Therefore $N_{DW}=1$ can be formed even if the PQ phase transition happened before inflation. 
 
 The simplest way to  explain this claim   is to analyze  the expression for vacuum energy \cite{Witten:1980sp,Witten:1998uka} in the limit of large number of colours  $N_c\rightarrow \infty$, though, it is known that  the arguments still hold for finite $N_c$ as well\footnote{For finite $N_c$ in some  simple models the  computations for the vacuum energy can be  exactly carried out. In many  cases  formula (\ref{E_vac}) assumes the form
$ E_{\rm vac}(\theta)\sim\min_{k}(-N_c^2)\cos\left(\frac{  \theta+2\pi k }{N_c}\right)$ which is obviously reduced to (\ref{E_vac}) in large $N_c$ limit.}. The main point is that the vacuum energy as a function of $\theta$ assumes the following form 
\be
 \label{E_vac}
 E_{\rm vac}(\theta)\sim\min_{k} \left(\theta+2\pi k\right)^2 +{\cal{O}}(\frac{1}{N_c}),
\ee
where $\theta$ in the present context plays the role of the axion field. 
This formula explicitly shows that for each given $\theta$ the vacuum state is unique. However, there is a  number of different branches, classified by parameter $k$ such that when $\theta=\pm \pi$ the system becomes double degenerate, and one branch replaces another branch at    $\theta=\pm \pi$. Precisely this pattern  provides the required $2\pi$ periodicity of the system. This picture of the $\theta$ dependence is commonly  accepted by the community, and in fact emerges in many different gauge theories where exact computations can be carried out, including the holographic description \cite{Witten:1998uka}.  

The key point in these arguments   is the presence of $k$ different branches
  which must be present in the system everywhere in space in order to provide the $2\pi$ periodicity of the vacuum energy (\ref{E_vac}). There is only one physical vacuum in the system, which however always accompanied by $k$ different branches. Inflation cannot remove different $k$ branches outside  the horizon because they are inherent elements of the system at each point in space. The domain wall solution with $N_{DW}=1$   corresponds to interpolation between different topological sectors $k=1$ and $k=0$ which always present in the system inside the same horizon. 
  
  Another argument which leads to the same conclusion goes as follows. 
 The $N_{DW}=1$ is  formed as a result of twisting of the axion field in configurational space when the axion field returns to its initial physical vacuum state   after making a full circle as explained above. Topologically it is identical to  creation of solitons in two dimensional  Sine-Gordon model $\sim \cos\phi$ in condensed matter physics when the $\phi$
 field interpolates between one and the same physical, but  topologically distinct  states $\phi=0$ and $\phi=\pm 2\pi$. In the dual picture the Sine -Gordon solitons can be thought as $\psi$ fermions, see section \ref{baryons} for references on this duality relation. In this dual picture  the production of solitons corresponds to production of the fermi $\psi$ fields. It is quite obvious that the production of the $\psi$ fields is perfectly allowed process at $T\neq 0$ irrespectively whether inflation happened before or after the PQ symmetry breaking occurred. Formally, the mere  existence of  $\psi$ field in the system is due to $k$ topological sectors in the theory when $\phi$  enters   the Lagrangian in combination  $(\phi+2\pi k)$.
 The inflation obviously cannot remove $k$ sectors from the system because it would   violate   the fundamental properties of the theory, such as duality between $\psi$ and $\phi$ descriptions.  
 
 For our system it implies that   the $N_{DW}=1$ corresponding to interpolation between $k=1$ and $k=0$ is allowed configuration, irrespectively to inflation,  as all $k$  topological sectors must be present in the system in every point of space-time.

\exclude{
The domain wall formation with $N_{DW}=1$  for example can proceed as follows. 
Let us assume that the axion field is trapped    at some excited branch $k\neq 0$ at some initial  moment, which of course could happen as all branches are degenerate at sufficiently high temperature. 
 The axion field $\theta$ starts to evolve toward a  lower energy  branch and   inevitably hits the point   $\theta=\pm \pi$ where the double degeneracy enforces the domain wall formation. When $\theta$ evolves even further to $\theta=0$ this domain wall becomes   $N_{DW}=1$ axion domain wall  when the path in configurational space  returns to the same unique vacuum state. When the temperature is not zero, the domain wall formation may even occur  before the axion field hits the point $\theta= \pi$ as there will be always two (almost) degenerate states corresponding to the different branches $k$.  
 }

 To conclude this section:  as we argue below the  generic consequence  of this framework (\ref{Omega}) is not very   sensitive to an absolute value of $\theta$ at the moment of the domain wall formation, see comment  in  footnote \ref{theta} on this matter. One can say that  the coherent axion field $\theta(x)\neq 0$, being numerically small,  plays the role of  the $\cal{CP}$ violating catalyst which determines a preferred direction  for {\it  separation of the baryon charges} on the Universe  scale. This role of $\cal{CP}$ violation in our proposal is quite  different  
 from  the role it plays in conventional  ``baryogenesis" mechanisms.

 \section{formation of the nuggets: accretion of  the baryon  charge  }\label{baryons}
 From now on and until section \ref{asymmetry}  we focus on the dynamics of a single closed bubble   produced during the domain wall formation as described in item {\bf 3} in section \ref{formation}. 
 The correlation length $\xi(T)$  is defined as an average distance between folded domain walls at temperature $T$.  We assume that initial size of the bubble  $\xi(T)$ is sufficiently large,   few times  larger  than the  axion domain wall width $\sim m_a^{-1}$, such that one can locally treat the surface of the closed bubble being flat. 
 
 The main goal of this section is 
  to demonstrate  that such a bubble will generically    acquire a
baryon (or anti-baryon) charge  in very much the same way  as the 
 $\eta'$ field was dynamically accreted  as originally discussed in \cite{FZ} and briefly explained   above  as item {\bf 2} in section \ref{formation}. In other words,  we shall argue in this section that the bubbles with baryon or anti-baryon charge will be copiously produced during the phase transition as they are very generic configurations of the system. In both cases the effect emerges as a result  of the  nontrivial boundary conditions formulated far away from the domain wall core 
 when the field assumes physically the same but topologically distinct  vacuum states on opposite sides of the axion domain wall.

 The technique we shall adopt  in this section has been previously used  to study the
generation of the magnetic field in the   domain wall background \cite{Forbes:2000gr}.    This method makes
the approximation that the domain wall is flat and that translational
and rotational symmetries are preserved in the plane of the wall (which
we take to be the $x$--$y$ plane).  These approximations are marginally justified  in
our  case because the initial curvature $R\sim \xi (T)$  is assumed to be few times   larger that the
width of the wall $\sim m_a^{-1}$.

Once this approximation is made, we can reformulate the problem in
$1+1$ dimensions ($z$ and $t$) and calculate the density of the
  bulk properties along the domain wall.  To regain the full
four-dimensional bulk properties, we shall estimate the density of the
particles in the $x$--$y$ plane to obtain the appropriate density and
degeneracy factors for the bulk density.   

We proceed to demonstrate this technique by computing   the accumulation  
of baryon charge  along the wall.  We take the standard form for the  
interaction between the pseudo-scalar   fields and the fermions 
which respect all relevant symmetries:
\begin{equation}
  \label{eq:L4d}
  \mathcal{L}_4=\bar{\Psi}\left(i{\not{\!\partial}}-
    m e^{i \left[\theta(z)-\phi(z)\right]\gamma_5}-\mu \gamma_0\right)\Psi.
\end{equation}
Here $\theta(z)$ and $ \phi(z)$ are the dimensionless axion and $\eta'$  domain wall solution.  Parameter $m$ is the the typical QCD scale of the problem, while $\mu$ is the typical chemical potential at a specific time in evolution of the system, see below with more precise explanations. We also simplify the problem by ignoring all flavour and colour indices, as well as an effective 4-fermi interactions,    as our main goal is to explain the basic idea with simplified setting. 

Parameter $m$  in eq.(\ref{eq:L4d}) should not be literally identified with the quark mass, nor with the nucleon mass. Instead, this dimensional parameter $m\sim \Lambda_{QCD} $ should be thought as an effective  coupling in our model when parameter $m$ effectively describes the interaction with fermi field  $\Psi$ in all phases during the formation time, including the quark gluon plasma as well as hadronic and CS phases\footnote{In quark gluon phase the colour singlet $\eta'$ field does not exist. However, the singlet phase which accompanied the quark field still present in the system. The coefficient $m$ in this phase can be computed using the  instanton liquid model.  At very high temperature the parameter $m$ is proportional to the quark masses and indeed very small. When temperature decreases  the instanton contribution grows very fast. At this point parameter $m$ is proportional to the vacuum expectation value of the 't Hooft determinant. When temperature further decreases the parameter $m$ is proportional to the diquark condensate in CS phase or the chiral condensate in the hadronic phase, see Fig.\ref{fig:phase_diagram}. We shall not elaborate along this line by assuming  $m\sim \Lambda_{QCD} $ for all our estimates which follow.}. The same comment also applies to a numerical value of the chemical potential $\mu$: it vanishes  during initial time and becomes very large when CS phase sets in inside the nugget.

    The strategy is to break (\ref{eq:L4d}) into two $1+1$
dimensional components by setting $\partial_x=\partial_y=0$ (this is
the approximation that the physics in the $z$ direction decouples from
the physics in the $x$--$y$ plane) and then by manipulating the system
of equations that result.

First, we introduce the following chiral components of the Dirac
spinors\footnote{We are using the standard representation here:
\begin{align*}
  \gamma_0&=\begin{pmatrix}
    {I} & {0}\\
    {0} & -{I}
  \end{pmatrix},&
  {\gamma}_j&=\begin{pmatrix}
    {0} & {\sigma}_j\\
    -{\sigma}_j & {0}
  \end{pmatrix},&
  {\gamma}_5&=\begin{pmatrix}
    {0} & {I}\\
    {I} & {0}
  \end{pmatrix},
\end{align*}
\begin{align*}
  \sigma_1&=\begin{pmatrix}
    0 & 1\\
    1 & 0
  \end{pmatrix},&
  \sigma_2&=\begin{pmatrix}
    0 & -i\\
    i & 0
  \end{pmatrix},&
  \sigma_3&=\begin{pmatrix}
    1 & 0\\
    0 & -1
  \end{pmatrix}.
\end{align*}
}
:
\begin{align}
  \Psi_+&=\frac{1}{\sqrt{S}}\begin{pmatrix}
    \chi_1\\
    \chi_2
  \end{pmatrix}, &
  \Psi_-&=\frac{1}{\sqrt{S}}\begin{pmatrix}
    \xi_1\\
    \xi_2
  \end{pmatrix},
\end{align}
\begin{equation}
  \Psi=\frac{1}{\sqrt{2S}}\begin{pmatrix}
    \chi_1+\xi_1\\
    \chi_2+\xi_2\\
    \chi_1-\xi_1\\
    \chi_2-\xi_2
  \end{pmatrix}=
  \frac{1}{\sqrt{2}}\begin{pmatrix}
    \Psi_++\Psi_-\\
    \Psi_+-\Psi_-
  \end{pmatrix},
\end{equation}
where $S$ is the area of the wall. This normalization factor cancels
the degeneracy factor proportional to $S$ added in the text below.

The associated Dirac equation is
\begin{subequations}
  \begin{align}
    \begin{pmatrix}
      -me^{i(\phi-\theta)}&i(\partial_t+\partial_z)-\mu\\
      i(\partial_t-\partial_z)-\mu&-me^{-i(\phi-\theta)}
    \end{pmatrix}
    \begin{pmatrix}
      \chi_1\\
      \xi_1
    \end{pmatrix}=0,\\
    \begin{pmatrix}
      -me^{i(\phi-\theta)}&i(\partial_t-\partial_z)-\mu\\
      i(\partial_t+\partial_z)-\mu&-me^{-i(\phi-\theta)}
    \end{pmatrix}
    \begin{pmatrix}
      \chi_2\\
      \xi_2
    \end{pmatrix}=0.
  \end{align}
\end{subequations}
where  we decouple the $z$ coordinates from $x$ and $y$ by setting
$\partial_x=\partial_y=0$.
Remember that we are looking for a two-dimensional Dirac equation,
thus we want the kinetic terms to look the same.  For this reason we
should flip the rows and columns of the second equation.  Doing this
and defining the two two-dimensional spinors
\begin{align}
  \Psi_{(1)}&=\begin{pmatrix}
    \chi_1\\
    \xi_1
  \end{pmatrix},&
  \Psi_{(2)}&=\begin{pmatrix}
    \xi_2\\
    \chi_2
  \end{pmatrix},
\end{align}
the equations have the following structure:
\begin{subequations}
  \label{eq:D2d}
  \begin{align}
    (i{\hat{\gamma}}^\nu\partial_\nu
    -me^{+i(\theta-\phi){\hat{\gamma}}_5}-\mu\hat{\gamma_0})\Psi_{(1)}&=0
    \\
    (i{\hat{\gamma}}^\nu\partial_\nu
    -me^{-i(\theta-\phi){\hat{\gamma}}_5}-\mu\hat{\gamma_0})\Psi_{(2)}&=0
  \end{align}
\end{subequations}
where the index $\nu\in\{t,z\}$, the Lorentz signature is $(1,-1)$ and
we define the following two-dimensional version of the gamma matrices:
\begin{align*}
  {\hat{\gamma}}_t&={\sigma}_1,&
  {\hat{\gamma}}_z&= -i{\sigma}_2, &
  {\hat{\gamma}}_5&= {\sigma}_3.
\end{align*}
These satisfy the proper two-dimensional relationships
$\hat{{\gamma}}_5 = \hat{{\gamma}}_t\hat{{\gamma}}_z$ and
${\hat{\gamma}}_\mu{\hat{\gamma}}_\nu =
g_{\mu\nu}+\epsilon_{\mu\nu}{\hat{\gamma}}_5$.  We can reproduce
equation (\ref{eq:D2d}) from the following effective two-dimensional
Lagrangian density,
\begin{align}
  \mathcal{L}_2=&\bar\Psi_{(1)}\left(i{\hat{\gamma}}^\mu\partial_\mu
    -me^{+i(\theta-\phi){{\hat{\gamma}}_5}}-\mu\hat{\gamma_0}\right)\Psi_{(1)}+\nonumber\\
  +&\bar\Psi_{(2)}\left(i{\hat{\gamma}}^\mu\partial_\mu
    -me^{-i(\theta-\phi){{\hat{\gamma}}_5}}-\mu\hat{\gamma_0}\right)\Psi_{(2)},
  \label{eq:L2d}
\end{align}
where two different species of fermion with opposite chiral charge
interact with the axion domain wall background determined by $\theta(z)$ and $\phi(z)$ fields.  Note that, due to
the normalization factor $1/\sqrt{S}$ we introduced above, the
two-dimensional fields $\Psi_{(i)}$ have the correct canonical
dimension $1/2$.

We have thus successfully reduced our problem to a two-dimensional
fermionic system.  It is known that for several systems in $1+1$
dimensions, the fermionic representation  is equivalent to a $1+1$
dimensional bosonic system through the following
equivalences\cite{Mandelstam:1975hb,Coleman:1975bu}:
\begin{subequations}
\label{bosonization}
  \begin{align}
    \bar\Psi_{(j)}i{\hat{\gamma}}^{\mu}\partial_\mu\Psi_{(j)}
    &\rightarrow
    \frac{1}{2}(\partial_\mu\theta_j)^2,\\
    \bar\Psi_{(j)}{\hat{\gamma}}_{\mu}\Psi_{(j)}
    &\rightarrow
    \frac{1}{\sqrt{\pi}}\epsilon_{\mu\nu}\partial^\nu\theta_j,\\
    \bar\Psi_{(j)}\Psi_{(j)}
    &\rightarrow
    -m_0\cos(2\sqrt{\pi}\theta_j),\\
    \bar\Psi_{(j)}i{\hat{\gamma}}_5\Psi_{(j)}
    &\rightarrow
    -m_0\sin(2\sqrt{\pi}\theta_j).
  \end{align}
\end{subequations}
The constant $m_0$ in the last two equations is
  a dimensional  parameter of order $m_0\sim m\sim \Lambda_{QCD}$.  The exact coefficient of this
  factor  depends on renormalization procedure and is only known for few exactly solvable
  systems but in all cases, is of order unity.

After making these replacements, we are left with the following
two-dimensional bosonic effective Lagrangian density describing the
two fields $\theta_1$ and $\theta_2$ in the domain wall background determined by 
$\phi(z)$ and $\theta(z)$:
\beq
\label{L}
  \mathcal{L}_2=\frac{1}{2}(\partial_\mu\theta_1)^2
  +\frac{1}{2}(\partial_\mu\theta_2)^2   -U(\theta_1,\theta_2) +\frac{\mu}{\sqrt{\pi}}\frac{\partial(\theta_2+\theta_1)}{\partial z} 
\eeq
where the effective potential is
\be
  \label{U}
  U(\theta_1,\theta_2)=&-&m m_0\left[\cos(2\sqrt{\pi}\theta_1-\phi +\theta)\right]
  \nonumber\\
    &-&m m_0\left[\cos(2\sqrt{\pi}\theta_2+\phi -\theta)\right].
      \ee
 The conventional procedure to study the system (\ref{L}) is to add the kinetic terms for the axion $\theta$ and the $\eta'$ field $\phi$  into (\ref{L}) and study a resulting  solution depending on four dynamical fields by specifying all possible boundary conditions when the potential energy (\ref{U}) assumes its minimal value\footnote{In fact it was precisely the procedure which has been adopted in \cite{FZ} for a similar problem of computing of the profile functions of the axion, $\pi$- meson and $\eta'$-domain wall described by   $\theta-\pi-\eta'$ fields.}. In other words, one should take into account the dynamics of the $\theta$ and $\phi$ fields together with $\theta_1, \theta_2$  because the typical scales for $\phi, \theta_1, \theta_2$ are roughly the same order of magnitude and of order of $\Lambda_{QCD}$. 
 Recapitulate it: one cannot study  the dynamics of $ \theta_1, \theta_2$ field by  neglecting their back reaction on the background axion and  $\phi$ fields. 
 
 For our present purposes, however,  we do not really need an explicit profile functions for large number of different domain walls  determined by various  boundary conditions controlled  by (\ref{U}). The only important element relevant for our future discussions is the observation that some of the domain walls  may carry the baryon (anti baryon) charge.
 Indeed, the domain walls which satisfy the boundary conditions
 \be
 \label{bc}
 2\sqrt{\pi}\theta_1 (z=+\infty)-2\sqrt{\pi}\theta_1(z=-\infty)=2\pi n_1\\
 2\sqrt{\pi}\theta_2 (z=+\infty)-2\sqrt{\pi}\theta_2(z=-\infty)=2\pi n_2 \nonumber
 \ee  
   carry the following baryon charge $N$  defined for one particle Dirac equation
      \be
   \label{N}
   N&=&\int d^3x \bar{\Psi}\gamma_0\Psi = \int d z\left(\bar{\Psi}_{1}\hat{\gamma_0}{\Psi}_{1}+\bar{\Psi}_{2}\hat{\gamma_0}\Psi_{2}\right)\nonumber\\
   &=&-\frac{1}{\sqrt{\pi}}\int^{+\infty}_{-\infty} dz \frac{\partial}{\partial z} (\theta_1+\theta_2)=-(n_1+n_2),
 \ee
   where we express the final formula in terms of the  auxiliary two-dimensional fields $\theta_1$ and $\theta_2$ and corresponding boundary conditions given by eq. (\ref{bc}). Factor $S$ also cancels with our normalization for four dimensional $\Psi$ field.

   To complete the computations for four dimensional baryon charge  $B$ accumulated on the domain wall we need to multiply (\ref{N}) by  the degeneracy factor in vicinity of the domain wall which can be estimated as follows
   \be
   \label{B}
   B=N \cdot g\cdot \int \frac{ d^2 x_{\perp}d^2k_{\perp}}{(2\pi)^2} \frac{1}{\exp(\frac{\epsilon-\mu}{T})+1},
   \ee
   where $g$ is appropriate degeneracy factor, e.g.  $g\simeq N_cN_f$ in CS phase. We note that an additional degeneracy  factor $2$ due to the spin is already accounted for by parameter $N$ defined  in eq. (\ref{N}). For   high chemical potential $\mu\gg T$ corresponding to CS phase the baryon charge per unit area accreted  in vicinity of the domain wall can be approximated as
   \be
   \label{B_1}
   \frac{B}{S}\simeq N  \cdot \frac{g\mu^2}{4\pi}.
   \ee
   In the opposite limit of high  temperature   $\mu\ll T$ which corresponds  to the quark gluon plasma phase, the corresponding  magnitude can be estimated as follows
   \be
   \label{B_2}
   \frac{B}{S}\simeq N  \cdot \frac{g \pi T^2}{24}. 
   \ee 
  
    It is instructive to compare the estimate (\ref{B_2}) with   number density ${\cal{N}}/ V$ of all degrees of freedom in vicinity of the domain wall. Assuming that the baryon charge in the domain wall background is mainly concentrated on distances of order $m^{-1}$ from the center of the domain wall we arrive to the following   estimate
    for the ratio of the baryon number density bound to the wall in comparison with the total number density of all degrees of freedom responsible for the thermodynamical equilibrium in this phase
    \be
    \label{ratio}
    r\sim \frac{(B/S)\cdot m}{{\cal{N}}/ V }\sim N \left(\frac{m }{T}\right) \left(\frac{ \pi^3 g }{ 18 \xi(3) g^*}\right),
    \ee
    where effective degeneracy factor $g^*$  for quark gluon plasma is $g^*\simeq \left[\frac{3}{4} 4N_cN_f+ 2 (N_c^2-1)\right]$ and $\xi(3)\simeq 1.2$ is the Riemann zeta function. 
   Ratio (\ref{ratio})  shows that the accreted quark density  bounded to the  domain wall at high temperature represents parametrically small contribution to  all  thermodynamical  observables mainly because of a small parameter $m/T\ll 1$   in this phase. The situation  drastically changes as we discuss in next section \ref{time-evolution} when  the temperature slowly decreases   due to expansion of the Universe and the system enters the hadronic or CS phase, as shown on Fig. \ref{fig:phase_diagram}. At this point 
   the baryon charge accumulation in the domain wall background becomes the major player of the system, which eventually leads to the formation of the CS nuggets or anti-nuggets when quarks (anti-quarks) fill entire volume of the nuggets (anti-nuggets). 
   
   We conclude this section with the  following important comments. 
   First, we argued  that the domain walls in general accrete the baryon (or anti baryon) charge in vicinity of the centre of the domain wall. 
   The effect in many respects is similar to fractional charge localization  on domain walls, while the rest of the charge is de-localized in the rest of  volume of the system as discussed in original paper \cite{Jackiw:1975fn}. 
   The effect  is also very similar to previously discussed phenomenon on dynamical generation of the $\eta'$ field in the domain wall background. The key point  is that at sufficiently high temperature   the $N_{DW}=1$ domain walls form by the usual Kibble-Zurek  mechanism as explained in section \ref{formation}. The periodic fields $\theta, \phi, \theta_1, \theta_2$   may assume  physically identical but topologically distinct vacuum values (\ref{N}) on opposite sides of the walls.   When the system cools down the corresponding fields inevitably form  the domain wall structure, similar to  analysis  in hadronic  \cite{FZ} and CS phases \cite{Son:2000fh}. 
    
    We advocate the picture that the closed bubbles   will be also  inevitably  formed as discussed in section \ref{formation}.    The collapse  of these bubbles halts as a result of Fermi pressure due to the quarks accumulated inside the nugget during the evolution of the domain wall network. Next  section \ref{time-evolution}
    is devoted precisely the question on time evolution of these closed bubbles  made of $N_{DW}=1$ domain wall.
    
$\bullet$  The most important lesson of this section  is that  there is a variety of acceptable boundary conditions  determined by    potential (\ref{U}) when the energy assumes its vacuum values. Some of the domain walls will cary zero baryon charge when the combination $(n_1+n_2)$ vanishes   according to (\ref{N}). However, generically the domain walls will acquire the baryon or anti-baryon charge.  This is because the domain wall tension 
    is mainly determined by the axion field while corrections due to QCD substructure will lead to a small correction of order $\sim m/f_a \ll 1$, similar to studies of the (axion -$\eta'$-$\pi$) domain wall \cite{FZ}. Therefore, the presence 
    of the QCD substructure with non vanishing $(n_1+n_2)\neq 0$  increases the domain wall tension only slightly. In other words, accumulation of the baryon charge  in vicinity of the wall  does not lead to any suppression during the formation stage. Consequently, this implies that the domain closed bubbles  carrying   the baryon or anti baryon charge will be copiously produced during the phase transition as they are very {\it  generic  configurations of  the system.} Furthermore, the baryon charge   cannot leave the system  during the evolution as it is strongly bound to the wall due to the topological reasons. The corresponding binding energy per quark is order of $\mu$ and increases with time  as we discuss in next section. 
    
    This phenomenon of ``separation of the baryon charge" can be interpreted  as a local version of spontaneous symmetry breaking of the baryon charge.  This symmetry breaking occurs not in the   entire volume in the ground state  determined by the potential (\ref{U}). Instead, the symmetry breaking occurs on scale $\xi(T)$ in vicinity of the  field configurations  which describe the interpolation
    between physically identical but topologically distinct vacuum states (\ref{bc}). One should add that a similar phenomenon occurs with accumulation of the $\eta'$ field in vicinity of the axion domain wall as described in \cite{FZ}. However, one could not term  that effect as a ``local spontaneous violation" of the $U(1)_A$ symmetry because the $U(1)_A$ symmetry is explicitly broken by anomaly, in contrast with our present studies when the baryon charge is the exact symmetry of  QCD. Nevertheless, the physics is the same in a sense that the closed bubble configurations generically acquire the axial as well as the baryon charge. This phenomenon as generic as formation of the topological domain walls themselves  when the periodic fields $ \phi, \theta_1, \theta_2$   may randomly assume  physically identical but topologically distinct vacuum values  on the correlation lengths of order  $ \xi$.
    
   Finally,   one should also mention   here that  very  similar  effect of the ``local $\cal{CP}$ violation"     can be experimentally tested  in heavy ion collisions in event by event basis where the so-called  induced $\theta_{\rm ind}$- domain with a specific sign in each given event can be formed. This leads to the ``charge separation effect"  which can be experimentally observed  in relativistic heavy ion collisions \cite{Kharzeev:2007tn}. This ``charge separation effect" in  all respects is very similar to the phenomenon discussed in the present section. In fact, the main motivation for 
   one of the authors (AZ) for studies \cite{Kharzeev:2007tn} was a possibility to test   the ideas advocated in this work  by performing a specific analysis in the  controllable    ``little Bang" heavy ion collision experiments, in contrast with ``Big Bang" which happened billion of years ago.   This field  of research initiated in \cite{Kharzeev:2007tn}  became the hot topic  in recent years as a result of many interesting theoretical and experimental advances,    see recent review papers \cite{Kharzeev:2009fn,Kharzeev:2013ffa,Kharzeev:2015znc} on the subject.

 \section{formation of the nuggets: time evolution} \label{time-evolution}
 We assume that a closed $N_{DW}=1$ domain wall has been formed as discussed in previous section \ref{formation}.
 Furthermore, we also assume that this domain wall is classified by non-vanishing baryon number $(n_1+n_2)$ according to eq.(\ref{N}). Our goal now is to study the time evolution of the obtained configuration. As we argue below  the contraction of the bubbles halts as a result of the Fermi pressure due to baryon  charge accreted during the evolution. As a result, the system comes to the equilibrium at some temperature $T_{\rm form}$ when the  nuggets complete their formation. We want to see precisely how it happens, and what are the typical time scales relevant for  these processes. 
  
We start with the following effective Lagrangian describing the time evolution of the closed spatially  symmetric domain wall of radius $R(t)$, 
  \be
\label{lagrangian}
L&=&\frac{4\pi\sigma R^2(t)}{2} \dot{R}^2(t)  - 4\pi\sigma R^2(t) \\
&+&\frac{4\pi R^3(t)}{3}\left[P_{\rm in}(\mu)-P_{\rm out}(t)\right]  +\left[{\rm other ~terms}\right]. \nonumber
\ee
Here $\sigma$ is the key dimensional parameter, the domain wall tension $\sigma \sim f_{\pi}m_{\pi}f_{a}\sim m_a^{-1}$   as reviewed in Introduction, see eq.(\ref{sigma}).   The tension $\sigma$, in principle,  is also time-dependent parameter because the axion mass depends on time,  but for qualitative analysis  of this section    we ignore this time dependence  for now.  We return to this question later in the text.  Parameters $P_{\rm in}[\mu(t)]$ and $P_{\rm out}(t)$
represent the pressure inside and outside the bubble. The outside pressure    in QGP phase at high temperature can be estimated  as
\be
\label{P_out}
P_{\rm out}&\simeq& \frac{\pi^2 g^{\rm out}}{90}T_{\rm out}^4,  ~~~ T_{\rm out}\simeq T_0\left(\frac{t_0}{t}\right)^{1/2},\nonumber\\
g^{\rm out}&\simeq&\left(\frac{7}{8}4 N_cN_f +2(N_c^2-1)\right)
\ee
where $g^{\rm out}$ is the degeneracy factor, while  $T_0\simeq 100$ MeV and $t_0\sim 10^{-4}$s represents initial temperature and time determined by the cosmological expansion. We also assume that the thermodynamical equilibrium is maintained at all times between inside and outside regions such that the temperature inside the bubble approximately follows the outside temperature $T_{\rm out}(t)\simeq T_{\rm in}(t)$. Very quick  equilibration indeed is known to take place even in much faster processes such as heavy ion collisions. The fast equilibration in our case can be justified  because the heat transport between the phases is mostly due to the light NG bosons which can easily penetrate   the domain wall with little on no interaction, in contrast with  quarks and baryons discussed in the previous section. This assumption will be justified a  posteriori, see (\ref{exchange})  
on flux exchange rate between interior and exterior regions.   Therefore, we believe  our  approximation $T_{\rm out}(t)\simeq T_{\rm in}(t)$  is sufficiently good,  at least for qualitative estimates which is the main goal of this work.  

 The expression for the pressure inside the bubble $P_{\rm in}(t)$ depends on a number of quite nontrivial features of QCD  such   as the bag vacuum energy, corrections due to the gap in CS phase and many other phenomena, to be   discussed later in the text. 
 
 The equation of motion which follows from (\ref{lagrangian}) is
\be
\sigma \ddot R(t) = - \frac{2\sigma}{R(t)} - \frac{\sigma \dot R^2(t)}{R(t)} + \Delta P(\mu) 
-4\eta\frac{\dot{R}(t)}{R(t)},
\label{equation}
\ee
where $\Delta P[\mu(t)] \equiv  \left[P_{\rm in}(\mu)-P_{\rm out}(t)\right]$. 
We also inserted an additional term (which cannot be expressed  in the  Lagrangian formulation (\ref{lagrangian})), the shear viscosity $\eta$ to the right hand side of the equation, which effectively describes the ``friction" of the system when the domain wall bubble moves in   ``unfriendly" environment\footnote{We use conventional  normalization factor of $4\eta {\dot{R}(t)}/{R(t)}$ for the viscous term. This normalization  factor is the same which  appears in  the Rayleigh-Plesset equation in the classical  hydrodynamics when the viscous term, the surface tension term $ {2\sigma}/{R(t)}$ and the pressure term $\Delta P $ enter the equation in a specific combination as presented in (\ref{equation}). One should emphasize that our equation 
(\ref{equation}) describes the dynamics of the 2d surface characterized by the same surface tension $\sigma$   in contrast  with classical equation of the Rayleigh-Plesset equation describing a 3d spherical bubble in a liquid of infinite volume. This   difference explains some distinctions between the   kinetic terms proportional to   factor $\sim \sigma$    in our case (\ref{equation}) in contrast with   the classical Rayleigh-Plesset equation.}. On the microscopical level this term effectively accounts for a large number  of different effects  which do occur during the time evolution. Such processes include, but not limited to different scattering process by quarks,  gluons or Nambu Goldstone Bosons in different phases. All these particles and quasiparticles interact between themselves  and also  with a moving domain wall. Furthermore, the annihilation processes which   take place inside the bubble and which result in production of a    large number of 
strongly interacting quasi-particles also contribute to $\eta$.

Having discussed an expression for $P_{\rm out}(T)$  and viscous term $\sim \eta$  we now wish  to discuss the structure of the internal pressure $P_{\rm in}(\mu)$ which  enters (\ref{equation}). 
It has a number of contributions which are originated from very different physics. We  represent $P_{\rm in}(\mu)$  as as a combination of three  terms to be discussed one by one in order, 
\be
\label{P_in}
P_{\rm in}(\mu)\simeq    P_{\rm in}^{\rm (Fermi)}(\mu)+P_{\rm in}^{\rm (bag ~const)}(\mu)+P_{\rm in}^{\rm (others)}.~~~~
 \ee
 In this formula $ P_{\rm in}^{\rm (Fermi)}$
 can be represented as follows
 \be
 \label{P_fermi}
 P_{\rm in}^{\rm (Fermi)}(\mu) =\frac{ E}{3V}=\frac{g^{\rm in} }{6\pi^2} \int_0^{\infty}  
 \frac{k^3dk}{\left[\exp(\frac{\epsilon(k)-\mu }{T})+1\right]},~~
 \ee
 where we assume that quarks are massless and the chemical potential $\mu (t)$  implicitly  depends on time as a result of   the bubble's  evolution (shrinking). The degeneracy factor in this formula is 
 \be
g^{\rm in}\simeq  2 N_cN_f , 
\ee
where we keep only the quark contribution by  neglecting  the antiquarks. In other words, we simplify the problem by ignoring  the time dependence of the degeneracy factor $g^{\rm in}(t)$ which effectively varies as a result of $\mu(t)$ variation. 

Now we are in position to discuss $P_{\rm in}^{\rm (bag ~constant)}$ from (\ref{P_in}) which can be represented as follows
\be
\label{P_bag}
P_{\rm in}^{\rm (bag ~const)}(\mu)\simeq - E_B\cdot\theta\left[\mu-\mu_1\right] \left[1-\frac{\mu_1^2}{\mu^2}\right],~~
\ee
where positive parameter $E_B\sim (150 {\rm ~MeV})^4 $ is the famous ``bag constant" from MIT bag model, see \cite{Zhitnitsky:2002qa} for references and numerical estimates for this parameter in the given context  of the nugget structure. The bag constant can be expressed in terms of the gluon and quark condensates in QCD. We shall not elaborate on this problem in the present work by  referring to 
\cite{Zhitnitsky:2002qa} with relevant studies in the given context. 
 
The bag ``constant" $E_B$  describes the differences of vacuum energies (and therefore, vacuum pressure)  in the interior and exterior regions of the nuggets. This difference occurs in our context because the phases realized outside   and inside of the nugget are drastically distinct. For example, at the end of formation the outside region of the nugget is in cold hadronic phase, while inside region is in CS phase. The vacuum energies in these two phases are known to be drastically different.  This term works as a ``squeezer",  similar to the role it plays in the MIT bag model, when the vacuum energy outside of the nugget is lower than the vacuum energy inside the nugget.   Therefore it enters with the same sign minus as the domain wall pressure. 

A specific $\mu$-  dependence  used in (\ref{P_bag}) is an attempt to model a known feature of QCD that the absolute value of the vacuum energy decreases when the chemical potential increases. This feature is well established and tested in conventional nuclear matter physics, and it was analytically derived in simplified version of QCD with number of colours $N_c=2$, see \cite{Zhitnitsky:2002qa} for references and details. Our parametrization (\ref{P_bag}) corresponds to the behaviour when 
$P_{\rm in}^{\rm (bag ~constant)}(\mu)=0$ for small chemical potentials $\mu\leq \mu_1$ when the vacuum energy inside and outside of the nuggets approximately equal. This term becomes very important  ``squeezer" at large chemical potential at $\mu\geq \mu_1$  when  the system outside is in the hadronic vacuum state while inside it  is in a CS phase. The numerical value for parameter $\mu_1$ can be estimated as $\mu_1\sim 330~ $MeV \cite{Zhitnitsky:2002qa} when the baryon density is close to the nuclear matter density. 

The last term entering (\ref{P_in}) and coined as $P_{\rm in}^{\rm (others)}(\mu)$
 is due to a large number of 
other effects which we ignore in present work. In particular, there is a conventional contribution due to the boson degrees of freedom which cancels the corresponding  portion of  $g^{\rm out}$ from (\ref{P_out}) at high temperature, $T\gg \mu$. It does not play any important role in our analysis because we are mainly concerned with analysis of fermion degrees of freedom and building the chemical potential inside the bubble.   Another effect which worth to be mentioned  is  the formation of the gap in CS phase  due to the quark pairing, similar to formation of the gap in  conventional superconductors. The generation of the gap obviously decreases the energy of the system.  There are many other phenomena which are known to occur in CS phase \cite{Alford:2007xm}. However, we expect that these effects  are less important in comparison with the dominating contributions which are explicitly written down,  (\ref{P_fermi}) and (\ref{P_bag}). 

	The equation (\ref{equation}) can be numerically solved for $R(t)$ if time variation of the chemical potential $\mu(t)$ entering (\ref{P_fermi}) and (\ref{P_bag}) is known. 
	To study   the corresponding time evolution for the chemical  potential  $\mu (t)$ we use expression (\ref{B})  for the baryon charge bounded to the domain wall.  We assume that the thermodynamical equilibrium is maintained between internal and external parts of the nugget such that $T_{\rm in}(t)\simeq T_{\rm out}(t)$.  This assumption will be justified a posterior, see discussions after eq.(\ref{exchange}). At the same time the chemical potential is quickly increasing with time   inside the nugget due to  decreasing of   the nugget's size.   We also assume a fast equilibration for the chemical potential within    the nugget in its  entire volume. In other words,    we describe the system using one and the same chemical potential in vicinity of  the wall and deep inside the bubble. 
	Justification for this assumption will be given later in the text. 
	
	With  this picture in mind, we proceed by differentiating eq.(\ref{B}) with respect to time to arrive to the following implicit equation relating $\mu(t)$ and $R(t)$ at fixed temperature $T$, 
\be
\label{flux}
&&\dot{B}=\frac{N g}{4\pi^2} \dot{S}(t)\int \frac{d^2 k_{\perp}}{\left[\exp(\frac{\epsilon-\mu (t)}{T})+1\right]}\\ &+&\frac{N g S}{4\pi^2} \frac{\dot{\mu}(t)}{T} \int \frac{d^2 k_{\perp}\left[\exp(\frac{\epsilon-\mu(t)}{T})\right]}{\left[\exp(\frac{\epsilon-\mu(t)}{T})+1\right]^2} +(\rm fluxes) =0, \nonumber
\ee
where term ``fluxes" in (\ref{flux}) describes the loss of baryonic matter  due to annihilation   and other processes  describing incoming and outgoing fluxes, to be discussed later in the text. The relation (\ref{flux}) gives an implicit relation between $\mu(t)$ and $R(t)$ which   can be used for numerical studies of  our equation (\ref{equation}) describing the  time evolution
of the  system.  

We shall discuss the physics related to  incoming and outgoing fluxes in Appendix \ref{appendix:flux}. 
If  we neglect this  term  which    describes the loss of baryonic matter  we  
can analytically solve  (\ref{flux}) for small $\mu\ll T$ when one can use the Taylor  expansion of the integrals entering (\ref{flux}). The result is
\be
\label{small_mu}
\left(\mu (t)-\mu_0\right)\simeq \frac{\pi^2 T}{6\ln2}\ln \left(\frac{R_0}{R(t)}\right), 
\ee
where $R_0$ is initial size of the system at $t=t_0$ while $\mu_0\simeq 0$ is initial chemical potential.
One can explicitly see that the chemical potential builds in very fast when the nugget  reduces its size only slightly. This formula (\ref{small_mu}) is only justified for very small $\mu (t)$. For larger values of $\mu$ one should use exact formula (\ref{flux}). 

Finally, one should note that at the end of formation at time $t\rightarrow \infty$ when  temperature $T \ll \mu$ the evolution stops,  in which case all derivatives vanish,     $ \ddot{R}_{\rm form} =  \dot{R}_{\rm form}=  \dot{\mu}_{\rm form}= 0 $.   At this point    the nugget assumes its final configuration with size ${R}\simeq {R}_{\rm form}$, and the equation (\ref{equation}) assumes the form 
\be
   \frac{2\sigma}{R_{\rm form}} =   P_{\rm in} =\frac{g^{\rm in} \mu^4}{24\pi^2}-E_B\left(1-\frac{\mu_1^2}{\mu^2}\right), ~~ \mu\geq \mu_1.~~
\label{equlibrium}
\ee
This condition is precisely  the equilibrium condition  studied in \cite{Zhitnitsky:2002qa} with few neglected contributions (such as the quark-quark interaction leading to the gap). This is of course expected result as the time evolution, which is the subject of the present work,  must lead to     the equilibrium configuration  when the free energy assumes its minimum determined by (\ref{equlibrium}).  

One should recall that  analysis of the equilibrium presented in  ref. \cite{Zhitnitsky:2002qa} with typical QCD parameters strongly suggests that the system indeed falls  into CS phase when the axion domain wall pressure 
$\sigma$ assumes its conventional value. At the same time, the equilibrium is not likely to emerge   with the same   typical QCD parameters without   an additional external pressure related to the axion domain wall. In this sense the axion domain wall with extra  pressure due to $\sigma\neq 0$  plays the role of an additional  ``squeezer" stabilizing the nuggets.

The key   element of this section  is  equation  (\ref{flux}) which is the direct  consequence  of a spontaneous accretion of the baryon (or antibaryon) charge in the domain wall background as discussed in section \ref{baryons}. Precisely this equation explicitly shows that the chemical potential $\mu(t)$  grows very fast when the domain wall   shrinks  as a result of the  domain wall pressure $\sigma$. The presence of a  non-vanishing   chemical potential in the vicinity of the domain wall obviously implies the generation of the   binding forces between the fermions and the domain wall, such that a typical bound energy of a single fermion  to the domain wall is of order of $\mu$. 

A generic solution of equations  (\ref{equation}) and (\ref{flux}),  as we discuss in next section,   shows an  oscillatory behaviour  of $R(t)$  with a slow damping of the amplitude  such that the system  eventually   settles  down  at the equilibrium point (\ref{equlibrium}).  However, even the very first oscillation with  initial  $\mu_0\approx 0$
leads to very fast growth of the chemical potential $\mu(t)\approx T$ as analytical estimates represented by eq.(\ref{small_mu}) shows.  In next section we develop a quantitative framework which allows us to analyze our basic equation  (\ref{equation})  for $R(t)$ where time dependence $\mu(t)$ is implicitly expressed in terms of the same variable $R(t)$ as determined by (\ref{flux}).

\section{Formation of the nuggets.   Qualitative analysis.}\label{analysis}
  Our goal in this section is to solve for $R(t)$ and therefore $\mu(t)$ by solving (\ref{equation}) and (\ref{flux}), which implicitly relate both variables. We shall observe  that a nugget  experiences  large number of oscillations during its evolution with  slow damping rate,
  and eventually   settles  down  at the equilibrium point (\ref{equlibrium}). This  behaviour of the system will be coined  as  ``underdamped oscillations".    In next  section \ref{algebraic}  we formulate some assumptions and present the  technical details, while  the interpretation of the obtained results will be presented   in section \ref{interpretation}. We want to make a number of simplifications in our analysis in present section to demonstrate the generic features  of these oscillations. The  numerical studies presented in Appendices \ref{appendix:flux}, \ref{numerics} and \ref{integral}
  support our basic picture of oscillatory behaviour advocated in this section. 
    
\subsection{Assumptions, approximations, simplifications }\label{algebraic} 
Exact analytical analysis  of either (\ref{equation}) or (\ref{flux}) can be obtained only during the first moment of the initial stage of   evolution of the system when $\mu$ is sufficiently small (\ref{small_mu}). We need to understand the behaviour of the system for a much longer period of time. Thus, we make  two important  technical simplifications to proceed with our qualitative  analysis.  The first one is to neglect the   term in (\ref{flux}) describing the fluxes. This assumption will be supported by some estimates presented  in Appendix \ref{appendix:flux} which show that incoming and outgoing fluxes cancel each other with very high accuracy, such that net flux is indeed quite small.   Hence, (\ref{flux}) is now simplified to:
\begin{equation}
\label{eq:6.B0}
\dot{B}=\frac{d}{dt}\left\{\frac{Ng}{4\pi^2}S
\int\frac{d^2k_\perp}{\operatorname{exp}(\frac{\epsilon-\mu}{T})+1
}\right\}=0
\end{equation}
which means in this approximation, the baryonic charge is roughly conserved in the domain wall background  at all times during the evolution of the system. 

As our  second simplification we neglect the mass of the fermions in comparison with temperature $T$ and the chemical potential $\mu$, i.e. we use the following dispersion relation $\epsilon=\sqrt{k_\perp^2+m^2}\simeq k_\perp$ in vicinity of the domain wall.  This approximation is somewhat  justified in QGP and CS phases, and therefore along the path 3 as shown  on Fig. \ref{fig:phase_diagram}. It is not literally justified for paths 1 and 2 as in the hadronic phase where the quark mass $m$ should be identified with the so-called  ``constituent" quark mass which is proportional to the chiral condensate.   Nevertheless, to simplify the problem we neglect  
the mass $m(T)$ for all paths in our qualitative analysis of the  time evolution  as we do not expect any  drastic changes  in our final outcome
as a result  of   this technical simplification. 
With these assumptions we can  approximate    the integral entering eq.  (\ref{eq:6.B0}) as follows,
\be 
\label{eq:6.integral}
& & \int_0^\infty\frac{dk_\perp
  \cdot k_\perp}{e^{\frac{\epsilon (k_{\perp})-\mu}{T}}+1} = T^2\cdot I(\frac{\mu}{T}) \\
& &I(\frac{\mu}{T})\simeq\frac{\pi^2}{6}+\frac{1}{2}\left(\frac{\mu}{T}\right)^2
-\frac{\pi^2}{12}e^{-\mu/T}+\mathcal{O}(\frac{\mu}{T}e^{-\mu/T}) \nonumber
\ee
where the omitted terms  $\sim \frac{\mu}{T}e^{-\mu/T}$ will be  neglected thereafter, as they  will never dominate in neither small nor large limit of $\mu$. One can numerically check   that   approximation (\ref{eq:6.integral})  describes the relevant integral $ I(\frac{\mu}{T}) $ sufficiently well  in the entire parametrical space of $\mu/T$, see Appendix \ref{integral} with corresponding analysis. As a quick test  of this approximation one can check that approximate expression   (\ref{eq:6.integral}) reproduces an exact (in the small $\mu$ limit) expression (\ref{small_mu})    with accuracy of order $15\%$, which is more than sufficient for our qualitative studies  of this section.

As mentioned above, if flux term (\ref{flux}) is neglected, the curly-bracket term in (\ref{eq:6.B0}) is a conserved quantity. Equating it to its initial values  where $S(t=0)=4\pi R_0^2$, $\mu(t=0)=\mu_0\simeq0$ one arrives to
\begin{equation}
T^2R^2\left[\frac{\pi^2}{6}+\frac{1}{2}
\left(\frac{\mu}{T}\right)^2-\frac{\pi^2}{12}e^{-\mu/T}\right]
=\frac{\pi^2}{12}T_0^2R_0^2.
\end{equation} 
In what follows we assume that the  thermodynamical equilibration is established very quickly such that  one can approximate $T\simeq T_0$ during the time evolution as we already discussed in previous section \ref{time-evolution}. To simplify further the system  
we  wish  to represent the equation relating $R$ and $\mu/T$ in the following form
\be
\label{eq:6.B1}
f(R)\equiv\frac{\pi^2}{6}\left[\frac{1}{2}
\left(\frac{R_0}{R}\right)^2-1\right]
=\frac{1}{2}\left(\frac{\mu}{T}\right)^2-\frac{\pi^2}{12}e^{-\mu/T},~~~~~
\ee
where we introduced  function $f(R)$ for convenience   of  the analysis which follows. 
Essentially, the   idea here is to simplify the basic equation (\ref{equation}) as much as possible to express the $\mu(t)-$ dependent terms entering  through the    pressure (\ref{P_in}) in terms of $R(t)$ such that the equation (\ref{equation}) would assume a  conventional differential equation form for  a single variable $R(t)$.

 Our next step is to simplify the expression for the Fermi pressure (\ref{P_fermi}) entering (\ref{P_in}) using the same  procedure we used to approximate  formula (\ref{eq:6.integral}), i.e.
\be 
\label{P_fermi1}
&&P_{\rm in}^{\rm (Fermi)}
=\frac{g^{\rm in}}{6\pi^2}\int_{0}^{\infty}
\frac{k^3dk}{\exp\left(\frac{\epsilon (k)-\mu}{T}\right)+1}   \\
&&\simeq\frac{g^{\rm in}T^4}{6\pi^2}
\left\{\frac{7\pi^4}{60}+\frac{\pi^2}{2}\left(\frac{\mu}{T}\right)^2
-\frac{7\pi^4}{120}e^{-\mu/T}
+\frac{1}{4}\left(\frac{\mu}{T}\right)^4\right.  \nonumber \\
&&\left.\qquad\qquad\quad
+\mathcal{O}(\frac{\mu}{T}e^{-\mu/T})
\right\}  \nonumber \\
&&\simeq\frac{g^{\rm in}T^4}{6}
\left\{\frac{7\pi^2}{60}
+\left[\frac{1}{2}\left(\frac{\mu}{T}\right)^2
-\frac{\pi^2}{12}e^{-\mu/T}\right]  
+\frac{1}{4\pi^2}\left(\frac{\mu}{T}\right)^4
\right\} \nonumber\\
&&+\frac{g^{\rm in}T^4}{6}\left\{ \frac{\pi^2}{40}e^{-\mu/T} + \mathcal{O}(\frac{\mu}{T}e^{-\mu/T})  \right\}. \nonumber
\ee
In what follows we neglect the last line in eq. (\ref{P_fermi1}). The justification for this procedure is the same as before:  
 it produces a  small contribution in entire region of  $\mu$ in comparison with accounted terms. The technical advantage for this procedure is the possibility to rewrite (\ref{P_fermi1})
 in terms of function of $R(t)$, rather than $\mu(t)$  using our relation (\ref{eq:6.B1}). 
 
The formula in the square bracket in (\ref{P_fermi1}) is just   $f(R)$   defined by (\ref{eq:6.B1}). The remaining $\left(\frac{\mu}{T}\right)^4$ term can be also expressed in terms  of $R$ by taking square of (\ref{eq:6.B1}):
\begin{equation}
\begin{aligned}\label{f}
 f^2(R) 
&=\left[\frac{1}{2}\left(\frac{\mu}{T}\right)^2
-\frac{\pi^2}{12}e^{-\mu/T}\right]^2  \\
&\simeq\frac{1}{4}\left(\frac{\mu}{T}\right)^4
+\left(\frac{\pi^2}{12}\right)^2
+\mathcal{O}\left(\frac{\mu}{T}e^{-\mu/T}\right)  \\
\end{aligned}
\end{equation}
where the correction term $\sim \mathcal{O}({\mu}e^{-\mu/T}) $ will be dropped in what follows, as before. Thus, we approximate  $P_{\rm in}^{\rm (Fermi)}$ in terms of $R(t)$ as follows
\begin{equation}
P_{\rm in}^{\rm (Fermi)}
\simeq\frac{g^{\rm in}T^4}{6}\left[
\frac{7\pi^2}{60} +f(R)+\frac{f^2(R)}{\pi^2}
-\frac{\pi^2}{144}
\right].
\end{equation}
The expression for the Fermi pressure $P_{\rm in}^{\rm (Fermi)} (R)$ now is expressed in terms of $R$ rather than in terms of $\mu$ as in the original  expression (\ref{P_fermi}).

We wish  to simplify the expression for  $P_{\rm in}^{\rm (bag~ const)}(\mu)$ entering (\ref{P_in}) in a similar manner to express $P_{\rm in}^{\rm (bag~ const)}$ in terms of $R$.
This contribution becomes important as discussed after eq. (\ref{P_bag}) only for sufficiently large $\mu$. In this region $f(R)$ can be well approximated as 
\begin{equation}
\label{eq:6.big-mu}
f(R)\simeq\frac{1}{2}\left(\frac{\mu}{T}\right)^2, ~~~ \mu\gg T
\end{equation}
so that we have
\be
P_{\rm in}^{\rm bag}
\simeq-E_B\cdot\theta\left(\sqrt{2f(R)}-\frac{\mu_1}{T}\right)
\left(1-\frac{\mu_1^2}{2T^2f(R)}
\right). ~~~
\ee
As a result of these simplifications and approximations   the pressure term which enters the basic equation (\ref{equation}), 
$\Delta P(\mu)\equiv [P_{\rm in}(\mu)-P_{\rm out}(t)]$ which was initially formulated in terms of the chemical potential $\mu$ inside the bubble can be now written entirely in terms of a single variable, the    size of the bubble $R(t)$:
\be
\label{eq:6.P0}
&&\Delta P[f(R)]
\simeq\frac{g^{\rm in}\pi^2}{6}T^4\left[
\frac{79}{720}-\frac{g^{\rm out}}{15g^{\rm in}}+\frac{f(R)}{\pi^2}+\frac{f^2(R)}{\pi^4}
\right]  \nonumber \\
&&\quad-E_B\cdot\theta\left(\sqrt{2f(R)}-\frac{\mu_1}{T}\right)
\left(1-\frac{\mu_1^2}{2T^2f(R)}
\right), 
\ee
where $f(R)$ is defined  by eq. (\ref{eq:6.B1}).
With these technical simplifications the basic equation  (\ref{equation}) can now be written as a second order differential equation entirely  in terms of $R(t)$ rather than $\mu$:
\begin{equation}
\label{eq:6.R0}
\sigma \ddot{R}(t)
=-\frac{2\sigma}{R}-\frac{\sigma\dot{R}^2}{R}
+\Delta P[f(R)]-4\eta\frac{\dot{R}}{R}, 
\end{equation}
with $\Delta P[f(R)]$ determined by eq. (\ref{eq:6.P0}).
 
This equation   can be solved numerically. In fact, it is precisely   the subject of  Appendix \ref{numerics}. 
However, the most important quantitative features  of the obtained solution can be understood without any numerical studies, but rather using a simplified  analytical analysis, which is precisely the subject of the next  section. 

\subsection{Time evolution. Qualitative analysis. }\label{interpretation}  

As we already mentioned a nugget assumes its   final form    at $ t\rightarrow \infty$   when all time derivatives vanish and the equation for the  equilibrium is given by (\ref{equlibrium})  at $T=0$. In this section we generalize this  equation  for the equilibrium by 
  defining   $R_{\rm form}(T)$ as the  solution of eq.(\ref{eq:6.Rform}), see below,  at $T\neq 0$. 
 In other words, the starting point of  the  present analysis at $T\neq 0$  is the equilibrium condition when the ``potential" energy assumes its minimal value. The corresponding minimum condition 
is determined by equation 
\begin{equation}
\label{eq:6.Rform}
\frac{2\sigma}{R_{\rm form}}=\Delta P(R_{\rm form}), 
 \end{equation}
 where  $\Delta P(R_{\rm form})$ is defined  by eq.(\ref{eq:6.P0}). This condition obviously reduces to eq. (\ref{equlibrium}) at $ t\rightarrow \infty$ when $ \mu\gg T$. 

We follow the conventional technique and  expand (\ref{eq:6.R0})  around the equilibrium value $R_{\rm form}(T)$ to arrive to an equation for a  simple damping oscillator:
\begin{equation}
\label{eq:6.R1}
\frac{d^2(\delta R)}{dt^2}+\frac{2}{\tau}\frac{d(\delta R)}{dt}+\omega^2(\delta R)=0, ~~~ 
\end{equation}
where $\delta R\equiv [R(t)-R_{\rm form}]$ describes the deviation from the equilibrium position, while new parameters   $\tau$ and $\omega$ 
describe the effective  damping coefficient  and   frequency of the oscillations. Both new coefficients   are    expressed in terms of  the original parameters entering (\ref{eq:6.R0}) and are given by  
\begin{subequations}
\label{eq:6.R1tw}
\begin{equation}
\label{eq:6.R1t}
\tau=\frac{\sigma}{2\eta}R_{\rm form}
\end{equation}
\begin{equation}
\label{eq:6.R1w}
\omega^2=\left.-\frac{1}{\sigma}\frac{d\Delta P (R)}{dR}\right|_{R_{\rm form}} -\frac{2}{R_{\rm form}^2}.
\end{equation}
\end{subequations}
The expansion (\ref{eq:6.R1}) is justified, of course, only  for small oscillations about the minimum determined by eq.(\ref{eq:6.Rform}), while the oscillations determined by original equation (\ref{eq:6.R0}) are obviously  not small.  
However, our  simple analytical treatment  (\ref{eq:6.R1}) is quite instructive and gives a good qualitative understanding of the system.
Our numerical studies presented in Appendix \ref{numerics} fully support the qualitative picture presented below.

We start our qualitative analysis with estimates of the parameter $\omega$  which depends on   $\frac{d\Delta P (R)}{dR}$ computed at $R=R_{\rm form}$ according to (\ref{eq:6.R1w}).
First of all, in this qualitative analysis we neglect  the  bag constant term  $P_{\rm in}^{\rm (bag ~constant)}$ because 
it only starts to play a role for sufficiently large 
  $\mu\geq \mu_1\sim 330~ $ MeV,  when formation is almost completed.  This term obviously cannot change the  qualitative behaviour 
  of the system discussed below.   Our numerical studies presented in Appendix \ref{numerics} (where the  bag constant term  $ \sim E_B$ is included in the analysis)  support this claim. 
  
 The key element  for our simplified analysis is the observation that  the ratio $(R_0/R_{\rm form} )^2\geq  14$ is expected to be numerically     large number.  This expectation   will be  soon    confirmed a posteriori. This observation  considerably simplifies our qualitative analysis because in this case $\Delta P(R_{\rm form})$ defined by (\ref{eq:6.P0})   can be approximated by a single term $\sim f^2(R)$  in square brackets in (\ref{eq:6.P0})  as this term  essentially saturates $\Delta P(R_{\rm form})$. This is because  the function  $f(R)/\pi^2$ becomes numerically large in the relevant region  $f(R)/\pi^2\sim (R_0/R_{\rm form})^2$
according to (\ref{eq:6.B1}). 

With these simplifications we can now estimate $\omega^2$ as follows
\begin{equation}
\label{eq:6.R1w2}
\omega^2
\approx \left(\frac{g^{\rm in}\pi^2}{216}\right)\cdot \left(\frac{T^4}{\sigma R_{\rm form}}\right)\cdot 
\left(\frac{R_0}{R_{\rm form}}\right)^4-\left(\frac{2}{R_{\rm form}^2}\right).
\end{equation}
To simplify  analysis further one can represent the last term as 
\be
\left(\frac{2}{R_{\rm form}^2}\right) = \left(\frac{1}{R_{\rm form}}\right) \cdot \left( \frac{\Delta P(R_{\rm form})}{\sigma}\right),
\ee
and keep the leading term $\sim f^2(R)$ in expression for $\Delta P(R_{\rm form})$. One can easily convince yourself that $\omega^2>0$ is always positive in this approximation such that the condition for a desired underdamped oscillations assumes a simple form
\be
\label{R_0}
\frac{f(R_{\rm form})}{\pi^2} \gtrsim 1 ~~ \Rightarrow ~~  \left(\frac{R_0}{R_{\rm form}}\right) \gtrsim \sqrt{14}
\ee
when $\Delta P(R_{\rm form})$ defined by (\ref{eq:6.P0})  is dominated by a single term $\sim (\frac{f}{\pi^2})^2$, which itself can be approximated by the leading quadratic term $\sim \left(\frac{R_0}{R}\right)^2$ according  to (\ref{eq:6.B1}). Our numerical studies presented in Appendix \ref{numerics}    support the numerical  estimate (\ref{R_0}). 

One can also check that if condition (\ref{R_0}) is not satisfied than  system shows an ``over-damped"  behaviour 
when very few oscillations occur before   complete collapse of the system, in which case the nuggets obviously  do not  form. 
These short-lived bubbles  will never get to a stage when the temperature drops below the critical value $T_{CS}$. Therefore, a  CS phase  cannot form in these ``short-lived" bubbles. It should be contrasted with ``long-lived" bubbles  with much longer  formation-time   of order $\tau$, see comments below.

The condition (\ref{R_0}) is extremely important for our analysis. It essentially states that the initial size of a closed bubble  $R_0$ must be sufficiently large for a successful formation of a nugget of size $R_{\rm form}$. On other hand, a formation of very large closed bubbles is strongly suppressed $\sim \exp[-(R_0/\xi)^2]$ by the KZ mechanism as  reviewed in section \ref{formation}.  This constraint will be important in our estimation of a suppression factor  in section \ref{asymmetry2}  due to necessity to form a sufficiently large   bubble (\ref{R_0})  during the initial stage of formation. 

Assuming that condition (\ref{R_0}) is satisfied we estimate a typical frequency oscillations as follows
\be
\label{omega}
\omega\sim \frac{1}{R_{\rm form}}\sim m_a, ~~~~ t_{\rm osc}\simeq \omega^{-1}\simeq m_a^{-1}
\ee
where we used the scaling properties (\ref{sigma}) to relate the nugget's size ${R_{\rm form}}$ with the axion mass $m_a$.
One should emphasize that the estimate (\ref{omega}) is not sensitive to any approximations and simplifications we have made in our qualitative treatment of the time evolution in this section. In fact, all parameters entering relation (\ref{omega}) are expressible in terms of the QCD scale $\Lambda_{\rm QCD}$
and a single ``external" parameter, the axion mass $m_a$, which we keep unspecified at this point. Of course we always assume  that the axion mass may  take  any value from the 
observationally allowed window $10^{-6} {\rm eV}\lesssim m_a \lesssim  10^{-3}{\rm eV}$. 

We now turn our attention to the damping coefficient defined in terms of the original parameters by eq. (\ref{eq:6.R1t}).
It is convenient to  estimate the dimensionless combination   $\omega\tau$ as follows
\be
\label{tau}
\omega \tau\simeq \frac{1}{R_{\rm form}}\cdot \left(\frac{\sigma}{2\eta}R_{\rm form} \right)\simeq \frac{\sigma}{2\eta}\sim \frac{m_{\pi}}{m_a}\sim 10^{11}, 
\ee
where we substituted $\omega\sim R^{-1}_{\rm form} $ according to (\ref{omega}) and assumed that $\eta\sim m_{\pi}^3$ has conventional QCD scale of order $\rm{fm}^{-3}$ while the wall tension $\sigma$ can be approximated with high accuracy as  $\sigma\simeq m_{\pi}^4/m_a$.
 This relation implies that the damping is extremely slow on the QCD scales. Therefore, the solution describing the time evolution of a ``long-lived"  bubble can be well approximated as follows  
 \begin{equation}
\label{eq:6.R2}
R(t)=R_{\rm form}+(R_0-R_{\rm form})e^{-t/\tau}\cos\omega t
\end{equation}
which is obviously a solution of the approximate equation (\ref{eq:6.R1}). This solution represents an ``under-damped" oscillating $R(t)$  with frequency $\omega\sim\frac{1}{R_{\rm form}}$ and damping time $\tau\sim\frac{\sigma}{2\eta}R_{\rm form}$. Precisely these ``long-lived"  bubbles will  eventually form the DM nuggets.  

The time scale (\ref{tau}) is very suggestive and implies that the damping term starts to play a role   on very large  scales when the cosmological expansion of the Universe with the typical scale $t_0\simeq 10^{-4} s$ must be taken into account. We have not included the corresponding temperature variation in our studies because on the QCD scales (which is the subject of the present studies)  the corresponding variations are negligible. However, the estimate (\ref{tau}) shows that for a proper analysis  of the   time scales $\tau$ the expansion of the Universe (and related to the expansion the temperature variation) must be included. The corresponding studies are  beyond the scope  of the present work. However, the important comment we would like to make here is that the emergent large time scale (\ref{tau}) is fully consistent with our anticipation that the temperature of the Universe drops 
approximately by a factor of $\sim 3 $ or so when a CS phase forms  in interior of  the nugget during the formation period. It is quite obvious that if the time scale (\ref{tau}) were considerably shorter than  the cosmological time scale $t_0\simeq 10^{-4} s$ than the temperature $T\sim t^{-1/2}$ inside the nugget could not drop sufficiently deep into the region where CS sets in as plotted on Fig.\ref{fig:phase_diagram}.  Fortunately, the timescale (\ref{tau}) is long enough and automatically satisfies this requirement. 

 Now we want to elaborate on one more element of the dynamics which is also important for a  successful formation of the nuggets. 
To be more specific, we want to discuss  the flux of particle exchange, which was ignored in our qualitative analysis in this section and which is estimated in Appendix \ref{appendix:flux}. This flux describes the rate of number of particle flowing between inside and outside the system, which can be appreciably large even if the net baryonic flux is negligibly small. To be more precise, there are two kinds of fluxes, both investigated in Appendix  \ref{appendix:flux}, that we are discussing in this paper: the net flux of baryonic charge $\Delta \Phi\equiv\Phi_\Rightarrow-\Phi_\Leftarrow$, and the average  flux of particle number $\la \Phi \ra\equiv\frac{1}{2}(\Phi_\Rightarrow+\Phi_\Leftarrow)$. The first one corresponds to the flux term  entering eq. (\ref{flux}); while the latter is important in understanding what is the typical time scale for  a complete ``refill" of the particles  during the time evolution. The last question  is important for understanding  of the  time scale for   thermal equilibration.

We start our analysis with discussions of an average flux   $\la \Phi \ra $ at small chemical potential. It is estimated to be $\la \Phi \ra\simeq 1~ {\rm fm}^{-3}$ according to Appendix \ref{appendix:flux}.   The magnitude of this flux can be fully appreciated   by computing the total number of particle exchange per one cycle of the oscillation 
\begin{equation}
\label{exchange}
\frac{2\pi}{\omega}\cdot  4\pi R_{\rm form}^2 \cdot \la \Phi \ra 
\sim R_{\rm form}^3\rm{fm}^{-3}\sim \left|B\right|,
\end{equation}
where $\omega$ is a typical frequency oscillation estimated in (\ref{omega}) while 
 $\left|B\right|$ is the total number of particles (quarks and antiquark) stored in the nugget.  The physical meaning of this estimate is that
  a nugget can in principle  entirely refill its interior  with ``fresh'' particles within a few cycles of exchange.  Similar estimate for the net baryon flux which includes $\Delta \Phi$ is suppressed, see Appendix \ref{appendix:flux}.  
  
  The main reason for emergence of this large scale in expression (\ref{exchange}) is a long time scale of a single cycle (\ref{omega}) which is determined by the axion mass $m_a$ rather than by QCD physics. Nevertheless, estimate (\ref{exchange}) is quite remarkable and shows that even very 
  low rate of chemical potential accretion  of (anti)quarks being tracked per oscillation, the high exchange rate (\ref{exchange})  is still sufficient enough to turn a baryonically neutral nugget into one completely filled with (anti)quarks. When the quarks become effectively massive as it happens in hadronic and CS phases, the flux for the exchange of the baryon charge  is drastically decreased  by a factor $\sim \exp(-{m}/{T})$.
  
  The same  estimate (\ref{exchange})  essentially holds  for exchange of almost massless Nambu-Goldstone bosons for sufficiently high temperature.  In fact, 
 the lightest degrees of freedom play the crucial role in cooling processes  of the interior of the nugget  as these particles can easily penetrate 
 the sharp domain wall structure. Therefore, the high  exchange rate  between exterior and interior of a nugget  essentially implies that the thermal equilibrium is maintained in our system with very high precision due to a huge    rate  per cycle (\ref{exchange}) when large number of degrees  of freedom  $\sim B$ have   a chance of order one to interact with ``fresh" particles from the exterior  during a single cycle.   Therefore, our assumption 
 on thermal equilibrium between interior and exterior is justified a posteriori. 
 
 We conclude this section with few  important comments. 
 The most important result of this section is that the nuggets can be formed   during the QCD phase transition provided the initial size of the nuggets is sufficiently large as stated in eq.(\ref{R_0}), in which case they survive the evolution. The key role in  this  successful formation plays, of course, the effect of ``local spontaneous violation" of the baryon symmetry as discussed in section   \ref{baryons} and explicitly expressed by eqs.(\ref{bc}) and (\ref{N}). One should emphasize that our qualitative analysis in this section  are fully supported by numerical studies presented in Appendices \ref{appendix:flux} and \ref{numerics}. Therefore, we do not expect that any   numerical simplifications  in our analysis may drastically change the  basic qualitative results  presented in this section. 
 
 Another important point is the observation (\ref{omega}) that a typical time scale for the oscillations is of order $t_{\rm osc}\simeq \omega^{-1}\simeq m_a^{-1}$. Both these estimates will be crucial elements  in our analysis presented in next section \ref{asymmetry}: equation (\ref{R_0}) will be important in estimate for efficiency of a bubble formation with a large size $\sim R_0$, while equation (\ref{omega}) will play a key role in our arguments suggesting a coherent preferential  formation   of one type of  nuggets (baryonic or antibaryonic) on largest possible  scale of the visible Universe.

\section{\label{asymmetry} Baryon  charge separation.  Correlation on cosmological  scales.}
Until this section we mostly concentrated on the   time evolution of a single nugget (or anti-nugget). 
The main lesson of our previous discussions is that such nuggets can be formed, remain stable configurations,  and therefore, can serve as the dark matter candidates. In other words, the focus of our previous studies was  a problem  of a  local  separation of charges  on small scales of order nugget's size. The key element of that separation of charges  is eq. (\ref{bc}) which can be thought as a local version of spontaneous symmetry breaking of the baryon charge as explained in section \ref{baryons}.  However, on a larger scale it is quite obvious that equal number of nuggets and anti-nuggets will be formed as a result of an  exact symmetry as we discuss below. 

This symmetry, however,  does not hold anymore on large scales if the axion $\cal{CP}$ -odd coupling is included into consideration, which eventually leads to 
very generic, essentially insensitive to most parameters,  consequence of this framework represented by eq.(\ref{Omega}), which is the subject of next subsections \ref{asymmetry1} , \ref{generic}.  
   The subsection \ref{asymmetry2} is devoted to some more specific and  model-dependent consequences of this framework.  In particular,  we want to estimate a suppression factor  related to a necessary to form   a large size  bubble (\ref{R_0}) in KZ mechanism.

\subsection{\label{asymmetry1} Coherent axion field as the source of $\cal {CP}$ violation}
First of all, let us show that the baryon charge hidden in nuggets on average is equal to the baryon charge hidden in  anti-nuggets, of course with sign minus.  
Indeed, the analysis of the anti-nuggets can be achieved by  flipping  the sign of  the chemical potential in eq. (\ref{L}), i.e. $\mu\rightarrow -\mu$. 
One can restore the original form of the $\mu$ term in Lagrangian (\ref{L}) by replacing $\theta_1\rightarrow -\theta_1$ and $\theta_2\rightarrow -\theta_2$. Finally, one should  change the signs for the axion $\theta$ and the pseudo-scalar singlet  $\eta'$ meson represented by  $\phi$ field 
  in the interaction  term (\ref{U}) to restore the original form of the Lagrangian. These  symmetry arguments imply that as long as the  pseudo-scalar  axion field fluctuates around zero as conventional pseudo-scalar fields (as $\pi, \eta'$ mesons, for example), the theory remains invariant under $\cal {P}$ and $\cal {CP}$ symmetries. Without this symmetry the number density and size distribution of the nuggets and anti-nuggets could be drastically different\footnote{If $\pi $ meson condensation were occur   in nuclear matter it would unambiguously  imply that the $\cal {CP}$ invariance is broken in such a phase. Some of the phases in CS systems indeed break the $\cal {CP}$ invariance as a result of condensation of a pseudo-scalar Nambu-Goldstone bosons.}.  

Therefore, the symmetry arguments suggest that on average an equal number of nuggets and anti-nuggets would form if the axion field is represented by a conventional quantum fluctuating field oscillating around zero point.  If it were the case, the   baryons and anti baryons   would continue to annihilate each other as well as annihilate with the nuggets and anti-nuggets in our framework. Eventually it would   lead to the Universe with large amount of dark matter in form of nuggets and anti-nuggets (they are far away from each other, therefore they do not annihilate each other) and no visible matter.
However, the axion dynamics which is determined by the axion field correlated on the  scale of the entire Universe
leads to a preferential formation of a specific type of nuggets on  the same large scales where the axion field is correlated as we argue below. Such coherent axion field emerges if the PQ phase transition occurs before or during inflation as discussed in items {\bf 1} and {\bf 5} in section \ref{formation}.

First of all we want to argue that the time dependent axion field  implies that there is an    additional coupling to fermions (\ref{mu_5}). 
Indeed, by making the time-dependent $U(1)_A$ chiral transformation in the path integral one can always represent the conventional $\theta$ term in the following form
\be
\label{mu_5}
\Delta {\cal{L}}_4=\mu_5(t)\bar{\Psi}\gamma_0\gamma_5\Psi ~~~~~~~ \mu_5\equiv \dot{\theta}.
\ee
In this formula $\mu_5\equiv \dot{\theta}$ can be thought as the chiral chemical potential. Many interesting properties emerge in the systems if  $\mu_5$ is generated. In fact, it has been an  active area of research  in recent years, mostly due  to very interesting  experimental data   suggesting that the  $\mu_5$ term  can be generated in heavy ion collisions,  see  original paper \cite{Kharzeev:2007tn} and recent reviews \cite{Kharzeev:2009fn,Kharzeev:2013ffa,Kharzeev:2015znc} for the details.  In the present context the $\mu_5$ term is generated as a result of the axion dynamics. As a matter of fact, the original studies  \cite{Kharzeev:2007tn} were motivated by the proposal that the separation of the baryon charges which may occur in early Universe, as advocated in this paper,  could be tested in  laboratory  experiments with heavy ion collisions. 

   Now we are prepared to formulate the  main claim of this section which can be stated  as follows. When  interaction (\ref{U}), (\ref{mu_5}) is  introduced into the system  there will be a {\it preferential evolution} in  the system of the nuggets versus anti-nuggets provided that nuggets and anti-nuggets had been already formed and chemical potential $\mu$ had been already generated locally  inside the nuggets as described in the previous section \ref{analysis}. As we already explained earlier, the generation of $\mu$ can be interpreted as a ``local violation" of $\cal{C}$ invariance in the system.  
  
  This {\it preferential evolution} is correlated with  the $\cal{CP}$-odd parameter  on the  scales where   the axion field $\theta(x)$ is coherent. In our arguments presented below 
we make a standard assumption that the initial value of $\theta(x)$ and its time derivative $\dot{\theta}(x)$
are correlated on the entire observable  Universe, such that  $\mu_5\equiv \dot{\theta}$ is also correlated
on the same large  scale.  Such large scale correlation is guaranteed if the PQ phase transition occurs before inflation, see items {\bf 1} and {\bf 5} in section \ref{formation} with details.
  This is the standard  assumption in most studies on axion physics when one computes   the present  density of axions due to   the misalignment mechanism,  see refs \cite{vanBibber:2006rb,   Sikivie:2008, Asztalos:2006kz,Raffelt:2006cw,Sikivie:2009fv,Rybka:2014cya,Rosenberg:2015kxa,Graham:2015ouw}.   

For our present studies the   key element is that the dynamics of the axion field until the QCD phase transition is determined by the  coherent state of axions at rest such that  \cite{vanBibber:2006rb, Sikivie:2008, Asztalos:2006kz,Raffelt:2006cw,Sikivie:2009fv,Rybka:2014cya,Rosenberg:2015kxa,Graham:2015ouw}:
\be
\label{axion}
\theta (t)\sim \frac{C}{t^{3/4}}\cos\int^t dt' \omega_a(t'), ~ \omega_a^2(t)=m_a^2(t)+\frac{3}{16t^2}, ~~~~
\ee
where $C$ is a constant, and $t=\frac{1}{2H}$ is the cosmic time. This formula suggests that for $m_a(t)t\gg 1$ when the axion potential is sufficiently strongly  tilted  the chiral chemical potential is essentially determined by the axion mass at time $t$ 
\be
\label{axion1}
\mu_5(t)=\dot{\theta} (t)\sim \omega_a (t)\simeq m_a(t). 
\ee
The  crucial point is  that $ \theta(t)$   is one and the same  in the entire Universe as it is correlated on the Universe size scale.  Another important remark is that the axion field $\theta(t)$ continues to oscillate with frequency (\ref{axion1}) until the QCD phase transition at $T_c$, though its absolute value $|\theta/\theta_0 | \sim  0.01$ might be few orders of magnitude lower at $T_c\simeq 170 $ MeV  than its original value $\theta_0$ at $T\simeq 1 $ GeV when the axion field only started to roll, see e.g. \cite{Wantz:2009it}.   As we discuss below, the relevant physics is not very sensitive to an absolute value of $|\theta (t)|$ in this regime, and therefore, we do not elaborate further  on this rather technical  and computational element of the axion dynamics, see footnote \ref{decoherence} below with comments on this matter.

In the context of the nugget's evolution (accretion of the baryon charge) this claim  implies that on  the entire Universe  size scale with one and the same sign of  ${\theta} (t)$  a specific single type of nuggets will prevail in terms of the number density and sizes.   Indeed, one can present the same arguments (see the beginning of this section)  with flipping the sign  $\mu\rightarrow -\mu$ with the only  difference is that the interaction (\ref{U}) prevents us from making the variable change $\theta_{(i)}\leftrightarrow -\theta_{(i)}$ for a given $\theta(t)$  because it  changes  its form   under $\theta_{(i)}\leftrightarrow -\theta_{(i)}$. 
In other words,  slow varying (on the QCD scale)   $\cal{CP}$ violating terms   (\ref{U}), (\ref{mu_5}) lead to a  preferential evolution of  the system for  a specific species of the nuggets with a given sign of $\mu$.   
 
 Indeed, it has been known for quite sometime, see e.g. \cite{Halperin:1998rc,Fugleberg:1998kk} that in the presence of $\theta\neq 0$ a large number of different 
 $\cal{CP}$ violating effects take place. In particular, the Nambu-Goldstone bosons  become a mixture of pseudo-scalar and scalar fields, their masses are drastically different from $\theta=0$ values. Furthermore, the  quark chiral $\la\bar{\psi}\psi\ra$ and the gluon $\la G^2\ra$ condensates  become the superposition with their pseudo-scalar counterparts $\la\bar{\psi}\gamma_5\psi\ra$ and $\la G\tilde{G}\ra$ such that entire hadron spectrum   and their interactions modify in the presence of $\theta\neq 0$. All these strong effects, of course, are proportional to $\theta$, and therefore numerically suppressed in case under consideration (\ref{axion}) by a factor $|\theta/\theta_0|\sim 10^{-2}$  in the vicinity of  the QCD phase transition. Naively, this small numerical factor $|\theta/\theta_0 | \sim  10^{-2}$ may lead only to   minor  effects $\sim 10^{-2}$.  However, the crucial point is that   while coupling   (\ref{U}) of the axion background field with   quarks is indeed relatively small  on the QCD scales, it is   nevertheless effectively long-ranged and long-lasting in contrast with conventional QCD interactions.   As a result,  this coherent   $\cal{CP}$ odd coupling   may produce large  effects of order of one as we argue  below.

Indeed,  as we discussed  in previous  section \ref{analysis}  a typical oscillation time $t_{\rm osc}$     when  the baryon charge accretes on the wall     is  of order $t_{\rm osc}\sim m_a^{-1}$
according to eq. (\ref{omega}).   But this  time scale $t_{\rm osc}\sim m_a^{-1}$ is precisely the time scale when $\dot{\theta}=m_a(t)$ varies  according to (\ref{axion1}).  Therefore, while the dynamical fermi fields $\theta_1, \theta_2$  defined by (\ref{bosonization})  fluctuate with typical scale of order $\Lambda_{\rm QCD}\gg m_a$, the coherent variation of these fields     will occur during a long (on the QCD scales)  coherent process when a  nugget makes a single cycle. These coherent corrections  are expected to  be different for nuggets (positive $\mu$) and anti-nuggets (negative $\mu$)    as a result of  many $\cal{C}$  and $\cal{CP}$ violating effects  such as scattering, transmission, reflection, annihilation, evaporation,  mixing of the scalar and pseudo-scalar condensates, etc which are all responsible for  the accretion of the baryon charge on a nugget during its long evolution.    

Important comment here is  that   each  quark   experiences a small difference  in interacting with the domain wall surrounding  nuggets or  anti-nuggets during every single QCD event  (mentioned above)  with typical QCD time scale $\Lambda_{\rm QCD}^{-1}$.   However, the number of the coherent QCD events  $n_{\rm coherent}$ during a   long single  cycle is very large
\be
 n_{\rm coherent}\sim \Lambda_{\rm QCD}t_{\rm osc}\sim \frac{\Lambda_{\rm QCD}}{m_a}\sim 10^{10} \gg 1. 
 \ee
Therefore, a net effect during every single cycle will be order of one, in spite of the fact that each given QCD event is proportional to the axion field $\theta (t)$ and could be quite small. 
 
The argument presented above holds as long as the axion field remains coherent, see also a comment at the very end of this subsection. In other words, a small but non vanishing coherent $\cal{CP}$ violating parameter $\theta(t)$ plays the role of catalyst which determines a preferred direction for separation of the baryon charges on the Universe scale, see   few comments in section \ref{formation} on justification of this assumption. This role 
of $\cal{CP}$ violation in our framework is very different from conventional ``baryogenesis" mechanisms  when $\cal{CP}$ violating parameter explicitly enters the final expression for the baryon charge production.

The corresponding large coherent corrections  during a single cycle $t_{\rm osc}$ imply that the   fast fluctuating fields $\theta_1, \theta_2$  
(which effectively describe the dynamics of the fermions living on the wall according to (\ref{bosonization}))  receive large  corrections  during every single  cycle 
\be
\label{phase1}
\Delta \theta_1(t) &\sim &\Delta \theta_2 (t)   \sim 1.
\ee
These  changes   of order one  of the  strongly interacting $\theta_1, \theta_2$- fields lead to 
modification   of the accreted baryon charge per single cycle per single degree of freedom  
\be
\label{phase2}
\Delta N\sim (\Delta \theta_1+\Delta \theta_2)\sim 1
\ee
 on the nuggets according to (\ref{N}).  One should emphasize that the corrections  (\ref{phase2}) are expected to be different for  nuggets and anti-nuggets because the interaction (\ref{mu_5}), (\ref{U}) which is responsible for these corrections (\ref{phase2})  breaks the symmetry between nuggets and anti-nuggets when $\mu\rightarrow -\mu$ as discussed above. 
 
 Precise computations of these coherent $\cal{CP}$ violating effects are hard to carry out explicitly as it requires 
 a solution of many-body problem of the coherent wall fermions with surrounding environment in the background of axion field (\ref{axion})
 when a large number of $\cal{C}$ and $\cal{CP}$ violating effects take place and  drastically modify evolution of nuggets versus anti-nuggets.
  A large number of cycles  of every individual nugget  (anti-nugget) also introduces a huge uncertainty in computations of $\Delta N$ during the time evolution  when a single  cycle  leads to the effect of order one, with possible opposite sign for a consequent  cycle.  In other words, it is very hard to predict what would be the final 
 outcome of the system after  a large number of cycles when each cycle produces the effect of order 1. We expect that the final result would be again of order one. Such a computation is beyond the scope of the present work.  Therefore, in what follows we introduce a phenomenological parameter $c(T)$ of order one  to account for these effects. All the  observables will be expressed in terms of this single phenomenological parameter $c(T)\sim1$, see eq. (\ref{ratio2}).

 Our  final comment in this subsection is as follows.
 The   charge separation effect on largest possible scales is only possible when the axion field (\ref{axion}) is coherent on the   scales of the Universe. This coherence  is known to  occur in conventional studies  on the dynamics of the axion field in the vicinity of  the QCD phase transition if the PQ phase transition occurs before inflation, see few comments in section \ref{formation} on this matter. 
  At the same time, soon after the QCD phase transition the dominant part of the axion field transfers its energy to  the free propagating on-shell axions (which is the subject of axion search experiments  \cite{vanBibber:2006rb, Sikivie:2008, Asztalos:2006kz,Raffelt:2006cw,Sikivie:2009fv,Rosenberg:2015kxa,Graham:2015ouw}). These randomly distributed  free axions are not in coherent state anymore. Therefore, the coherent accumulation effect which leads to a preferential formation of one species of nuggets, as discussed above, ceases to be  operational at the moment of decoherence  $t_{\rm dec}$ when the description in terms of the coherent axion field (\ref{axion})  breaks down\footnote{\label{decoherence}The decoherence time $t_{\rm dec}$ is not  entirely  determined by   absolute value of amplitude of the axion field  (\ref{axion}). In fact, the amplitude could be quite small, but the field remains  coherent on large scales. The computation of the 
 decoherence time $t_{\rm dec}$ is a hard problem of QFT, similar to a problem  in quantum  optics when initially coherent light becomes de-coherent superposition of uncorrelated photons.}. The baryon asymmetry we observe today  in this framework is a result of accumulation of the    charge separation effect  from the beginning of the nugget's formation  until this very last  ``freeze-out" moment  determined by $t_{\rm dec}$. 
 
 \subsection{Nuggets vs anti-nuggets on the large scale. Generic consequences. }\label{generic}
 As we already mentioned to  
 make any precise dynamical computations of $\Delta N\sim 1$ due to the coherent axion field (\ref{axion})     is a hard problem of strongly coupled QCD at $\theta\neq 0$.   In order to  effectively  account for these coherent effects   one can  introduce an unknown coefficient $c(T)$ of order one
 as follows
 \be
 \label{ratio2}
 B_{\rm antinuggets}=c(T) \cdot   B_{\rm nuggets}  ,~~{\rm where  } ~~ |c(T)| \sim 1,~~
  \ee
  where $c(T)$ is obviously a negative constant of order one. 
 We emphasize that the main claim of this section represented by eq.  (\ref{ratio2}) is not very sensitive to the axion mass $m_a(T)$ nor to the magnitude of $\theta(T)$ at the QCD phase transition when the bubbles start to oscillate and slowly accrete the baryon charge. The only crucial factor in our arguments is that the typical variation of $\theta(t)$ is determined by the axion mass (\ref{axion1}), which is the same order of magnitude as  $t^{-1}_{\rm  osc}$, and furthermore, this variation is correlated on the   scale where the axion field (\ref{axion}) can be represented by the coherent superposition of the axions at rest.

 The key relation of this framework (\ref{ratio2}) unambiguously implies that
 the baryon charge in form of the visible matter can be also expressed in terms of the same coefficient $c(T)\sim 1$
 as follows
 \be
 \label{ratio3}
B_{\rm visible} =- B_{\rm antinuggets}    - B_{\rm nuggets}. 
  \ee
Using eq. (\ref{ratio2}) it  can be rewritten as 
  \be
 \label{ratio4}
   &&B_{\rm visible}\equiv \left(B_{\rm baryons}+B_{\rm antibaryons}\right)\\
&=&  -\left[1+c(T)\right] B_{\rm nuggets} =-\left[1+\frac{1}{c(T)}\right] B_{\rm antinuggets}. ~~~\nonumber
  \ee
  The same relation can be also represented  in terms of the measured observables  $\Omega_{\rm visible}$ and $\Omega_{\rm dark}$ at later times  when only the baryons (and not anti-baryons) contribute to the visible component\footnote{\label{omega1}In eq. (\ref{ratio_omega}) we neglect   the differences (due to different gaps) between the energy per baryon charge in hadronic and CS phases to simplify notations. The corresponding corrections   in energy per baryon charge in hadronic and CS phases, in principle,  can be explicitly computed from the first principles. However, we ignore these modifications in the present work. This correction obviously does not change the main claim of this proposal stating that  $\Omega_{\rm visible}\approx\Omega_{\rm dark}$.} 
  \be
  \label{ratio_omega}
  \Omega_{\rm dark}\simeq \left(\frac{1+|c(T)|}{\left|1+c(T)\right|}\right)\cdot \Omega_{\rm visible} ~~ {\rm at} ~~ T\leq T_{\rm form}.
  \ee
One should emphasize that the relation (\ref{ratio4}) holds as long as the thermal equilibrium is maintained, which  we assume to be the case.
Another important comment   is that each individual contribution  $|B_{\rm baryons}|\sim |B_{\rm antibaryons}|$ entering  (\ref{ratio4})     is  many orders of magnitude greater  than the baryon charge hidden in the form of the nuggets and anti-nuggets at  earlier  times when $T_c>T> T_{\rm form}$. It is just their total baryon charge which is labeled as $B_{\rm visible}$ and representing the net baryon charge of the visible matter is the same order of magnitude (at all times) as the net baryon charge hidden in the form of the nuggets and anti-nuggets according to (\ref{ratio3}).

The baryons continue to annihilate each other (as well as baryon charge hidden in the nuggets) until the temperature reaches $T_{\rm form}$ when all visible anti baryons get annihilated, while visible baryons remain in the system and represent the visible matter we observe today. It    corresponds to $c(T_{\rm form})\simeq -1.5$ as estimated below if one neglects the differences in gaps in CS and hadronic phases, see footnote \ref{omega1}. After this temperature the nuggets essentially assume their final form, and do not loose or gain much of the baryon charge from outside. The rare events of the annihilation between anti-nuggets and visible baryons continue to occur. In fact, the observational excess of radiation in different frequency  bands,    reviewed in section \ref{nuggets}, is a result of  these rare annihilation events at present time.  

The  generic consequence of this framework  represented by eqs. (\ref{ratio2}), (\ref{ratio4}), (\ref{ratio_omega})  takes  the following form at this time $T_{\rm form}$ for   $c(T_{\rm form})\simeq -1.5$ which corresponds to the case when the nuggets saturate entire dark matter density:
\be
\label{ratio5}
B_{\rm visible}&\simeq& \frac{1}{2} B_{\rm nuggets}\simeq -\frac{1}{3}B_{\rm antinuggets}, \nonumber \\
\Omega_{\rm dark}&\simeq &5\cdot\Omega_{\rm visible}
\ee
which is identically the same relation (\ref{ratio1}) presented in Introduction. The relation  (\ref{ratio5})  emerges  due to the fact 
that  all components of matter, visible and dark,  proportional to one and the same dimensional parameter $\Lambda_{\rm QCD}$, see footnote \ref{omega1} with a comment on this approximation. In formula (\ref{ratio5})  $B_{\rm nuggets}$ and  $B_{\rm antinuggets}$ contribute to $\Omega_{\rm dark}$, while $B_{\rm visible}$ obviously contributes to $\Omega_{\rm visible}$. The coefficient $\sim 5$ in relation  $\Omega_{\rm dark}\simeq 5\cdot\Omega_{\rm visible}$ is obviously not universal, but relation (\ref{Omega}) is universal, and 
  very generic consequence of the entire framework, which   was the main motivation for the proposal \cite{Zhitnitsky:2002qa,Oaknin:2003uv}. 
  
  For example, if $c(T_{\rm form})\simeq -2 $ then the corresponding relation (\ref{ratio_omega}) between the dark  matter and the visible matter would assume the form    $\Omega_{\rm dark}\simeq 3\cdot\Omega_{\rm visible}$. Such a relation implies that    there is a plenty of room for other types of dark matter to saturate the observed ratio $\Omega^{\rm observed}_{\rm dark}\simeq 5\cdot\Omega^{\rm observed}_{\rm visible}$.  This comment will be  quite important 
  in our discussions in section \ref{axion-search} where we comment on  implications of this framework for other axion search experiments. 

One should emphasize once again that the  generic consequences of the framework  represented by (\ref{Omega}), (\ref{ratio_omega}) are not   sensitive to any specific parameters  such   as efficiency of the domain wall production or the magnitude of $\theta$ at the QCD phase transition, which could be quite small, see footnote \ref{decoherence} with few comments on that. Nevertheless, precisely the coupling with the coherent $\cal{CP}$ odd axion field plays a  crucial role in generation  of $|c(T)|\neq  1$, i.e. the axion plays the role of catalyst  in   the baryon charge separation effect on the largest possible scales.  Some other observables which are sensitive to the dynamical characteristics  (e.g. efficiency of the domain wall production) will be discussed below. 

\subsection{\label{asymmetry2}$n_B/n_{\gamma}$ ratio. Model dependent estimates. } 
The time evolution of the dark matter within this framework is amazingly simple. The relations (\ref{ratio2}),  (\ref{ratio3}), (\ref{ratio4}) hold at all times.
The baryon charges of the nuggets and anti-nuggets vary until its radius $R(T)$ assumes its equilibrium value as described in sections
\ref{time-evolution}, \ref{analysis}. It happens approximately at time when the CS phase forms  in interior of the nuggets, which can be estimated as  $T_{CS}\simeq 0.6 \Delta\simeq 60$ MeV, where $\Delta\simeq 100$ MeV is the gap of the CS phase. After this temperature the nuggets essentially assume their final form, with very little variation in size (and baryon charge). The rare events of the annihilation of course continue to occur even for lower tempearures. In fact, the observational consequences   reviewed in section \ref{nuggets} is a result of these annihilation events at present time.

The variation of the visible matter $B_{\rm visible}$ demonstrates   much more drastic changes after the QCD phase transition at $T_c$ because the corresponding number density is proportional to  $ \exp(-m_N/T)$ such that at the moment of formation $T_{\rm form}\approx 40$ MeV  the baryon to entropy ratio assumes its present value (\ref{eta}) which we express  as follows
 \be
\label{eta1}
\eta\equiv\frac{n_B}{n_{\gamma}}\simeq \frac{B_{\rm visible}/V}{n_{\gamma}}\sim 10^{-10} , ~~ n_B\equiv\frac{B_{\rm visible}}{V} .
\ee
 If the nuggets and anti-nuggets were not present at this temperature  the conventional baryons 
and anti-baryons would continue to annihilate each other until the density would be 9 orders of magnitude smaller 
than observed (\ref{eta1}) when  the temperature will be around 
  $T\simeq 22$ MeV.  Conventional baryogenesis resolves this ``annihilation catastrophe" by producing extra baryons in early times, see e.g. review \cite{Dolgov:1997qr}, while in our framework extra baryons and anti baryons are hidden in form of the macroscopically large nuggets.   

In our framework the ratio (\ref{eta1}) can be rewritten in terms of the nugget's density as  the baryon charge in form of the visible matter and in form of the nuggets   are related to each other according to (\ref{ratio4}).
This relation allows us to infer what efficiency is required for the bubbles to be formed and survive until the present time when  observed ratio is measured (\ref{eta1}). 

One should emphasize that any small factors which normally enter the  computations in conventional baryogenesis 
(such as $\cal{C}$ and $\cal{CP}$ violating  parameters) do not enter in the   estimates presented below  in our framework as  result of two effects. 
First, the $\cal{C}$ violation enters the computation  as a result of generation of the chemical potential $\mu$ as described in section \ref{baryons}.
It is expressed in terms of spontaneous accretion  of the baryon charge on the surface of the nuggets as given by eq.  (\ref{N}) which effectively generates the chemical potential (\ref{small_mu}), which can be thought as the local violation of the symmetry on the scale of a single nugget.
Secondly, the $\cal{CP}$ violation enters the computation in form of the coupling with the coherent axion field (\ref{mu_5}). Precisely this coupling as we argued above leads to removing of the degeneracy between nuggets and anti-nuggets formally expressed  as $c(T)\sim1$ in eq. (\ref{ratio2}).
Therefore, the only small parameter we anticipate in our estimates below is due to some  suppression of   the closed bubbles which must be  formed with sufficiently large sizes  during the QCD phase transition. 

We cannot compute the probability for the bubble formation as it obviously requires the numerical simulations, which is beyond the  scope  of the present work. Instead, we go backward and ask the question: What should be the efficiency of the bubble formation at the QCD phase transition in order to accommodate the observed ratio (\ref{eta1})?

With these comments in mind we proceed with our estimates as follows. 
First, from (\ref{ratio4}), (\ref{ratio5}) we infer that the baryon charge hidden in the nuggets and anti-nuggets is the same order of magnitude as the baryon charge of the visible baryons at $T_{\rm form}$ at the end of the formation, i.e. 
\be
\label{n_1}
 \frac{ B_{\rm nuggets}/V}{n_{\gamma}}\gtrsim  {10^{-10}}, 
\ee
where we use sign $\gtrsim$ instead of $\approx$ used in eq (\ref{eta1}) to emphasize that there is long time for equilibration between
 the moment $T_{\rm CS}\simeq 0.6 \Delta\simeq 60$ MeV when CS phase forms   in  interior of  the nuggets and $T_{\rm form}\simeq 40$ MeV when all anti baryons of the visible matter get annihilated, corresponding to the present observed value (\ref{eta1}). During this period the equilibrium between the visible matter and the baryons from nuggets is maintained, and some portion of the nugget's baryon charge might be annihilated by the visible matter.  
 It explains our sign $\gtrsim$ used in eq. (\ref{n_1}).
 
 The relation (\ref{n_1}) implies that the number   density of nuggets and anti-nuggets can be estimated as 
 \be
 \label{n_2}
 \frac{ \la B\ra  n_{\rm nuggets}}{n_{\gamma}}\gtrsim 10^{-10} , ~~~ \la B\ra  n_{\rm nuggets}\equiv  \frac{B_{\rm nuggets}}{V}, ~~~~
\ee
where $\la B\ra   $ is the average baryon charge of a single nugget at $T_{\rm form}$.

Now we want to estimate the same ratio   (\ref{n_2}) using the  Kibble-Zurek (KZ) mechanism\cite{KZ,KZ-review,Vilenkin} reviewed in section \ref{formation}. The basic idea of the KZ  mechanism is that the total area of the crumpled, twisted  and folded domain wall is proportional to the volume of the system, and can be estimated as follows: 
\be
\label{S}
S_{\rm (total~ DW)}=\frac{V}{\xi (T)} , 
\ee
where $\xi(T)$ is the correlation length which is defined as an average distance between crumpled  domain walls at temperature $T$.
Largest part of the wall belongs to the percolated large cluster. It is known that some   closed walls (bubbles) with typical size $\xi(T)$ will be also formed. These   bubbles with sufficiently large size $R\sim \xi(T)$ will eventually become nuggets.  We introduce parameter $\gamma$ to account for the suppression related to the   closed bubble  formation. In other words, we define
\be
\label{S1}
S_{\rm nuggets}= \gamma S_{\rm (total~ DW)} = \frac{\gamma V}{\xi (T)} , ~~~~  \gamma\ll 1.
\ee
At the same time total area of the nuggets $S_{\rm nuggets}$ can be estimated as 
\be
\label{S2}
S_{\rm nuggets}= 4\pi R_0^2 (T)\left[V\cdot n_{\rm nuggets}\right], 
\ee
where $R_0$ is the size of a nuggets at initial   time, while $\left[V\cdot n_{\rm nuggets}\right]$ represents the total number of nuggets in volume $V$.
Comparison (\ref{S1}) with (\ref{S2}) gives  the following estimate for the nugget's density when bubbles just formed, 
\be
\label{n_3}
n_{\rm nuggets}\simeq \frac{\gamma}{4\pi R_0^2 \xi}.
\ee
The last step in our estimates is the computation of the average baryon charge of a nugget at $T_{\rm CS}$  when CS sets in inside the nugget.
The corresponding estimates have been worked out long ago \cite{Zhitnitsky:2002qa} and reproduced in section \ref{time-evolution} in the course of the time  evolution by taking $t\rightarrow\infty$, see (\ref{equlibrium}). 
The baryon number density inside the nuggets depends on a model being used \cite{Zhitnitsky:2002qa}, but typically it is few times the nuclear saturation density $n_0\simeq 
(108~ {\rm MeV})^3$ which is consistent with  conventional computations  for the baryon density in CS phases. Therefore, we arrive to 
\be
\label{n_4}
\la B\ra \simeq (2-6) n_0\cdot \frac{4\pi R^3_{\rm form}}{3},
\ee
where $R_{\rm form}$ is the final size of the nuggets. By substituting (\ref{n_4}) and (\ref{n_3})    to (\ref{n_2})  
  we arrive to the following   constraint    on  efficiency of the bubble formation  represented by parameter $\gamma$  
\be
\label{constraint}
 (2-6)\cdot \frac{\gamma}{3} \left(\frac{R_{\rm form}}{\xi (T)} \right)   \left(\frac{R_{\rm form}}{R_0} \right)^2 \left( \frac{ n_0 }{n_{\gamma}}\right) \gtrsim 10^{-10},
\ee
where expression for  $n_{\gamma}(T) $ should be taken  at the formation time 
\be
\label{n_gamma}
n_{\gamma}=\frac{2\xi(3)}{\pi^2} T_{\rm form}^3, ~~~ \xi(3)\simeq 1.2, 
\ee
while the correlation length $\xi(T)$ should be evaluated at much earlier times, close to $T_c$   when domain wall network only started to form.
Typically bubbles form with $R_0\sim \xi$. However, the bubbles  shrink approximately  3-5 times according to (\ref{R_0}) before they reach equilibrium during the time evolution as discussed in section \ref{analysis}. Therefore,  to be on a safe side, we make very conservative assumption that  
\be 
  \frac{R_{\rm form}}{R_0}  \sim 0.1, ~~~R_0\simeq\xi.
  \ee
To proceed with numerical estimates, it is convenient to separate $\gamma$ on two pieces, 
\be
\gamma\equiv\gamma_{\rm formation}\cdot\gamma_{\rm evolution}, ~~~~~  \gamma_{\rm formation}\sim 0.1, 
\ee
 where the first part, $\gamma_{\rm formation}  \sim 0. 1$ has been estimated using  numerical simulations, see textbook \cite{Vilenkin} for review.   
 The second suppression factor  $\gamma_{\rm evolution}$ is unknown, and includes a  large number of different effects. 
 In particular, many small closed bubbles with $R_0\leq  \xi$ are very likely to be formed but may  not survive the evolution as we discussed in section \ref{analysis}. 
 Furthermore, there are many effects such as evaporation, annihilation inside the nuggets which may also lead to collapse of relatively small nuggets.
 Furthermore,  the formation probability of large closed bubbles with $R_0\gg  \xi$ (which most likely to survive) is highly suppressed $\sim \exp(-R^2_0/\xi^2)$. 
 All these effects are included in unknown parameter $\gamma_{\rm evolution}$.
 Our constraint (from observations on $n_B/n_{\gamma}$ within  our mechanism)   can be inferred  from (\ref{constraint}) 
 \be
 \label{constraint_final}
 \gamma_{\rm evolution} (T_{\rm form})  \gtrsim 10^{-7}. 
 \ee
 One suppression factor which obviously contributes to   suppression  (\ref{constraint_final}) is related to necessity to produce a sufficiently large initial bubble 
 for successful nugget formation as given by eq. (\ref{R_0}). 
 
 Now we can interpret the estimate (\ref{constraint_final}) in two complimentary ways\footnote{We are thankful to anonymous referee who hinted on possibility of the first interpretation. The second interpretation of estimate (\ref{constraint_final}) is our original and preferable interpretation.}.
 First interpretation of  estimate (\ref{constraint_final})  is as follows. Small numerical value (\ref{constraint_final}) implies that only sufficiently large nuggets  survive the evolution in unfriendly environment mentioned above. It is hard to estimate all the QCD effects mentioned above
 (evaporation, annihilation inside the nuggets etc), but the dominant suppression factor is related to  formation suppression $\sim \exp(-R^2_0/\xi^2)$. The observed abundance (\ref{eta1})  can be interpreted  in this case as a specific value for the formation size $R_0$ which satisfies the constraint (\ref{constraint_final}). There is an exponential sensitivity to $R_0$ within this interpretation. In particular  if  $R_0\sim (3-4)\xi$, 
  \be 
  \label{R_0-estimate}
 \exp\left(-\frac{R^2_0}{\xi^2}\right)\sim (10^{-4}-10^{-7}). 
 \ee
This estimate is   consistent with the observational constraint  as unaccounted QCD effects mentioned above may saturate (\ref{constraint_final}).    

We do not consider this   sensitivity to $R_0$ as a fine tuning problem. Indeed, in many cases the physics is highly sensitive to some parameters of the theory, which however cannot be  interpreted as a fine tuning problem.
In particular, in the context of this paper  the  conventional formula (\ref{dm_axion}) for the dark matter density resulted form the misalignment mechanism is highly sensitive to the axion mass $m_a$. However, we do not interpret this dependence as a fine tuning problem.  

Our second (and preferable) interpretation can be explained as follows. The observed ratio (\ref{eta1}) is highly sensitive to $T_{\rm form}\approx 40$ MeV
due to exponential dependence of the baryon number density   $\sim \exp(-m_N/T)$. This formation temperature in our framework is defined as the temperature when the   nuggets complete their formation. This value for the temperature is very reasonable as it lies  slightly below $T_{\rm CS}$ when the  CS phase sets in inside the nuggets. Obviously, we do not interpret this sensitivity to the formation temperature  $T_{\rm form}\approx 40$ MeV  as a fine tuning problem. 

The crucial point here is that the saturation of the observed ratio (\ref{eta1}) can be interpreted in terms of $T_{\rm form}$, or it can be interpreted in terms of $R_0$ which is the key parameter in  our first interpretation. Small variation of $T_{\rm form}$ can be thought  as small variations of 
$R_0$, which however lead to very large changes of the observed ratio (\ref{eta1}) due to the exponential sensitivity.  In other words, small increase of $T_{\rm form}$ when nuggets complete the formation can be interpreted as small decrease of survival size $R_0$ in our first interpretation given above.

 We do not call this effect as a    fine tuning.     This is because the equilibration of the baryon charge from the nuggets with the visible baryons always lead to the result (\ref{Omega}), (\ref{ratio4})    when all  contributions  are the same order of magnitude. A small observed  ratio (\ref{eta1})    is determined by a precise  and specific  moment in   evolution of the Universe   when  the nuggets complete their    formation at temperature $T_{\rm form}\sim \Lambda_{\rm QCD}$, which is again, perfectly consistent with the main paradigm  of the entire framework that all dimensional parameters are order of $\Lambda_{\rm QCD}$. This   $T_{\rm form}$ corresponds to a very  specific value $R_0$ for nuggets to complete their formation at time $T_{\rm form}$. 
 
 How one can understand the result (\ref{constraint_final}) which essentially states that even very tiny probability of the formation of the   closed bubbles is still sufficient to saturate the observed ratio (\ref{eta1})?
 The answer lies in the observation that the baryon density $n_B\simeq n_{\bar{B}}$ was 10 orders of magnitude larger at the moment of the bubble  formation. Therefore, even a tiny probability at the moment of formation of a closed bubble with sufficiently large size   will lead to effects of order one at the moment when  the baryon number density drops 10 order in magnitude. Another reason 
  why very tiny probability of the formation of the   closed bubbles nevertheless is  sufficient to saturate the observed ratio (\ref{eta1}) is that  typical  ``small factors" which normally accompany the conventional baryogenesis mechanisms such as $\cal{CP}$ and $\cal{C}$ odd couplings do not appear in estimate (\ref{constraint_final}) due to the reasons already explained  after eq (\ref{eta1}).
  
 $\bullet$ We  conclude this section with the following comment: The basic consequences of this framework represented by eqs. (\ref{Omega}), (\ref{ratio4}), (\ref{ratio_omega}) are very generic. These features are not  very sensitive to efficiency of the closed domain wall formation  nor to the absolute value of $\theta$ as long as coherence is maintained, see footnote \ref{decoherence}. These generic features  hold for arbitrary value of the  axion mass $10^{-6} {\rm eV} \leq m_a\leq 10^{-3} {\rm eV}$, in contrast with conventional treatment of the axion as the dark matter candidate, when $\Omega_{\rm DM}$ can be saturated by the axions only when the axion mass assumes a very specific and definite value $m_a\simeq 10^{-6}$ eV, see next section with details.  
  
   The  derivation of  the observed ratio (\ref{eta1}) from the first principles (which is determined by parameter $R_0$ in  first interpretations or parameter   $T_{\rm form}$ in the second interpretation)  is a hard    computational problem of  strongly coupled QCD when  all elements
  such as cooling rate, annihilation rate, charge separation rate, damping rate, evaporation rate and many other effects are equally contribute to $T_{\rm form}$. However, it is important that the ``observational" value $T_{\rm form}\simeq 40 $MeV  lies precisely in the region where it should be: 
    $T_{\rm form}< T_{\rm CS}$, i.e. slightly below the  temperature where  CS sets in. Therefore, any fine-tuning procedures have never been required in this framework   to accommodate the observed ratio presented by eq.(\ref{Omega}).

\section{ Implications for the axion search experiments}\label{axion-search}
The goal of this section is to   comment on relation of our framework and the   direct axion search experiments  \cite{vanBibber:2006rb, Sikivie:2008, Asztalos:2006kz,Raffelt:2006cw,Sikivie:2009fv,Rybka:2014cya,Rosenberg:2015kxa,Graham:2015ouw}. We start with the following   comment we made in   section \ref{nuggets}:  
   this model which has a single  fundamental parameters  (a mean baryon number of a nugget $\la B\ra \sim 10^{25}$    entering  all the computations) is   consistent with all known astrophysical, cosmological, satellite  and ground based constraints as reviewed  in section \ref{nuggets}.  For discussions of this section it is convenient  to express this single  normalization parameter $\la B\ra \sim 10^{25}$  in terms of the  axion mass $m_a\sim 10^{-4}$ eV as these two parameters directly related  according to the scaling relations (\ref{sigma}). 
   The corresponding relation between these two parameters occur because the axion mass $m_a$ determines the wall tension $\sigma\sim m_a^{-1}$ which itself enters the expression for the equilibrium value of the size of the nuggets, $R_{\rm form}$ at the end of the formation. One should emphasize  that it is quite nontrivial  that the cosmological constraints on the nuggets as shown on Fig. \ref{gorham} and formulated in terms of $\la B\ra$ are compatible  with known upper limit on the axion mass $m_a < 10^{-3}$eV within our framework. One could regard this compatibility as a nontrivial consistency check for this proposal. 
   
   The lower limit on the axion mass, as it is well known, is  
   determined by the requirement that  the axion contribution to the dark matter density does not  exceed the observed value $\Omega_{\rm dark}\approx 0.23$. There is a number of uncertainties in the corresponding estimates. We shall not comment on these  subtleties by referring to the   review papers\cite{vanBibber:2006rb, Sikivie:2008, Asztalos:2006kz,Raffelt:2006cw,Sikivie:2009fv,Rybka:2014cya,Rosenberg:2015kxa,Graham:2015ouw}. The corresponding uncertainties are mostly due to the  remaining discrepancies between different groups on the computations  of the  axion production  rates due to the  different mechanisms such as  misalignment mechanism versus  domain wall/string decays.  In what follows to be  more concrete in our estimates we shall use   the following  expression for the dark matter density in terms of the axion mass  resulted from the misalignment mechanism \cite{Graham:2015ouw}:
 \be
 \label{dm_axion}
 \Omega_{\rm (DM ~axion)}\simeq \left(\frac{6\cdot 10^{-6} {\rm eV}}{m_a}\right)^{\frac{7}{6}}
 \ee 
 This formula essentially states that the axion of mass $m_a\simeq 2\cdot 10^{-5}$ eV saturates the dark matter density observed today, while the axion mass in the range of $m_a\geq  10^{-4}$ eV contributes very little to the dark matter density. This claim, of course, is entirely based on estimate (\ref{dm_axion}) which accounts only for the axions directly produced by the misalignment mechanism suggested originally in \cite{misalignment}.
 
 There is another mechanism of the axion production when the Peccei-Quinn symmetry is broken after inflation.
 In this case the string-domain wall network produces a large number of axions such that the axion mass $m_a \simeq   10^{-4}$ eV may saturate the dark matter density,  see  relatively  recent estimates \cite{Wantz:2009it,Hiramatsu:2012gg,Kawasaki:2014sqa} with some comments and references on previous papers.    The corresponding formula  from refs. \cite{Wantz:2009it,Hiramatsu:2012gg,Kawasaki:2014sqa}   is also highly sensitive to the axion mass with  $m_a$- dependence being very  similar to eq. (\ref{dm_axion}).  
  \begin{figure}
\centering
\includegraphics[width=1.0\linewidth]{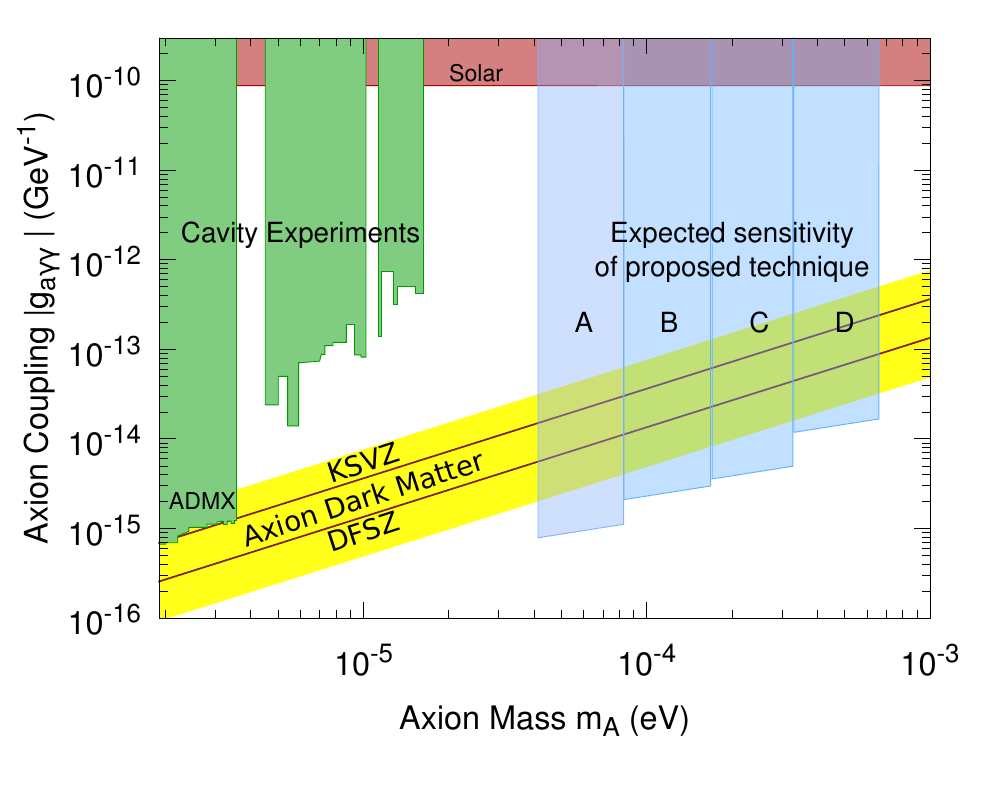}
\caption{ Cavity / ADMX experimental constraints on the axion mass shown in green.  The expected sensitivity for the  Orpheus axion search experiment \cite{Rybka:2014cya} is shown by blue regions ``A", ``B", ``C" and ``D".  In particular, experiment ``B",  covers   the most interesting region of the parametrical space with $m_a\simeq  10^{-4}$ eV  corresponding to the  nuggets with mean baryon charge $\la B\ra\simeq 10^{25}$ which itself satisfies all  known astrophysical, cosmological, satellite  and ground based constraints, see Fig.\ref{gorham}. The plot is  taken from   \cite{Rybka:2014cya}. 
  }
\label{rybka}
\end{figure}

 The main lesson to be learnt from  the present work  is   that in addition to these well established mechanisms previously discussed in the literature there is an additional  contribution to the dark matter density also  related to the axion field. However, the mechanism which is advocated in the present work contributes to the dark matter density    through formation of the nuggets, rather than through the direct axion production.  The corresponding mechanism as argued in section \ref{generic} always satisfies the relation  $\Omega_{\rm dark} \approx \Omega_{\rm visible}$, and, in principle is capable to  saturate the dark matter density $\Omega_{\rm dark}\approx 5  \Omega_{\rm visible}$ by itself for  arbitrary magnitude of the axion mass $m_a$ as the corresponding contribution is not sensitive to the axion mass  in contrast with conventional mechanisms mentioned above.  A precise coefficient in ratio  $\Omega_{\rm dark} \approx \Omega_{\rm visible}$ is determined by a parameter  of order one, $|c(T)|\sim 1$, which unfortunately is very hard to compute from the first principles,   as discussed   in section \ref{generic}. 
 
   Our choice for $m_a\simeq  10^{-4}$ eV  which corresponds  to $\la B\ra \sim 10^{25}$ is entirely  motivated by our previous analysis of astrophysical, cosmological, satellite  and ground based constraints as reviewed  in Section \ref{nuggets}. As we mentioned  in Section \ref{nuggets}  there is a number of frequency bands where some excess of emission was observed, and this model may explain some portion, or even entire excess of the observed radiation in these frequency bands. Our  normalization $\la B\ra \sim 10^{25}$   was fixed by   eq.(\ref{flux1}) with  assumption  that the observed dark matter is saturated by the nuggets. The relaxing this assumption obviously modifies  the coefficient $c(T)$ as well as $\la B\ra$.

    Interestingly enough, this range of the axion mass $m_a\simeq  10^{-4}$ eV  is perfectly consistent with recent  claim \cite{Beck},\cite{Beck1}
  that the previously observed small signal in resonant S/N/S Josephson junction  \cite{Hoffmann} is a result of the dark matter axions with the mass $m_a\simeq 1.1\cdot 10^{-4}$ eV. Furthermore, it has been also claimed that   similar   anomalies have been observed in 
   other experiments \cite{Golikova, He, Bae} which all point towards an axion mass $m_a\simeq 1.1\cdot 10^{-4}$ eV if interpreted within framework \cite{Beck},\cite{Beck1}. The only   comment we would like to make here is that
  if the interpretation \cite{Beck},\cite{Beck1} of the observed anomalies  \cite{Hoffmann,Golikova, He, Bae} is indeed due to the dark matter axions, then the corresponding axion mass is perfectly consistent with our  estimates (based on cosmological observations) of the average baryon charge of the nuggets $\la B\ra\simeq 10^{25}$ as reviewed in section \ref{nuggets}. 

We  conclude this section on   optimistic  note with  a remark  that the most interesting region of the parametric space corresponding to the  nuggets with mean baryon charge $\la B\ra\simeq 10^{25}$ might be   tested by the  Orpheus axion search experiment \cite{Rybka:2014cya} as shown on Fig. \ref{rybka}.

  \section*{ Conclusion. Future directions.} 
 First, we want to list the main results of the present studies, while the comments on possible future developments will be presented at the end of this Conclusion.
 
{\bf 1.}  First key element of this proposal is the observation (\ref{N}) that the closed axion domain walls  are copiously produced and generically  will acquire the baryon or anti-baryon charge.
 This phenomenon of ``separation of the baryon charge" can be interpreted as a local version of spontaneous symmetry breaking.
 This symmetry breaking occurs not in the entire volume of the system, but on the correlation length $\xi(T)\sim m_a^{-1}$ which is determined by the  folded and crumpled axion domain wall  during the formation stage. Precisely this local charge separation eventually leads to the formation of the nuggets and anti-nuggets  serving in this framework as the dark matter component  $\Omega_{\rm dark}$. 
 
  {\bf 2.} Number density of nuggets and anti-nuggets will not be identically the same  as a result of the coherent (on the scale of the Universe) axion     $\cal{CP}$ -odd field.  We parameterize the corresponding effects  of order one by phenomenological constant $c(T)\sim 1$. 
It is important to emphasize that  this parameter of order one  is not fundamental constant of the theory, but,  calculable from the first principles.  In practice, however,  such a computation could be quite a challenging problem   when even the QCD phase diagram is not  known. The fundamental consequence  of this framework, $\Omega_{\rm dark} \approx \Omega_{\rm visible}$, which is given by (\ref{Omega}) is {\it universal}, and not sensitive to any parameters as both components are proportional to $\Lambda_{\rm QCD}$.
The observed ratio (\ref{ratio1}), (\ref{ratio5}) corresponds to a specific value of $c(T_{\rm form})\simeq -1.5$ as discussed in section \ref{generic}.

{\bf 3.} Another  consequence of the proposal is a natural explanation of the ratio (\ref{eta}) in terms of the formation temperature $T_{\rm form}\simeq 40$ MeV, rather  than in terms of    specific   coupling constants which normally enter  conventional ``baryogenesis" computations. This observed  ratio is expressed in our framework in terms of a single parameter   $T_{\rm form}$ when nuggets complete their formation. 
This    parameter   is not fundamental constant of the theory, and as such is   calculable from the first principles.   In practice, however,  the  computation of $T_{\rm form}$ is  quite a challenging problem as explained in section \ref{asymmetry2}.
Numerically, the observed ratio 
(\ref{eta}) corresponds to $T_{\rm form}\simeq 40$ MeV which is indeed slightly below  the critical temperature $T_{CS}\simeq 60~ $MeV where the colour superconductivity sets in. 

The relation $T_{\rm form}\lesssim T_{CS}\sim \Lambda_{\rm QCD}$  is {\it universal} in   this framework as   both parameters are proportional to 
$\Lambda_{\rm QCD}$. As such, the universality of this framework  is similar to the universality $\Omega_{\rm dark} \approx \Omega_{\rm visible}$ mentioned in previous item. 
  At the same time, the ratio   (\ref{eta})  is not universal itself as it is exponentially  sensitive to precise value of $T_{\rm form}$  due to conventional  suppression factor $\sim\exp(-m_p/T)$.  

 {\bf 4.}  The only new fundamental parameter of this framework is  the axion mass $m_a$. Most of our computations (related to the cosmological observations, see section \ref{nuggets} and Fig. \ref{gorham}), however, are expressed in terms of the mean baryon number of nuggets $\la B\ra$   rather than in  terms of the axion mass. However, these two parameters are unambiguously related   according to the scaling relations (\ref{sigma}).  Our claim is that all {\it universal} properties of this framework listed above  still hold for any $m_a$.   In other words, there is no any fine tuning in the entire construction with respect to $m_a$. The  constraints (and possible cosmological observations) from section \ref{nuggets} strongly suggest $\la B\ra\simeq 10^{25}$ which can be translated into preferred  value for the axion mass $m_a\simeq 10^{-4}$ eV.
 
  {\bf 5.} 
  This region of the axion mass $m_a\simeq 10^{-4}$ eV  corresponding to  average size of the nuggets $\la B\ra\simeq 10^{25}$       can be tested in   the Orpheus axion search experiment \cite{Rybka:2014cya} as shown on Fig. \ref{rybka}.

 We conclude with few thoughts on future directions within our framework. 
It is quite obvious that future progress cannot be made without a much deeper understanding of the QCD phase diagram at $\theta\neq 0$.
In other words, we  need to understand the structure of possible  phases along the third dimension parametrized by $\theta$     on Fig \ref{fig:phase_diagram}. 

Presently,  very few results are available  regarding  the phase structure at $\theta\neq 0$. First of all, the phase structure is understood in simplified version of QCD with two colours,   $N_c=2$ at $T=0, \mu\neq 0$, see \cite{Metlitski:2005db}. In fact, the studies  \cite{Metlitski:2005db} were mostly motivated  by the subject of the present work and related to the problem of  formation of the quark nuggets during the QCD phase transition in early Universe with non vanishing $\theta$.  
With few additional assumptions the phase diagram can be also conjectured 
 for the system with  large number of colours $N_c=\infty$, at non vanishing $T, \mu, \theta$,  see \cite{Parnachev:2008fy,Zhitnitsky:2008ha}.

Due to the known ``sign problem", see footnote \ref{N=1},  the  conventional lattice simulations cannot be used at $\theta\neq 0$.  The corresponding analysis of the phase diagram for non vanishing  $\theta$  started just  recently by using  some newly invented   technical tricks \cite{D'Elia:2013eua,Bonati:2013tt,Bonati:2015uga,Bonati:2015sqt}. 

Another possible development from the ``wish list"    is a deeper understanding of the closed bubble formation. Presently, very few results are available on this topic. The most relevant for our studies is the observation  made in \cite{Sikivie:2008} that a small number of  closed bubbles  are indeed observed  in numerical simulations. However, their detail properties (their fate, size distribution, etc) have not been studied yet. 
A number of related questions such as an estimation of correlation length  $\xi(T)$, the generation of the   structure inside the domain walls, the baryon charge accretion on the bubble,  etc, hopefully can be also studied in such numerical simulations. 

One more possible direction for future studies  from the ``wish list"   is a development some QCD-based  models where 
a number of hard questions   such as:  
evolution of the nuggets,  cooling rates, evaporation rates, annihilation rates, viscosity of the environment, transmission/reflection coefficients,   etc in unfriendly environment  with non-vanishing  $T, \mu, \theta$ can be addressed, and hopefully  answered.
All these and many other    effects are, in general, equally contribute  to our parameters $T_{\rm form}$ and $c(T)$  at the $\Lambda_{\rm QCD}$  scale in strongly coupled QCD. 
Precisely these numerical factors  eventually determine  the coefficients in the observed relations:  $\Omega_{\rm dark} \approx \Omega_{\rm visible}$ given by eq. (\ref{ratio_omega})   and   $n_B/n_{\gamma}$ expressed by  eq. (\ref{eta1}).  
 
 Last but not least:  the  discovery  of the axion in the Orpheus experiment  \cite{Rybka:2014cya}  would  conclude  a long   and fascinating  journey 
  of searches for this unique and amazing particle conjectured almost 40 years ago.    Such a discovery  would be a strong motivation for related searches  of ``something else" as the axion mass $m_a\simeq 10^{-4}$ is unlikely to saturate the  dark matter density observed today.    
   We advocate the idea  that this  ``something else"  is the ``quark nuggets"  (where the axion plays the  key role in entire construction)  which could  provide the principle contribution to  dark matter of  the Universe as the relation $\Omega_{\rm dark} \approx \Omega_{\rm visible}$ in this framework  is not sensitive to the axion mass.

\section*{Acknowledgments}
 
This work was supported in part by the National Science and Engineering
Research Council of Canada. 

\appendix
\section{Estimation of fluxes}\label{appendix:flux}
The main goal of this Appendix is to argue that the  approximation in eq. (\ref{flux}) which was adopted in the text by    neglecting extra term ``fluxes"
 is    justified, at least on qualitative level. In other words, while these ``flux"  terms obviously present in the system, they, nevertheless,  do not drastically change a key technical element   (an implicit  relation  between $R(t)$ and $\mu(t)$) which this equation provides.  Precisely this implicit relation between $R(t)$ and $\mu(t)$ eventually allows us to express the $\mu$-dependent  pressure $\Delta P[\mu]$ in terms of $R$ dependent function $\Delta P[f(R)]$ such that the basic equation (\ref{eq:6.R0}) describing the time evolution of the nuggets is reduced to a differential equation on  a single variable $R(t)$. 
 
Our starting point is the observation that   the relevant flux which enters equation 
(\ref{flux}) is  $\Delta\Phi =(\Phi_\Rightarrow -\Phi_\Leftarrow)$,    counting the net baryon charge transfer and sensitive to the chemical potential difference, rather than total flux $\la\Phi\ra$ 
which counts the exchange of all the particles, including bosons\footnote{The dominant contribution to the fluxes normally comes from the lightest degrees of freedom which are the Nambu-Goldstone bosons in hadronic and CS phases. These contributions are crucial for maintaining   the thermodynamical equilibrium between exterior and interior,  but they do not play any role 
in the baryon fluxes which enter eq.(\ref{flux}).}.   In fact, if the average flux $\la\Phi\ra$
were entering equation (\ref{flux}) one could  explicitly check that  this term  would be the same order of magnitude as
two other terms of the equation. However, the key  point is that the baryon charge transfer $\Delta\Phi $ is numerically suppressed, i.e.   $\Delta\Phi \ll
\la\Phi\ra$. In fact, $\Delta\Phi$ identically 
  vanishes for $\mu=0$. Furthermore, one can use the same technique which has been used in section  \ref{algebraic} to argue that $\Delta\Phi \ll
\la\Phi\ra$  in entire region of $\mu$. Numerical analysis supports this claim. 

To reiterate this claim:  while a typical flux defined as 
 \begin{equation}
\Phi 
=\frac{g^{\rm in} }{(2\pi)^3}
\int\frac{v_z d^3k}{\exp(\frac{k -\mu}{T})+1} +(\rm bosons)
  \sim (\rm fm)^{-3}
\end{equation}
assumes a conventional QCD value, the net baryonic flux $\Delta\Phi \cdot S $ through surface $S$  is numerically suppressed, and   can be  neglected in eq. (\ref{flux}).

One can explain   this result as follows. Consider  a single oscillation of the domain wall  evolution. To be more specific, consider a   squeezing portion of this evolution when $R(t)$ decreases. During this process    the chemical potential (and the baryon charge density) locally grow as we discussed in section \ref{algebraic}.  The major portion of this growth is resulted from  the baryon charge which was already  bound to the domain wall, rather than from the  baryon charge 
 which enters  the system as a result of the baryonic flux transfer. 

On an  intuitive level   the dominance of the bound charges (accounted in eq. (\ref{flux})) in comparison with flux-contribution (neglected in eq. (\ref{flux})) can be explained using   pure geometrical arguments. 
Indeed, the chemical potential increases very fast as a result of rapid shrinking of the bubble
with speed $v\simeq c$. The corresponding  contraction    of a  bubble  leads to proportionally rapid  increase of the chemical potential on the domain wall. This happens   because 
the  baryon charges strongly bound to the wall, and   cannot leave the system due to the topological reasons as the boundary conditions effectively lock the charge to the macroscopically large domain wall.
As a result of this   evolution  the binding energy of a quark $\sim \mu$ increases when the bubble contracts. This process  represents a highly efficient mechanism
of very rapid growth of the chemical potential due to the domain wall dynamics. It is vey hard to achieve a similar efficiency with the  flux-contribution when probability for a reflection from the domain wall is typically much higher than probability for a transmission. Furthermore, a non-vanishing quark mass make suppression  even stronger $\sim \exp(-m/T)$.

To conclude: we do expect that an accounting for  the flux- contribution   modifies  our equations relating $\mu(t)$ and $R(t)$ as expressed by eqs. (\ref{flux}), (\ref{eq:6.B1}).
However, we do not expect that this modification may drastically change the basic qualitative  features  of eqs. (\ref{flux}), (\ref{eq:6.B1}) which have been heavily employed  in  this work. 

 \section{Formation of the nuggets: numerical analysis}\label{numerics} 
This appendix is devoted to exact numerical computation in contrast with    analytical qualitative  arguments presented  in section \ref{analysis}. The basic lesson of this Appendix is that a number of simplifications which have been made in section \ref{analysis}
are justified, at least, on a qualitative level. 

Before we proceed with numerical computations we want to make few comments on parameters entering the basic dynamical equation  (\ref{eq:6.R0}).
  In the previous sections, $\sigma$  was treated as  a constant in order to simplify the analysis. This approximation is justified 
  as long as a typical curvature of the domain wall is much smaller than the width of the domain wall, i.e. $R\gg m_a^{-1}$.
  This condition is only marginally  justified as a typical radius of the bubble is of order $m_a^{-1}$, which is the same order of magnitude as the width of the wall.   At the same time, the width of the QCD substructure of the domain wall (including the $\eta'$ substructure and the baryon substructure)  is very small in comparison with the curvature, and it does satisfy the criteria of a thin wall approximation as 
  $m^{-1}\ll R\sim m_a^{-1}$.  Precisely this QCD substructure plays a crucial role in our analysis in section \ref{baryons} where we studied 
  the ``local violation" of the baryon charge in the presence of the domain walls. The broad structure of the domain wall  due to the axion field with the width $m_a^{-1}$ does not play any role. However, precisely this structure determines the large  tension $\sigma\sim m_a^{-1}$ of the domain wall.

   We want to effectively account for this physics by assuming that $\sigma(R)$ effectively depends on the radius of the bubble $R$.   On the physical grounds we expect that  $\sigma (R)$ approaches its asymptotic vale at large $R$ when the domain wall is almost flat, $\sigma (R\rightarrow\infty)\rightarrow \sigma_0$, while $\sigma$ reduces its value   at  smaller $R$, and eventually vanishes at some cutoff $R_{\rm cut}$. A natural choice is $R_{\rm cut}\simeq0.24R_0$ which corresponds to large $\mu_{\rm cut}\lesssim500\rm MeV$ from (\ref{eq:6.B1}), when the chemical potential  assumes  its typical CS value.  To introduce such an infrared cutoff smoothly, it is convenient to parametrize $\sigma$ as follows
\begin{equation}
\label{eq:B.sigma}
\sigma(R)=\sigma_0 e^{-r_0/2(R-R_{\rm cut})}
\end{equation}
where $\sigma_0\simeq9f_a^2m_a$  is the conventional domain wall tension, see e.g. \cite{Sikivie:2008}, while $r_0$ is a free phenomenological  parameter,   $0<r_0\lesssim R_0$ as we expect $\sigma(R_0)\simeq\sigma_0$. In our numerical studies  we verified that the physical   results (such as formation size $R_{\rm form}$) are not very sensitive to parameter $r_0$. 

Another parameter which requires some comments is the viscosity $\eta$. In the context of the present work, the viscosity accounts for a number of different QCD effects which lead to dissipation and ``friction". Such effects include, but not limited to  
   different scattering process by quarks, gluons or Nambu Goldstone Bosons in different phases.  Furthermore, the annihilation processes which take place inside the bubble and which result in production of a large number of strongly interacting quasi-particles also contribute to $\eta$. The viscosity can be computed from the first principles in weakly coupled quark-gluon phase \cite{Arnold:2000dr}.
   However, we are more interested in behaviour of $\eta$ below $T_c$. In this case the computations \cite{Chen:2007} based on chiral perturbation theory suggest that    $\eta\sim m_{\pi}^3$. This numerical value  is quite reasonable in all respects, and  consistent with simple  dimensional arguments.  It is also known that $\eta(T)$ depends on temperature  \cite{Chen:2007}. However,  we neglect this dependence in our estimates which follow. 

Now we can proceed with our numerical studies. 
   Since $\sigma(R)$ is a function of $R$ as explained above, we should start with a modified differential  equation for $R(t)$:
\be
\label{eq:B.Bsigma}
\sigma (R) \ddot{R}(t)
&=&-\frac{2\sigma(R)}{R}-\frac{\sigma(R) \dot{R}^2}{R}+\Delta P(R) \\
&-&(\frac{1}{2}\dot{R}^2+1)\frac{d\sigma(R)}{dR}-4\eta\frac{\dot{R}}{R}.\nonumber
\ee
This equation is not identically the same as   equation (\ref{eq:6.R0}) discussed in section \ref{analysis}. This is due to the fact that  the tension $\sigma(R)$ is now become  $R$ dependent function  as we discussed above.  
The equation (\ref{eq:B.Bsigma}) has been solved numerically using   parameters  listed in Table \ref{table}. The numerical values of these parameters can be  obviously  somewhat modified. However, the basic qualitative features presented in section \ref{analysis} do not drastically change when  the QCD parameters are varied  within reasonable parametrical region.  Our  numerical studies, as we discuss below,  do support the analytical qualitative results presented in section \ref{analysis}. 

\begin{figure}
\centering
\includegraphics[width=0.8\linewidth]{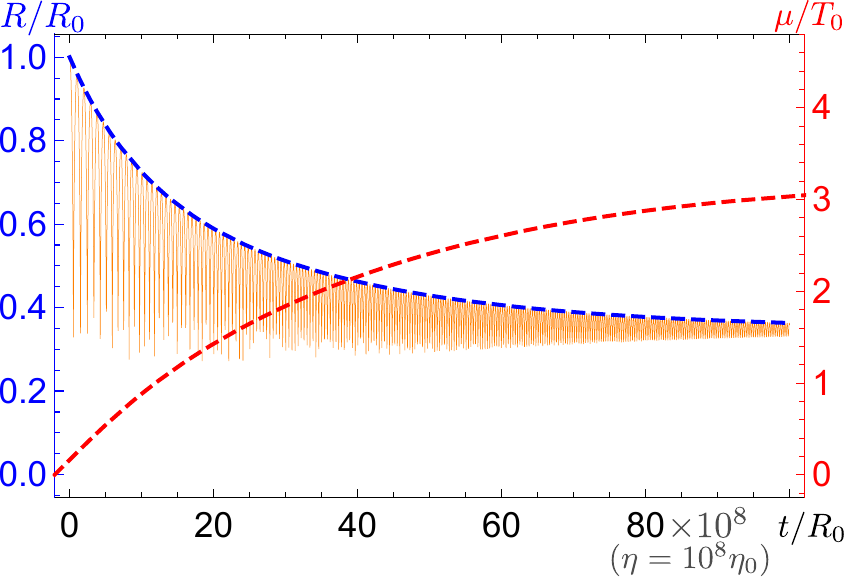}
\caption{Typical underdamped solution of $R(t)$ and $\mu(t)$. The oscillations with frequencies $\sim m_a^{-1}$ shown in orange, the modulation of $R(t)$ is shown in blue. The chemical potential $\mu(t)$ shown in red. The initial  $R_0=10^{11}\rm fm$ and $r_0=0.5R_0$. }
\label{fig:underdamp}
\end{figure}

We start our short description with Fig.\ref{fig:underdamp}. It shows a typical evolution of  a bubble with time. 
The frequencies of oscillations are determined by the axion mass $m_a^{-1}$, while typical damping time is determined by parameter $\tau$ as discussed in section \ref{analysis}. To make the pattern of oscillations visible, the viscosity has been rescaled\footnote{\label{footnote:eta}In this plot we use $\eta\simeq 10^{8}\eta_0$, which is eight orders of magnitude larger than  $\eta_0\simeq m_{\pi}^3$.  We did it on purpose:  First, it  simplifies the numerics.  Indeed, the  $\eta$ parameter  determines the dumping time scale (\ref{tau}) which  is  many orders of magnitude  longer than any other scales of the problem. Secondly, we use $\eta\simeq 10^{8}\eta_0$ for the demonstration purposes. Indeed,  a typical  oscillation time $\omega^{-1}$ and the damping time $\tau$  are characterized by drastically different scales.   If we take $\eta$ according to its proper QCD value than the  time scale on plots Fig. \ref{fig:underdamp} would be eight orders of magnitude longer than  shown. }. At large times, $t\rightarrow\infty$, the solution settles at $R_0/R_{\rm form}\simeq2.9$ and  $\mu_{\rm form}\simeq330{\rm MeV}\sim\mu_1$, consistent with qualitative analysis  of section \ref{analysis}.

\begin{figure}
\centering
\includegraphics[width=0.8\linewidth]{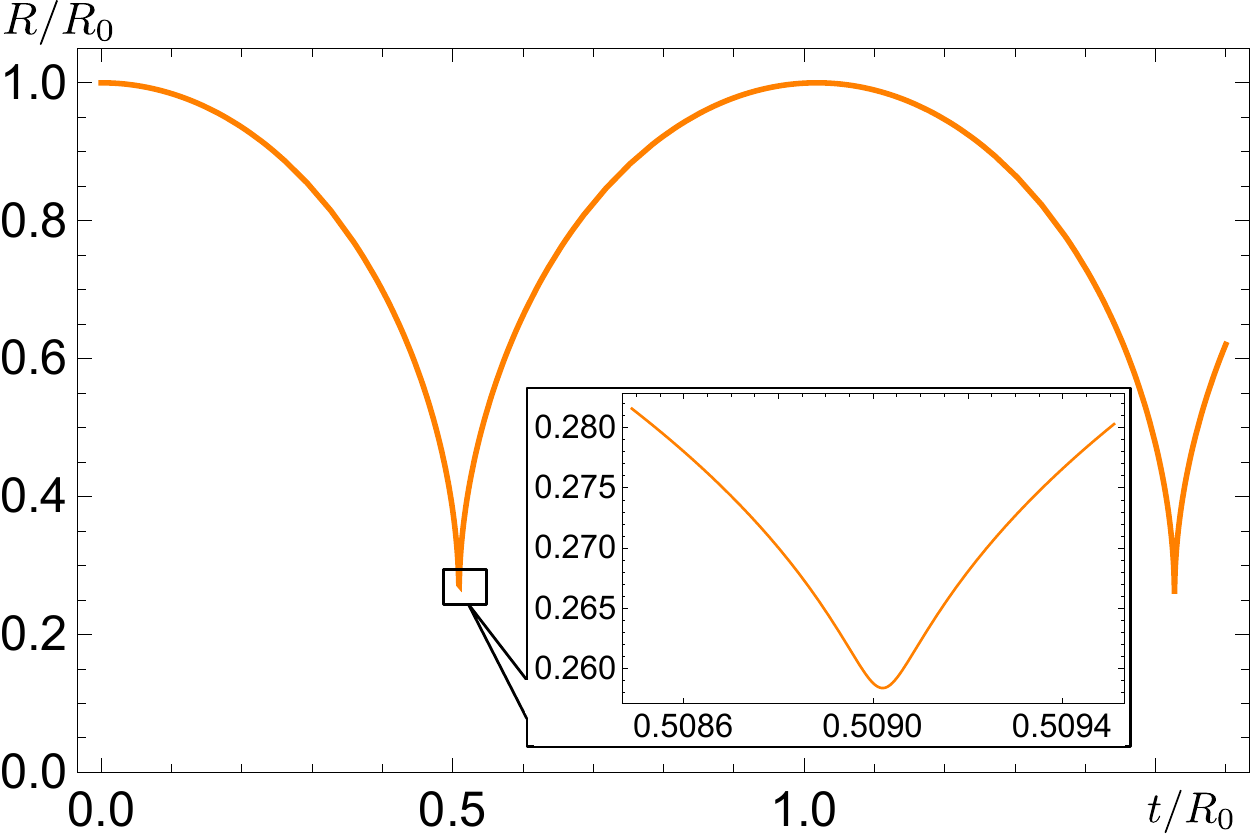}
\caption{The first few oscillations of an underdamped solution  shown on Fig. \ref{fig:underdamp}. }
\label{fig:cusp}
\end{figure}

We now describe Fig. \ref{fig:cusp} where we zoom-in  first few oscillations of a typical   solution shown on previous plot Fig. \ref{fig:underdamp}.
We want to emphasize that a  seeming   cusp singularity is actually a smooth   function near $R_{\rm min}$. It looks ``cuspy" as a result of a large time scale on Fig. \ref{fig:underdamp}. The ``cusp'' is relatively narrow comparing to macroscopic period of oscillation  ($\delta t_{\rm cusp}\sim10^{-3}R_0$).  Nevertheless it  actually  lasts  much longer in comparison with  a typical  QCD scale ($\delta t_{\rm cusp}\gg \Lambda_{\rm QCD}^{-1}$).

\begin{figure}
\centering
\includegraphics[width=0.8\linewidth]{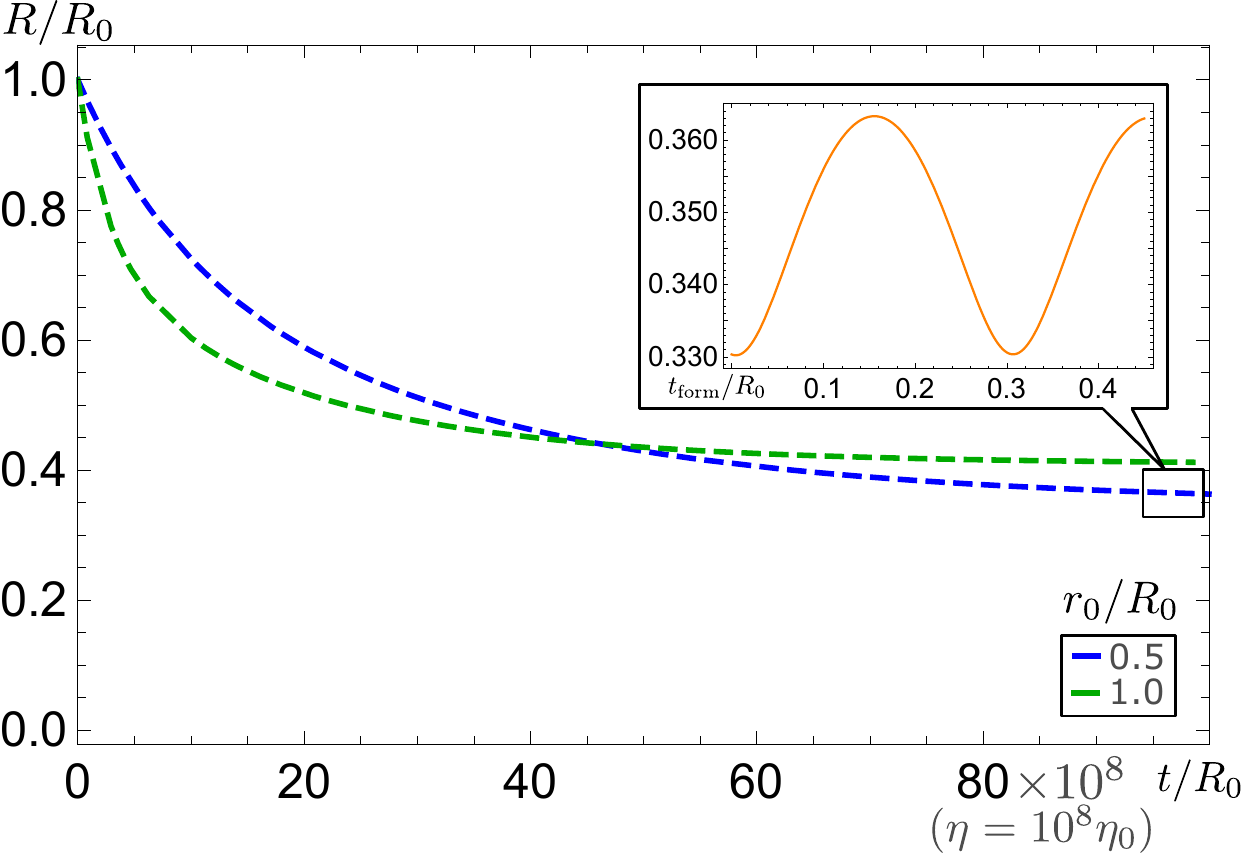}
\caption{Dependence on  parameter $r_0$ as defined by eq. (\ref{eq:B.sigma}). Zoom-in shows small oscillations during the final stage of formation.   }
\label{fig:Rs}
\end{figure}

On Fig. \ref{fig:Rs} we demonstrate   a (non)sensitivity of the system to parameter $r_0$ introduced in eq. (\ref{eq:B.sigma}).   
One can explicitly see that the  initial evolution is indeed quite sensitive to ad hoc parameter $r_0$. However,   the final stage of the evolution is not sensitive to $r_0$. In other words, the physical parameters   $R_{\rm form}$ and $\tau$ are not sensitive to ad hoc parameter $r_0$.
 Note that estimation of damping time $\tau$ and period of oscillation $t_{\rm osc}$ agree well with qualitative  estimations presented in  section \ref{analysis}.

\exclude{
\begin{figure}
\centering
\includegraphics[width=0.8\linewidth]{overdamp}
\caption{Typical over-damped solution of $R(t)$ (orange, blue) and $\mu(t)$ (red) with $R_0=10^{11}$fm and $r_0=10R_0$. Note that $r_0$ is taken out of the legitimate region $0<r_0\lesssim R_0$. The solution settles to equilibrium very fast ($10^7$ times faster than the typical underdamp) after limited oscillations. Near equilibrium, $R_0/R_{\rm form}\simeq2.2<\sqrt{14}$ and $\mu_{\rm form}\simeq220{\rm MeV}<\mu_1$. Thus, over-damped solutions may not even be able to form CS nuggets. 
}
\label{fig:overdamp}
\end{figure}
}

\begin{table}[]
\centering
\caption{Table for some numerical parameters}
\label{table}
\begin{tabular}{lccc}
\hline\hline
Quantity & Symbol & Value & \begin{tabular}[c]{@{}c@{}}QCD units\\ ($m_\pi=1$)\end{tabular} \\ \hline
flavours & $N_f$ & 2 & 2 \\
colors & $N_c$ & 3 & 3 \\
degeneracy factor (in) (\ref{P_fermi})& $g^{\rm in} $ & 12 & 12 \\
degeneracy factor (out) (\ref{P_out})& $g^{\rm out}$ & 37 & 37 \\
baryon charge on DW (\ref{N}) & $N$ & 2 & 2 \\
axion decay constant & $f_a$ & $10^{10}$GeV & $7\times10^{10}$ \\
mass of axion & $m_a$ & $6\times10^{-4}$eV & $~~4\times10^{-12}$ \\
domain wall tension & $\sigma_0$ & $5\times10^8 \rm{GeV}^3$ & $2\times10^{11}$ \\
bag constant (\ref{P_bag}) & $E_B$ & $(150 \rm{MeV})^4$ & $1.5$ \\
``squeezer'' parameter (\ref{P_bag}) & $\mu_1$ & 330 MeV & 2.4 \\
cosmological time scale & $t_0$ & $10^{-4}$s & $10^{19}$ \\
initial $\mu$ & $\mu_0$ & 1.35 MeV & 0.01 \\
initial radius & $R_0$ & $10^{-2}$cm & $10^{11}$ \\
initial temperature & $T_0$ & 100 MeV & 0.74 \\
QCD viscosity\cite{Chen:2007} & $\eta_0$ & 0.002 GeV$^3$ & 1 \\ \hline\hline
\end{tabular}
\end{table}

\section{Evaluation of Fermi-Dirac integrals}
\label{integral}
\begin{figure}
\centering
\includegraphics[width=1\linewidth]{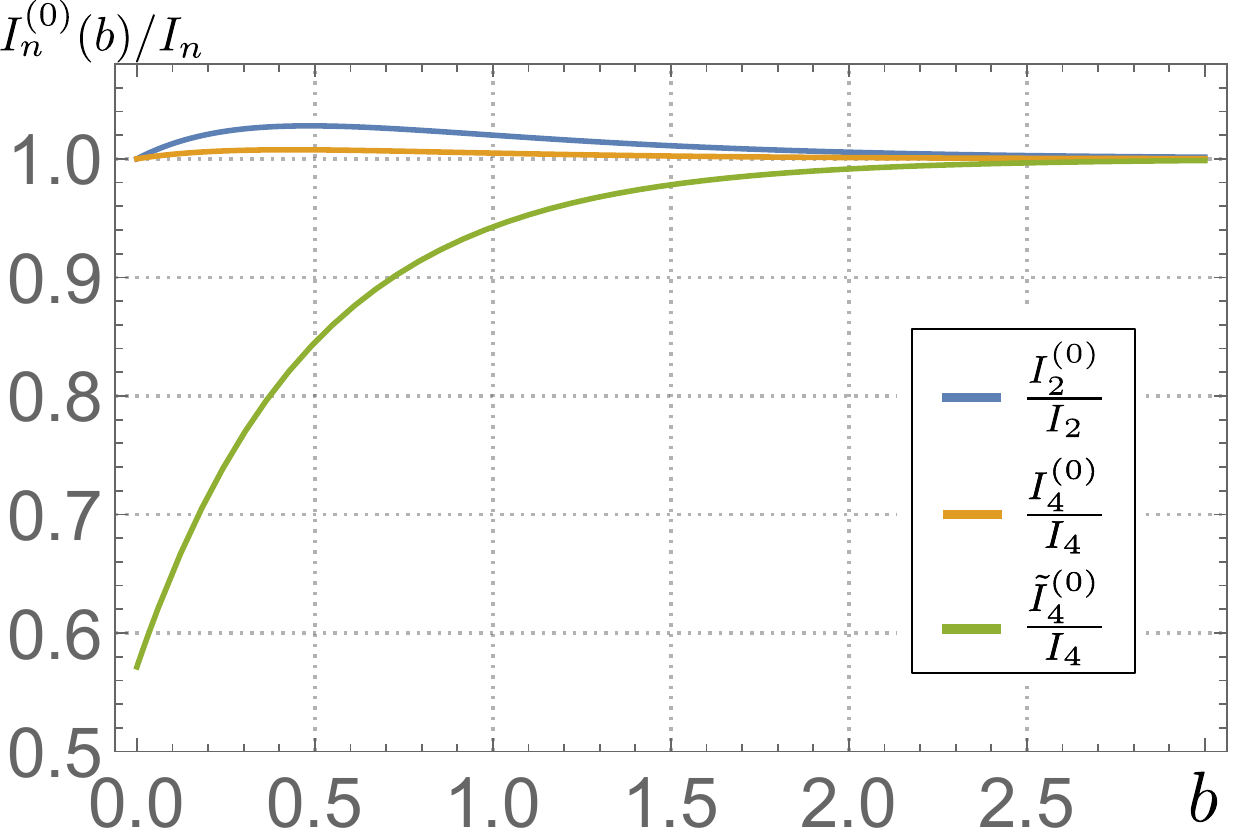}
\caption{Comparison of $I_n^{(0)}$ to $I_n$ with different values of $b$.  
}
\label{fig:Integral}
\end{figure}

The main goal of this Appendix is to present some supporting arguments suggesting  that the approximation we have used   
in section \ref{algebraic} and which involves the Fermi-Dirac integrals  are qualitatively justified. Indeed, the relevant integrals which enter 
 eqs. (\ref{eq:6.integral}), (\ref{P_fermi1}) have the form
\begin{equation}
I_n(b)\equiv\int_0^\infty\frac{dx\cdot x^{n-1}}{e^{x-b}+1}, \quad b=\frac{\mu}{T}>0, 
\end{equation}
where $n=2$ appears in integral  (\ref{eq:6.integral}), while  $n=4$ appears in (\ref{P_fermi1}). We will hence focus on evaluation of $I_2$ and $I_4$ in this appendix. They can be exactly evaluated as 
\begin{subequations}
\label{eq:C.I}
\begin{equation}
\label{eq:C.I2}
I_2(b)=\frac{\pi^2}{6}+\frac{1}{2}b^2+\operatorname{Li}_2(-e^{-b})
\end{equation}
\begin{equation}
\label{eq:C.I4}
I_4(b)=\frac{7\pi^4}{60}+\frac{\pi^2}{2}b^2+\frac{1}{4}b^4
+6\operatorname{Li}_4(-e^{-b}) 
\end{equation}
\end{subequations}
where $\operatorname{Li}_2(-z)$ and $\operatorname{Li}_4(-z)$ are the polylogarithm functions of order 2 and 4, respectively. Polylogarithm functions are commonly known to represent  the Fermi-Dirac and Bose-Eisterin integrals.  The Polylogarithm functions are  defined as 
\begin{equation}
\operatorname{Li}_n(-z)=\sum\limits_{k=1}^{\infty}\frac{(-1)^k}{k^n}z^k.
\end{equation}
Note that $\left|z\right|=e^{-b}\leq1$ for any positive $b$. In this case $\operatorname{Li}_n(-z)$ is evidently fast-converging, so that we can efficiently estimate it by extracting the leading exponent $e^{-b}$ then using the Taylor expansion for  the remaining piece: 
\begin{subequations}
\begin{equation}
\operatorname{Li}_2(-e^{-b})\simeq e^{-b}\left[
-\frac{\pi^2}{12}+(\ln2-\frac{\pi^2}{12}) b+\mathcal{O}(b^2)\right] \nonumber
\end{equation}
\begin{equation}
\operatorname{Li}_4(-e^{-b})\simeq e^{-b}\left[
-\frac{\pi^4}{720}+(\frac{3\zeta(3)}{4}-\frac{7\pi^4}{720})b+\mathcal{O}(b^2)\right], \nonumber
\end{equation}
\end{subequations}
where $\zeta(3)\simeq1.202$ is the Riemann zeta function. Neglecting the terms  of order $ \mathcal{O}(be^{-b}) $ which are small in both limits, at large and small chemical potentials, one can approximate $I_2$ and $I_4$  as follows 
\begin{subequations}
\label{eq:C.I0}
\begin{equation}
\label{eq:C.I20}
I_2^{(0)}\simeq\frac{\pi^2}{6}+\frac{1}{2}b^2-\frac{\pi^2}{12}e^{-b}
+\mathcal{O}(be^{-b})
\end{equation}
\begin{equation}
\label{eq:C.I40}
I_4^{(0)}\simeq\frac{7\pi^4}{60}+\frac{\pi^2}{2}b^2+\frac{1}{4}b^4
-\frac{7\pi^4}{120}e^{-b}+\mathcal{O}(be^{-b}).
\end{equation}
\end{subequations}
We test our approximation by comparing our approximate expressions (\ref{eq:C.I20}),  (\ref{eq:C.I40}) with exact formulae 
(\ref{eq:C.I2}), (\ref{eq:C.I4}). As one can see from Fig. \ref{fig:Integral}, our  approximation shown in blue ($I_2^{(0)}/I_2$)  and orange ($I_4^{(0)}/I_4$)  is  very good  with errors less than 3\% in the entire range of $b>0$. 

On the same plot we also show 
the approximation  $\tilde{I}_4^{(0)}$ for approximate expression $I_4^{(0)}$ used in the main text in eq. (\ref{P_fermi1}) 
\begin{equation}
\tilde{I}_4^{(0)}\simeq\frac{7\pi^4}{60}+\frac{\pi^2}{2}b^2+\frac{1}{4}b^4
-\frac{\pi^4}{12}e^{-b}.
\end{equation}
The error for $\tilde{I}_4^{(0)}$   is quite large for very small chemical potential $b\ll 0.5$, on the level of 40\%, shown in green. 
The error becomes much   smaller after  short period of time  when  $b={\mu}/T\geq 0.5$ becomes  sufficiently large.   
To conclude: the approximations of the integrals in section \ref{algebraic} are sufficiently good  for qualitative analysis 
presented in that section.

\end{document}